\documentclass[twocolumn, superscriptaddress, secnumarabic, amssymb, nobibnotes, nofootinbib, aps, prd]{revtex4}
\usepackage{amsmath}
\usepackage{amsfonts}
\usepackage{amssymb}
\usepackage{comment}
\usepackage{float}
\usepackage{caption}
\usepackage{subcaption}
\usepackage{graphicx}
\usepackage{appendix}
\usepackage{color}
\usepackage{tikz}
\usepackage{soul}
\usepackage{cases}
\usepackage{ulem}
\usepackage{diagbox}
\usepackage{array}
\usetikzlibrary{quotes, angles}
\usepackage[
unicode=true,
bookmarks=false,
breaklinks=false,
pdfborder={0 0 1},
backref=false,
colorlinks=true
]{hyperref}
\setstcolor{blue}

\begin{document}
	
\title{Magnetic levitation and spatial superposition of a nanodiamond with a current-carrying chip}

\date{\today}
\newcommand{\affone}{Canadian Institute for Theoretical Astrophysics, University of Toronto, 60 St George St, Toronto, ON M5S 3H8, Canada}
\newcommand{\afftwo}{Department of Physics and Astronomy, University College London, Gower Street, WC1E 6BT London, United Kingdom.}
\newcommand{\affthree}{Van Swinderen Institute, University of Groningen, 9747 AG Groningen, The Netherlands.}
\newcommand{\afffour}{Center for Fundamental Physics, Department of Physics and Astronomy, Northwestern University, 2145 Sheridan Road, Evanston, IL, USA}

\author{Qian Xiang}
\affiliation{\affthree}
\author{Shafaq Gulzar Elahi}
\affiliation{\afffour}
\author{Andrew Geraci}
\affiliation{\afffour}
\author{Sougato Bose}
\affiliation{\afftwo}
\author{Anupam Mazumdar}
\affiliation{\affthree}
\affiliation{\affone}

\begin{abstract}
We propose a current-carrying-chip scheme for generating spatial quantum superpositions using a levitating nanodiamond with a built-in nitrogen-vacancy (NV) centre defect. Our setup is versatile, for a mass range of $10^{-19}~{\rm kg}\leq m\leq 10^{-17}~{\rm kg}$ a superposition size of range ${\cal O}(100){\rm nm} < \Delta x < {\cal O}(10) {\rm  \mu m}$ is achievable in $t\leq 0.1$s, depending on the position we launch from the centre of the diamagnetic trap. Using numerical simulations, we provide an in-depth analysis of the chip that can support stable trapping of the particle in all three dimensions as well as create a spatial superposition along the $x$-axis, while producing a very tight trap in the transverse directions. This setup enables the creation of a Schr\"odinger cat state with a diamagnetically levitated nanoparticle in a table-top experiment. Although the mass and superposition size considered here remain below those required for a QGEM test, the proposed device provides a building block toward future QGEM experiments.

\end{abstract}

\maketitle	
\section{Introduction} 

The quest to understand the quantum nature of gravity in a lab is a monumental challenge at the interface of theoretical physics and experimental science \cite {Bose:2017nin}, see also~\cite{marletto2017gravitationally}. There are beliefs that gravity could be entirely classical, or that the wavefunction of a macroscopic object collapses due to gravity or other Standard Model interactions; see ~\cite{Penrose:1996cv,PhysRevA.40.1165,Pearle:1988uh,Bassi:2012bg,PhysRevLett.110.160403}. However, a promising avenue now exists to test such a hypothesis, which is known as the 
Quantum Gravity-induced Entanglement of Masses (QGEM) protocol\cite{Bose:2017nin,Marshman:2019sne,Bose:2022uxe}. 
The heart of the protocol lies with the fact that any quantum interactions will lead to entanglement, and if gravitational interactions with matter are quantum, then it must yield quantum entanglement between the two quantum objects mediated via a graviton exchange~\cite{Bose:2017nin,Marshman:2019sne,Bose:2022uxe}, see also~\cite{Christodoulou:2022vte,Carney_2019,Danielson:2021egj}. The QGEM protocol can test alternative theories of gravity that have infinite derivatives and are ghost-free~\cite{Vinckers:2023grv}, massive graviton~\cite{Elahi:2023ozf}, test the quantum weak equivalence principle~\cite{Bose:2022czr,Chakraborty:2023kel}, post-Newtonian theories of gravity~\cite{Toros:2024ozf}, fifth force~\cite{Barker:2022mdz}, and confirm the spin-2 nature of the massless graviton in the analogue light-bending due to gravity experiment via witnessing the entanglement between matter-and-photon~\cite{Biswas:2022qto}.

However, the biggest challenge is to perform an experiment that creates a macroscopic quantum superposition in a lab, see~\cite{Bose:2017nin}. To test QGEM, we will need to create a superposition of order $\Delta x\sim {\cal O}(10-50){\rm \mu m}$ for masses $m\sim 10^{-15}-10^{-14}$kg depending on the various rates of decoherence~\cite{Bose:2017nin,Tilly:2021qef,vandeKamp:2020rqh} and the witness, see~\cite{Chevalier:2020uvv,
Schut:2023hsy,Schut:2025blz}.

This paper aims to develop a dedicated chip-based matter-wave interferometer for a diamagnetic nanoparticle, such as a nanodiamond containing a single nitrogen-vacancy (NV) centre, as envisaged in \cite{DUrso16_GM,Wan16_GM,Scala13_GM, Bose:2017nin, marshman2020mesoscopic,Marshman:2021wyk,pedernales2020motional,Zhou:2022frl,Zhou:2022jug,Zhou:2022epb,Zhou:2024voj}, but now with on-chip superposition. All these protocols exploit creating spatial superposition via the Stern-Gerlach interferometer (SGI)~\cite{Folman2013,Folman2018,folman2019,Margalit:2020qcy, Keil_2021}, which utilises a magnetic field gradient to couple the particle's internal spin state to its centre-of-mass (c.o.m.) motional degrees of freedom. This coupling drives the particle's c.o.m. into a spatial superposition along a specific direction, effectively ``splitting'' its wavefunction. Current matter-wave interferometers have successfully implemented SGI configuration on atom chip protocols, see~\cite{Folman2013,folman2019,Margalit:2020qcy}. Following the methodology of atom chips, see~\cite{Folman:2000}, a scheme for creating spatial superpositions of more massive objects ($\sim 10^{6}$ carbon atoms) has been proposed in \cite{Margalit:2020qcy}. The pursuit of the next generation of SGI for higher mass is currently under extensive theoretical and numerical investigation, see~\cite{Wan16_GM,Scala13_GM, Bose:2017nin, marshman2020mesoscopic,Marshman:2021wyk,pedernales2020motional,Zhou:2022frl,Zhou:2022jug,Zhou:2022epb,Zhou:2024voj}.

As proposed in Ref.~\cite{Schut:2023hsy}, in the context of the QGEM protocol, a nanodiamond should not only be levitated and brought into a spatial superposition, but its transverse motion should also be tightly confined. More recently, Ref.~\cite{Elahi:2024dbb} proposed a Z-wire-based chip design that provides diamagnetic levitation, transverse confinement, and cooling of the motional state, while leaving a long and clean trapping region along the intended superposition axis.
Similar to the preceding proposal, the present work considers a current-carrying chip design that simultaneously provides diamagnetic levitation of the nanodiamond, confinement of its transverse motion, and, most importantly, the generation of a spatial Schr\"odinger cat state along the superposition axis. We will need another device to cool the nanoparticle. However, our setup may be useful in hybrid scenarios, where cooling can be performed in an ion trap (e.g., a Paul trap~\cite{Gupta:2025egh,dania2024ultrahigh,dania2021optical,dania2022position}), while in a diamagnetic magnetic trap, we create a spatial superposition. Such a paradigm is feasible and has recently been discussed in the literature, see~\cite{Elahi:2024dbb,Muretova:2025gnn}.

Our proposed setup comprises two integral components. The levitation assembly consists of four wires parallel to the $x$-axis, generating a magnetic quadrupole field ~\cite{perez2013does,reichel2011atom} in the $y-z$ plane to levitate the particle in the $z$-axis (which is the direction of Earth's gravity). While the separation assembly, which creates the spatial superposition, consists of four long wires parallel to the $z$-axis, it also produces another quadrupole field in the $x-y$ plane. Together, the two assemblies can confine the particle's $y$-motion throughout the experiment. These wires are grouped into pairs and printed on four separate chips, see Fig.~\ref{spdevice}.

A key advantage of this configuration is that it creates approximately harmonic potential wells in all three spatial directions. Consequently, during the creation of the spatial superposition, the wavepacket remains confined in the directions orthogonal to the superposition axis, while the c.o.m. motion along each interferometric arm remains well controlled, a feature crucial for maintaining interferometric visibility. Furthermore, with our chosen parameters, the particle's three-dimensional motional modes within the device are nearly decoupled. This enables precise control over the superposition dynamics, allowing us to tailor the interferometer sequence along the desired axis ($x$) while maintaining stability in the $y$ and $z$ directions. However, note that we assume the nanodiamond to be spherical and assume that the NV centre is located near the centre of the sphere.
We note that the rotational dynamics of the nanodiamond are important and have been studied in the context of SGI interferometry~\cite{Japha:2022phw,Rizaldy:2024viw}. Within the scope of the present study, however, we will explore the form that such an SGI setup should take and therefore focus exclusively on the translational centre-of-mass motion of the particle. The rotational degrees of freedom will be incorporated in a subsequent study addressing both translational and rotational dynamics. We also note that, within the constraints of our chip design, extremely large spatial superpositions are not achievable for heavier particles. Under the chosen parameters, superposition sizes of $\Delta x\sim{\cal O}(10$--$20)\mu\mathrm{m}$ for $m=10^{-19}\mathrm{kg}$ and $\Delta x\sim{\cal O}(100)\mathrm{nm}$ for $m=10^{-17}\mathrm{kg}$ can be generated for a one-loop interferometer within $t\sim 0.1\mathrm{s}$. In addition, we investigate both the decoherence induced by fluctuations in the device parameters, particularly the currents, and the Humpty--Dumpty effect~\cite{Englert1988-ENGISC,schwinger1988spin} associated with imperfect phase-space closure in a single full-loop operation.

This paper is structured as follows. In Sec.~\ref{modelanddevice}, we introduce the magnetic-field configuration required for levitation, transverse confinement, and spatial splitting. In Sec.~\ref{levitation}, we present the corresponding current-carrying-chip implementation and discuss its operating parameters and feasibility. In Sec.~\ref{seperation}, we describe the three-stage interferometric protocol and analyse dephasing induced by separation-current and bias-field fluctuations, as well as the recombination contrast under finite current switching, finite longitudinal motional occupation, and independent stage-dependent current deviations. Additional details of the magnetic-field calculations, superconducting critical current density, surface forces, spin-state mixing, trajectory dynamics, Gaussian covariance evolution, and motional heating are provided in the Appendices.


\section{Model}\label{modelanddevice}

We consider using a nanoscale diamond embedded with a nitrogen-vacancy (NV) centre spin. When it interacts with the magnetic field in space, the Hamiltonian is given by~\cite{pedernales2020motional,Marshman:2021wyk,Zhou:2022frl},
\begin{equation}
    \begin{aligned}
         H=\frac{p_x^2+p_y^2+p_z^2}{2m}-\frac{\chi_\rho m}{2\mu_0}\mathbf{B}^2+\hbar \gamma_e \mathbf{\hat{S}}\cdot \mathbf{B}+\hbar D \hat{S}_x^2+m \mathbf{g} \mathbf{z}\label{3dHamiltonian}
    \end{aligned}
\end{equation}
where the first term is the kinematic energy of the diamond whose mass is $m$. The second term represents the diamagnetic property of the nanodiamond, where $\chi_\rho\approx-6.2\times 10^{-9}~\text{m}^3/\text{kg}$~\cite{PhysRevB.49.15122}, and $\mu_0= 4\pi\times 10^{-7}$ T$\cdot$m/A \cite{tiesinga2021codata} are the mass magnetic susceptibility and the vacuum permeability, respectively. The third term is the key to achieving the separation of the nanodiamond wavepacket. It describes the interaction between the NV-centre spin and the external magnetic field, where $\gamma_e=2\pi\times28~\mathrm{GHz/T}$ \cite{tiesinga2021codata} is the electronic gyromagnetic ratio and $\hbar\approx 1.05\times 10^{-34}$ J$\cdot$s is the reduced Planck constant. $\hbar D \hat{S}_x^2$ is the NV zero-field splitting term, where $D=2\pi\times 2.87$ GHz.
The last term represents the gravitational potential energy of the levitated nanodiamond, where the gravitational acceleration constant $g\approx 9.8$ m/$\text{s}^2$ is taken. 

\begin{figure*}[t!]
	\begin{subfigure}[]{0.85\textwidth}
		\includegraphics[width=\textwidth]{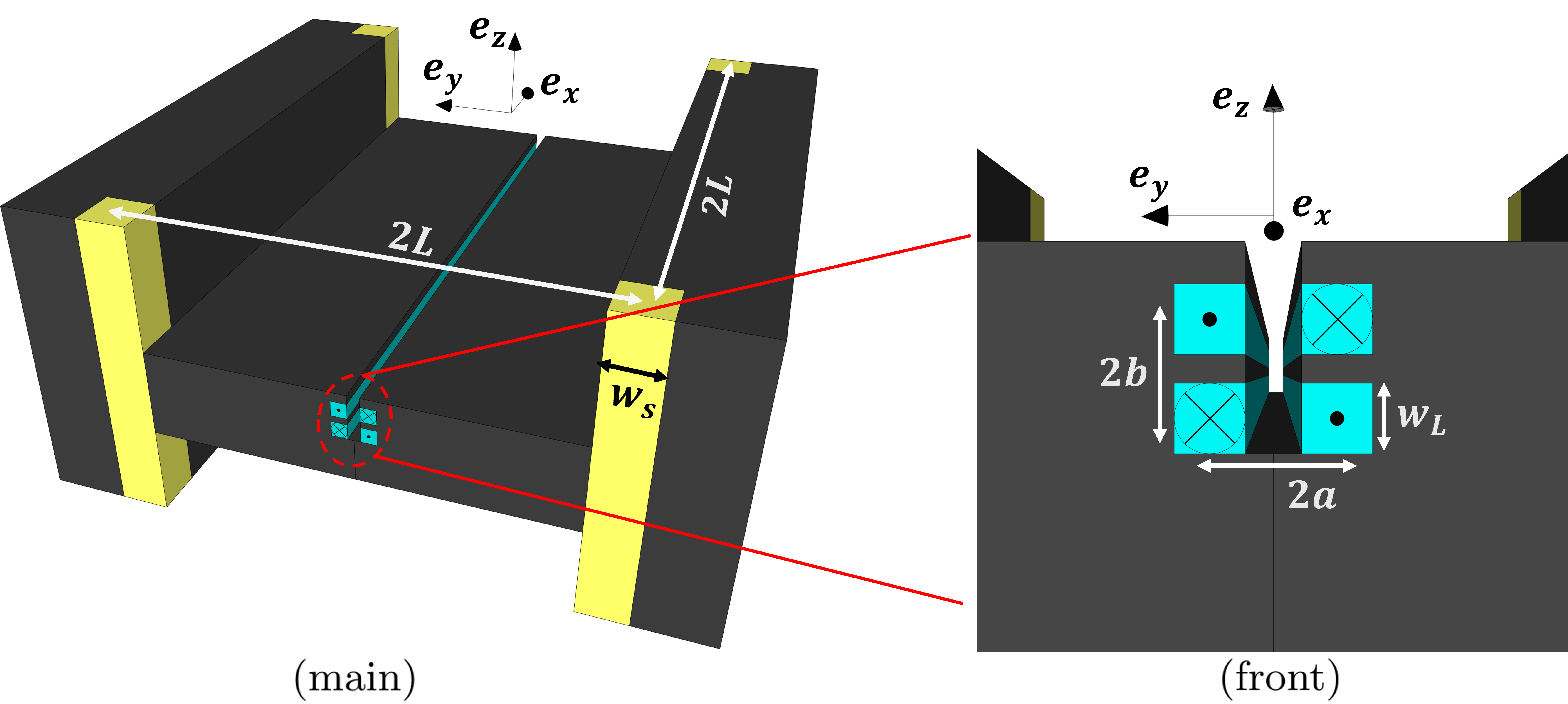} 
	\end{subfigure}\\
    \begin{subfigure}[]{.45\textwidth}
		\includegraphics[width=\textwidth]{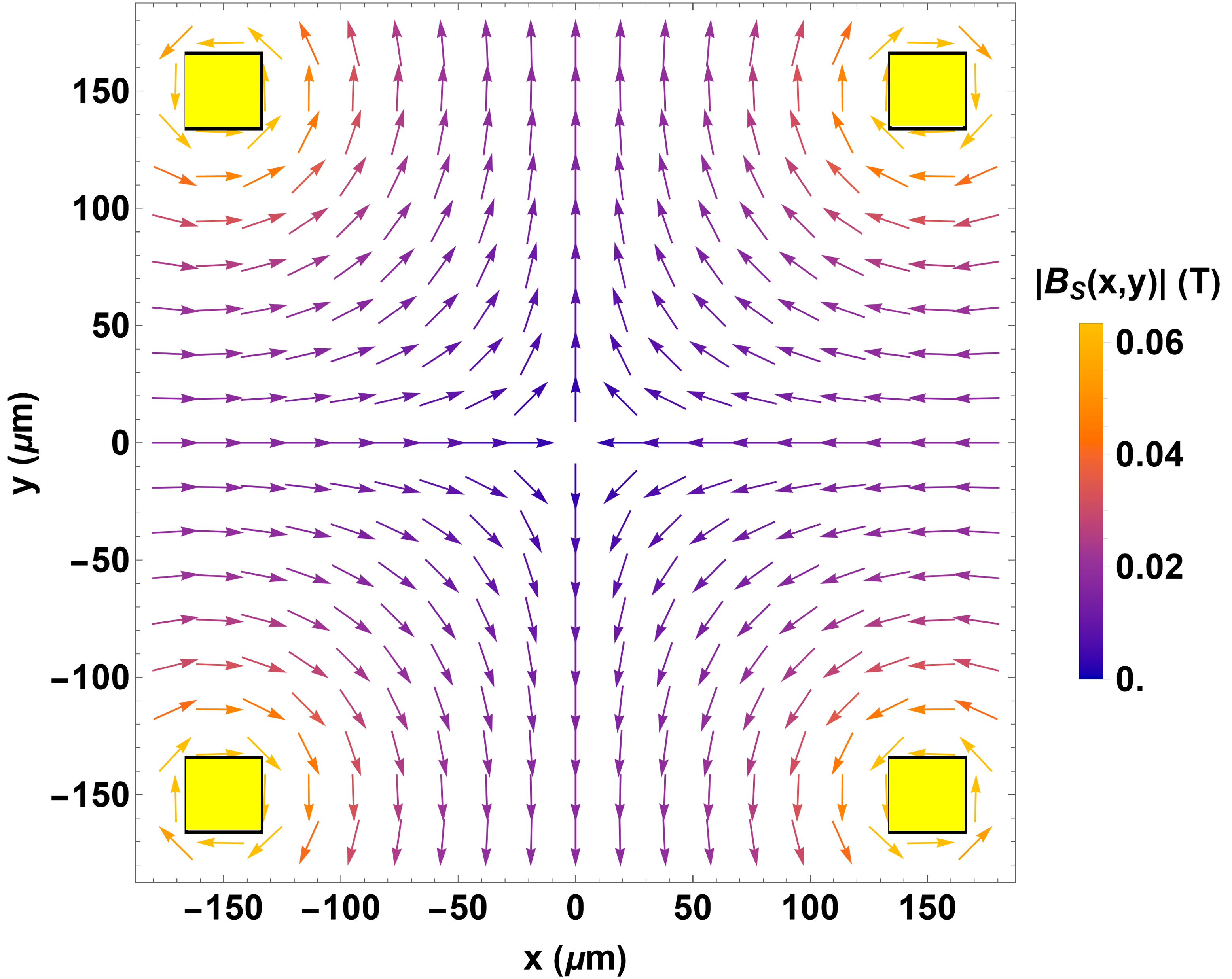} 
		\subcaption*{(a)}
	\end{subfigure}\hspace{0.2cm}
        \begin{subfigure}[]{.45\textwidth}
		\includegraphics[width=\textwidth]{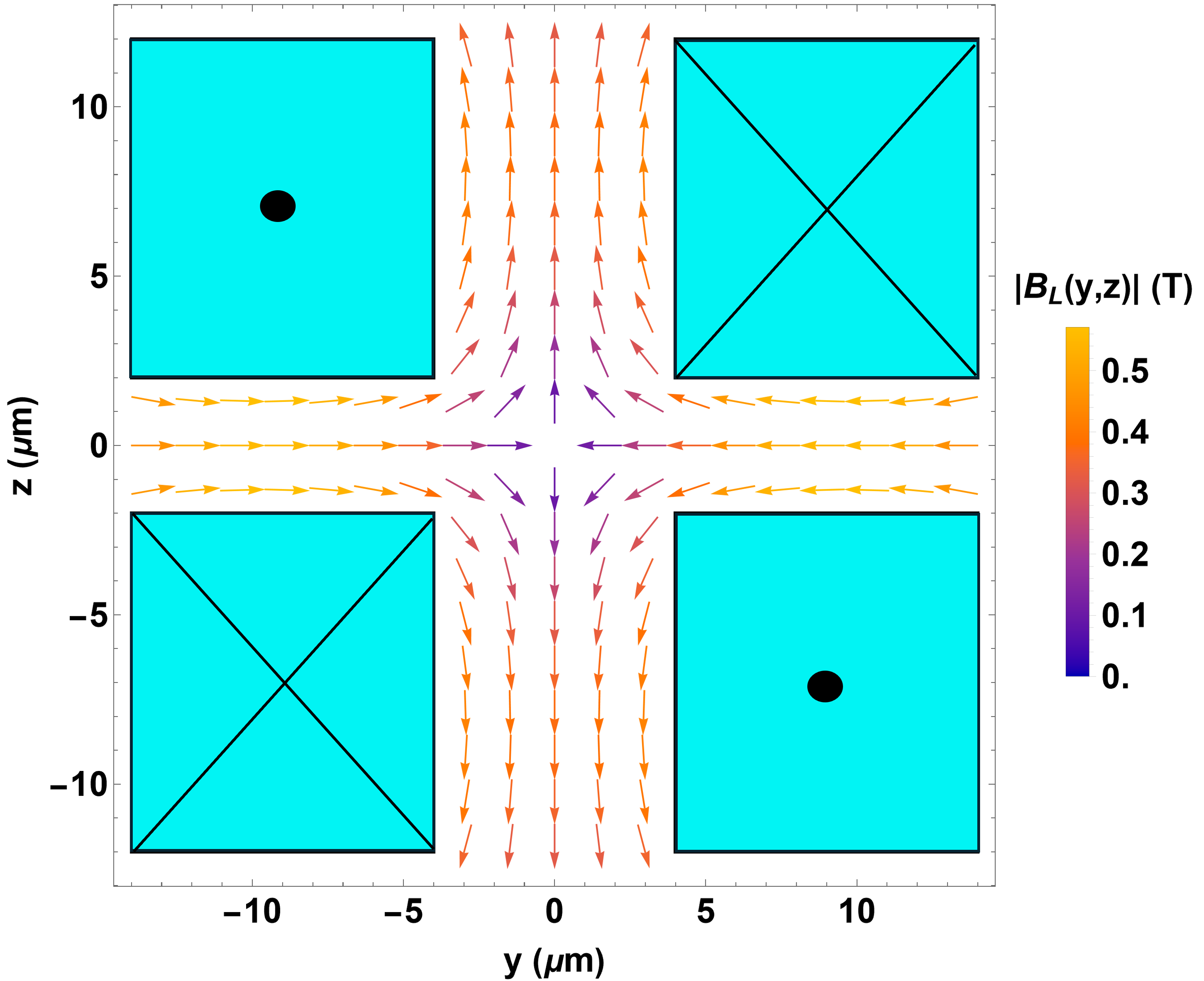} 
		\subcaption*{(b)}
	\end{subfigure}
	\caption{
The geometry of the I-Cat chip and representative magnetic-field profiles are shown above. The upper panels show the chip assembly in the main and front views. In both the main and front views, the wires parallel to the $z$ axis are considered to be gold wires, with both their cross-sectional width and thickness taken as $w_\text{S}=32~\mu\mathrm{m}$. Any two adjacent wires carry currents $|I_\text{S}|$ in opposite directions with spacing $2L=300~\mu$m, generating the gradient magnetic field $\mathbf{B}_\text{S}=\eta_\text{S}x\mathbf{e}_x-\eta_\text{S}y\mathbf{e}_y$.
As shown in subplot (a), this magnetic field is used to create the spatial superposition of the nanodiamond. These wires are therefore labelled as separation wires. By changing the combination of current directions in the separation wires, one can switch the sign of $\eta_\text{S}$ and thereby control the trajectory of the spatial superposition state. In the main and front views, the four blue wires parallel to the $x$ axis are primarily used for magnetic levitation. Their cross-sectional width and thickness are both taken as $w_\text{L}=10~\mu\mathrm{m}$, with spacings $2b=14~\mu\mathrm{m}$ and $2a=18~\mu\mathrm{m}$. The nanodiamond is loaded along the geometric central axis enclosed by the four blue wires. Owing to the proximity of the nanodiamond to the superconducting wires, the corresponding Casimir--Polder force and patch potential force are estimated in Appendix~\ref{cpforce} and \ref{patch}. Currents carried by these blue wires have the magnitude $|I_\text{L}|=14~\mathrm{A}$, whose directions are indicated by dots and crosses. Throughout the experiment, currents $|I_\text{L}|$ in the levitation wires are assumed to be constant, providing a stable magnetic-levitation field, $\mathbf{B}_\text{L}=\eta_\text{L}z\mathbf{e}_z-\eta_\text{L}y\mathbf{e}_y,$ as shown in subplot (b). Therefore, the entire device provides the total magnetic field
$\mathbf{B}= (\eta_\text{S}x + B_0)\mathbf{e}_x-(\eta_\text{L}+\eta_\text{S})y\mathbf{e}_y+\eta_\text{L}z\mathbf{e}_z ,$
where the bias field $B_0$ is supplied by external components (not shown in the figure). In the main and front views, the black structures are assumed to be made of silicon.
	} \label{spdevice}
\end{figure*}

We consider a tabletop scenario in which the diamond is magnetically levitated along the $z$ direction to counteract gravity and provide contact-free confinement. Such a levitated configuration may also reduce noise associated with mechanical supports, see Refs.~\cite{Toros:2020dbf,Wu:2024tcr,Wu:2024bzd,Wu:2025wwh}. To simultaneously realise the creation of a spatial superposition state and magnetic levitation, while avoiding interference between these two types of dynamics, one can consider the magnetic field $\mathbf{B}$ in Eq.~(\ref{3dHamiltonian}) as consisting of two parts,
\begin{align}
    \mathbf{B}_\text{L}(y,z) &= \eta_\text{L}z\mathbf{e}_z - \eta_\text{L} y\mathbf{e}_y, \label{Bfield2} \\
    \mathbf{B}_\text{S}(x,y) &= (\eta_\text{S}x + B_0)\mathbf{e}_x - \eta_\text{S} y\mathbf{e}_y. \label{Bfield3}
\end{align}
The total magnetic field is then,
\begin{equation}
	\begin{aligned}
    \mathbf{B} &= \mathbf{B}_\text{L} + \mathbf{B}_\text{S} \label{Bfield1} \\
    &= (\eta_\text{S}x + B_0)\mathbf{e}_x-(\eta_\text{L}+\eta_\text{S})y\mathbf{e}_y+\eta_\text{L}z\mathbf{e}_z.
    \end{aligned}
\end{equation}
Here, $B_0$ denotes a uniform bias field applied solely along the $x$-direction to ensure the $B_xS_x$ term dominates in the Zeeman interaction. This uniform bias magnetic field can be generated using devices such as Helmholtz coils~\cite{panofsky2012classical}, which is not depicted in Fig.~\ref{spdevice}. The parameters $\eta_\text{L}$ and $\eta_\text{S}$ represent magnetic field gradients (T/m). The value of $\eta_\text{S}$ assumes three distinct values during different stages to satisfy the trajectory closure requirement: $-\eta$, $\eta$, and $-\eta$, with $\eta>0$.  This part of the discussion is given in section \ref{seperation}.

The motion of the nanodiamond in the $z$- and $y$-directions will be governed mainly by the diamagnetic and gravitational terms
\begin{align}
-\dot{p}_i=\partial_{i} \left(-\frac{\chi_\rho m}{2\mu_0} \mathbf{B}^2+mgz\right),\qquad i=y,z.
\end{align}

Consequently, in the $y$-direction, the overlap of two linear magnetic field components naturally forms a harmonic trap, with the equation of motion,
\begin{align}
	&\frac{\text{d}^2y}{\text{d}t^2}+\omega^2_y~y=0,\label{ytrap}\\
	&\omega_y= (\eta_\text{L}+\eta_\text{S})\sqrt{\frac{-\chi_\rho}{\mu_0}}.\label{omegay}
\end{align}
If the particle is initialized at $y=0$ with zero velocity, the centre of its motion will remain at $y=0$. More generally, an initial Gaussian wavepacket will remain confined by the harmonic potential around $y=0$. As we shall see, the relevant parameters allow for a large trapping frequency $\omega_y$, and hence a tight transverse confinement of the wavepacket. This behaviour can also be directly inferred from the form of the magnetic field given by 
Eqs.~(\ref{Bfield2}, \ref{Bfield3}) where the minimum of the total magnetic field in the $y$-$z$ plane consistently lies along the axis at $y=0$, $z=0$. Consequently, the diamagnetic property of the particle will confine it to this region where the $y$-component of the magnetic field experienced by the particle remains zero at all times.

Similarly, in the $z$ direction, the equation of motion can be written as
\begin{align}
&-\frac{\text{d}^2z}{\text{d}t^2}=\omega_z^2 z+g \label{zeom}\\
&\omega_z=\eta_\text{L}\sqrt{\frac{-\chi_\rho}{\mu_0}}.\label{omegaz}
\end{align}
It can be observed that since $\mathbf{B}_\text{S}$ does not contain a $z$-component, it does not contribute to the levitation. Therefore, the levitation of the particle is solely provided by $\mathbf{B}_\text{L}$, while its motion along the $x$-direction and the subsequent c.o.m. separation induced by the spin–magnetic field coupling are exclusively governed by $\mathbf{B}_\text{S}$ (discussed in section \ref{seperation}). 

In order to realise stable levitation, the diamagnetic force must balance gravity in Eq.~\eqref{zeom}. Thus, the levitation height can be written as,
\begin{equation}
	\begin{aligned}
		z_\text{L}= -\frac{g}{\omega_z^2},
	\end{aligned}\label{levitationheight}
\end{equation}
where $z_\text{L}<0$. The Gaussian wavepacket in the $z$-direction remains Gaussian and peaked at a value $z=z_{\rm L}$, as discussed below. Again, this direction is an equally tight trap with a large $\omega_{z}$~
\footnote{The $x$-direction is nearly a flat direction, where $\omega_x$ will be given by Eq.~\ref{potential}. For the values we set up, $\eta\approx 106.7$ T/m leads to $\omega_x/(2\pi)\sim 1.2$~Hz (each stage). As we will see below for our setup we will have $\eta_{\rm L}\gg \eta_{\rm S}$, and $\eta_{\rm L}\sim 10^{5}$~T/m, so, we get, $\omega_y/(2\pi)\sim \omega_z/(2\pi)\sim 10^{3}$~Hz, showing the hierarchy in the trap frequencies.}.

Substitute $z_\text{L}$ into Eq.~(\ref{Bfield1}), the residual $z$-component of the total magnetic field can be written as
\begin{equation}
	\begin{aligned}
		&B_z(x,y,z_L)=\frac{g\mu_0}{\chi_\rho\eta_L}\approx-\frac{2\times10^{3}~\left(\mathrm{T^2/m}\right)} {\eta_L}.
	\end{aligned}\label{Bz0}
\end{equation}
This residual field is a nonzero quantity. In Eq.~\eqref{Bz0}, $\eta_L$ is expressed in units of $\mathrm{T/m}$, so that $B_z$ is obtained in tesla. In a ground-based experiment, it can only be reduced, but cannot be arbitrarily set to a small value. Therefore, if the libration mode of the NV spin has already been cooled to be parallel to the $x$ direction, and one intends to use a superposition of the spin-up and spin-down states to generate a spatial superposition state~\footnote{For the analysis of spin stability, see Refs \cite{Rizaldy:2024viw,zhou2025gyroscopic}}, then a sufficiently large bias field is required to ensure that the spin-up and spin-down superposition is not mixed into zero state by the residual field.


\begin{figure*}[t!]
	\begin{subfigure}[]{0.45\textwidth}
		\includegraphics[width=\textwidth]{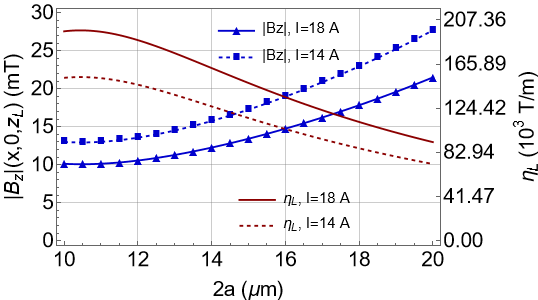}
		\subcaption*{(a)}
	\end{subfigure}
	\hspace{0.045\textwidth}
	\begin{subfigure}[]{0.4\textwidth}
		\includegraphics[width=\textwidth]{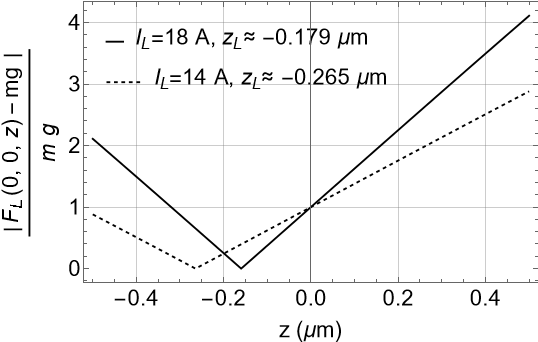}
		\subcaption*{(b)}
	\end{subfigure}
	\caption{
		(a) We show the dependence of the levitation gradient $\eta_\mathrm{L}$ and the residual field at the levitation height, $|B_z(x,0,z_\mathrm{L})|$, on the horizontal wire separation $2a$, for a fixed vertical separation $2b=14~\mu\mathrm{m}$. The blue curves and symbols, read from the left axis, show $|B_z(x,0,z_\mathrm{L})|$, while the red curves, read from the right axis, show $\eta_\mathrm{L}$. The solid and dashed curves correspond to $|I_\mathrm{L}|=18$ and $14~\mathrm{A}$, respectively. As the separation between the levitation wires increases, the levitation gradient $\eta_\mathrm{L}$ decreases. Since the levitation force scales as $F_\mathrm{L}\propto \mathbf{B}\cdot\nabla\mathbf{B}$, a larger residual field $|B_z(x,0,z_\mathrm{L})|$ is therefore required to maintain the levitation equilibrium.
		(b) Corresponding vertical force-balance condition. We plot $|F_{\mathrm{L},z}(0,0,z)/(mg)-1|$ as a function of $z$; the minima determine the levitation heights $z_\mathrm{L}$. The same line styles denote the same values of $|I_\mathrm{L}|$ as in panel (a). }\label{figure2}
\end{figure*}


\section{I-Cat Chip}\label{levitation}

The chip design shown in Fig.~\ref{spdevice} (main) and (front) is a current-controlled device intended to generate a Schr\"odinger-cat state, and we therefore call it the I-Cat Chip. It is designed primarily to achieve two purposes. The first is to use the four blue levitation wires to generate  $\mathbf{B}_{\text{L}}=\eta_\text{L}(z\mathbf{e}_z-y\mathbf{e}_y)$ along the $x$-axis in the $y$-$z$ plane, see Eq.(\ref{Bfield2}). The second purpose is to use the four yellow separation wires to generate  $\mathbf{B}_\text{S}=(x\eta_\text{S}+B_0)\mathbf{e}_x-y\eta_\text{S}\mathbf{e}_y$ in the $x-y$ plane, see Eq.(\ref{Bfield3}). Consequently, in the vacuum region of the central groove of the ``H''-shaped device, the total magnetic field produced by the device takes the form $\mathbf{B}=(x\eta_\text{S}+B_0)\mathbf{e}_x-y(\eta_\text{S}+\eta_\text{L})\mathbf{e}_y+ z\eta_\text{L}\mathbf{e}_z$. For the convenience of subsequent discussion, we define the origin of the coordinate system to be fixed at the geometric centre of this symmetric setup, with the positive directions of $\mathbf{e}_x$, $\mathbf{e}_y$, and $\mathbf{e}_z$ as indicated in Fig. \ref{spdevice}. We also define that currents flowing in the positive directions of $\mathbf{e}_x$ and $\mathbf{e}_z$ are considered positive, indicated by a dot on the wire cross-section, while currents flowing in the directions of $-\mathbf{e}_x$ and $-\mathbf{e}_z$ are considered negative, indicated by a cross on the wire cross-section.

The four yellow wires carry currents with the same magnitude, $|I_\text{S}|$, and the current directions of any two adjacent wires are opposite. The separations between these wires in the $x$ and $y$ directions are $2L=300~\mu\mathrm{m}$. The width and thickness of the four yellow wires are the same as $w_\text{s}= 32~\mu$m. This generates a quadrupole magnetic field Eq. (\ref{Bfield3}) in the $x-y$ plane, as shown in the subplot (a). By changing the combination of current directions in these wires, the sign of the magnetic-field gradient can be reversed, thereby enabling control over the trajectory of the spatial superposition state.

Similarly, the levitation assembly consists of four blue wires arranged in a rectangular configuration in the $y$-$z$ plane. The currents in adjacent wires have equal magnitudes $|I_\text{L}|$ and opposite directions. The horizontal and vertical separations between the wires are denoted by $2a$ and $2b$, respectively. Owing to diamagnetism, the diamond is confined within the vacuum region enclosed by the four wires, as discussed in Eqs. (\ref{ytrap}-\ref{omegaz}). 
To suppress field inhomogeneities arising from the wire ends, the eight wires must extend sufficiently far in the $x$ and $z$-direction. In the numerical calculations, the length of levitation wires (central blue wires in Fig. \ref{spdevice}) is set to $400~\mu\mathrm{m}$. Since the diamond is trapped close to the wires, their finite cross section is also included, with equal width and thickness $w_\text{L}=10~\mu\mathrm{m}$. For the separation wires, their length is not subject to a strict constraint and may be chosen according to the experimental conditions. In the numerical calculations, we take their length to be $300~\mu\mathrm{m}$.

Near the geometric central axis $(x,0,0)$, the gradient of $\mathbf{B}_\text{L}$ can be approximated as~\footnote{This expression is derived using an infinitely long and thin wire model; details are provided in Appendix~\ref{secsbpantaylor}. It is used only to simplify the discussion.}
\begin{equation}
\eta_\text{L}=\frac{4\mu_0 I_\text{L}}{\pi}\frac{ab}{(a^2+b^2)^2}.
\label{gradient}
\end{equation}
 \begin{figure*}[t!]
	\begin{subfigure}[]{0.325\textwidth}
		\includegraphics[width=1\linewidth]{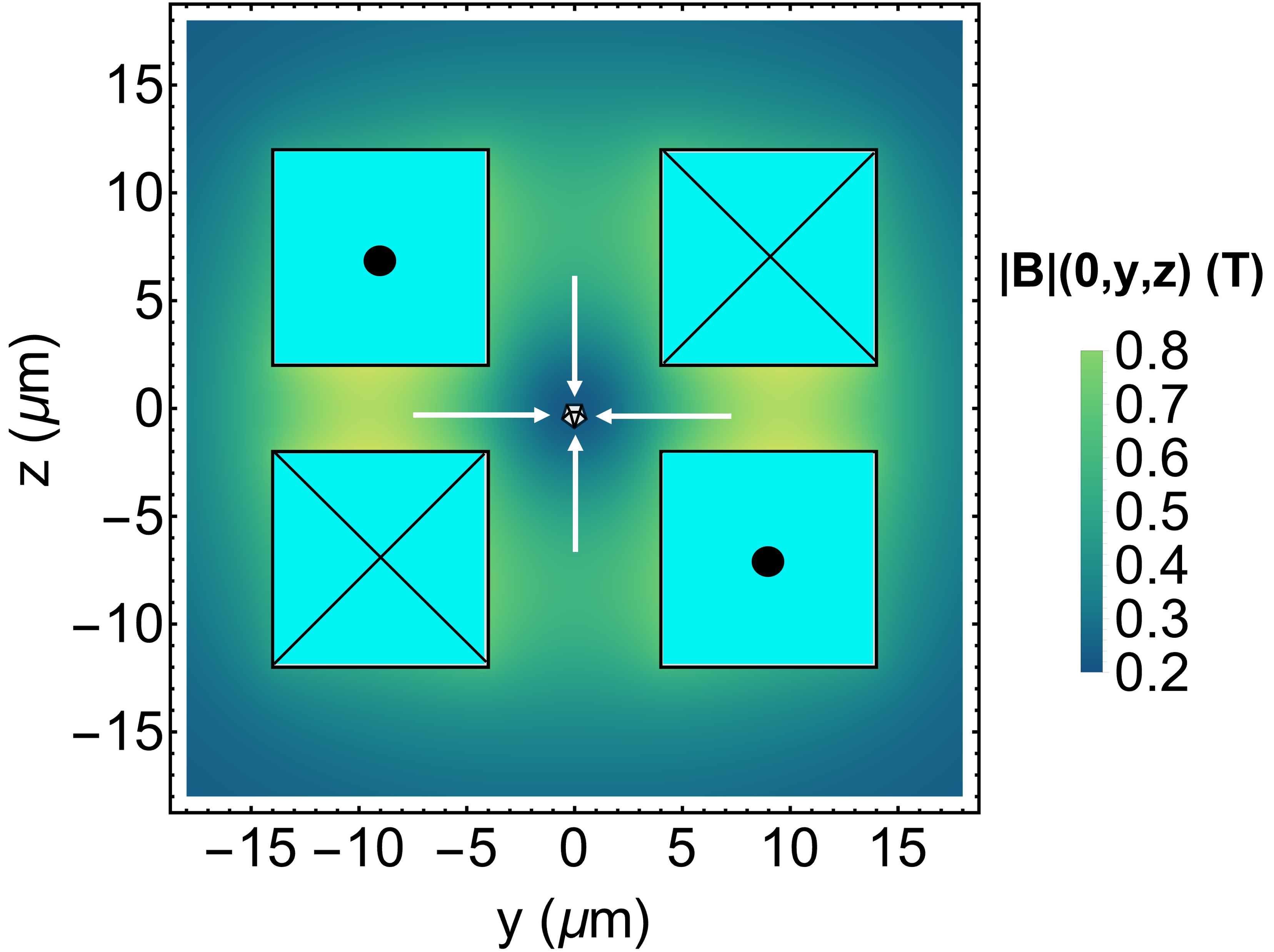}
		\subcaption*{(a)}
	\end{subfigure}\hspace{0.1cm}
	\begin{subfigure}[]{0.325\textwidth}
		\includegraphics[width=1\linewidth]{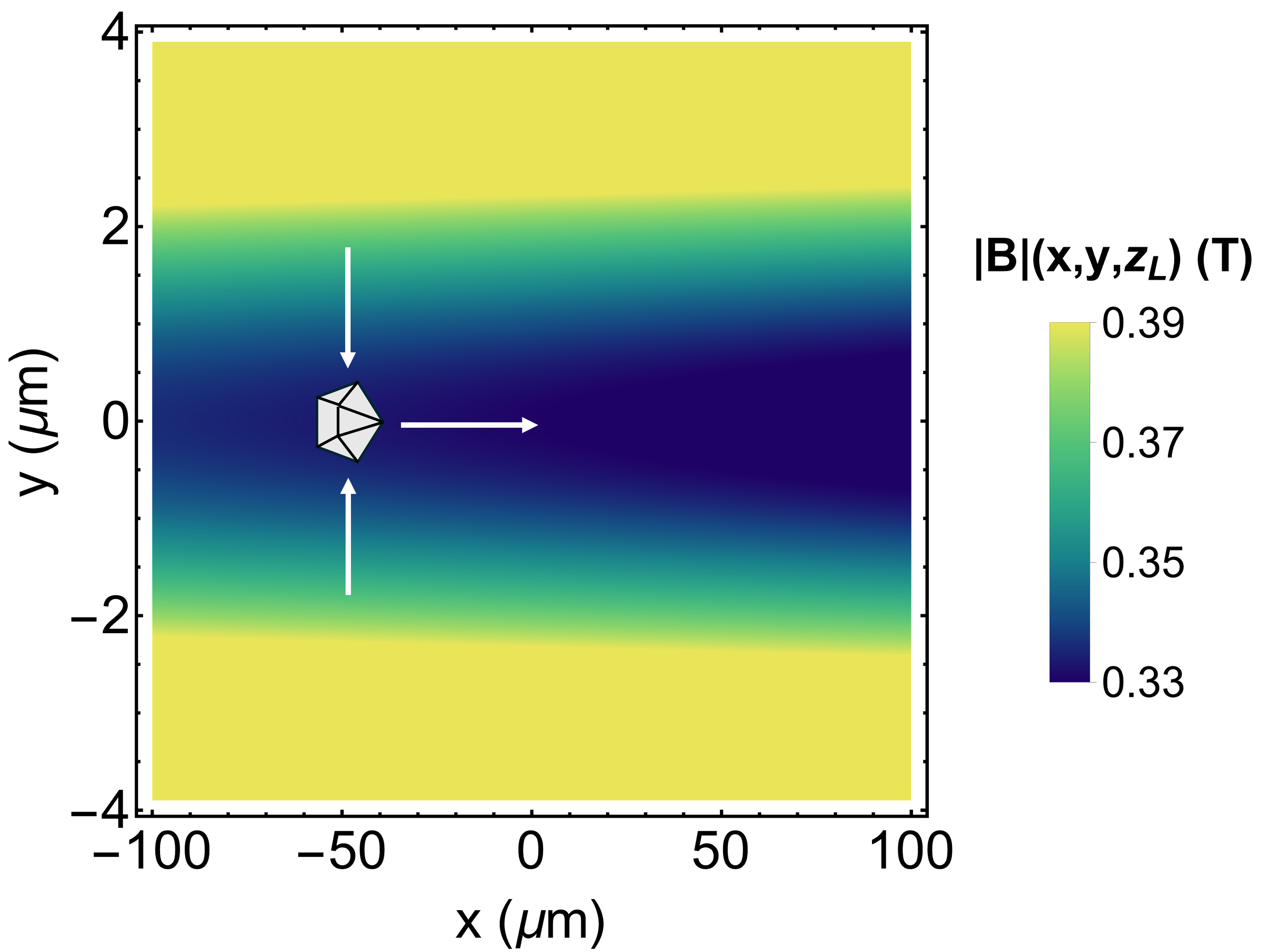}
		\subcaption*{(b)}
	\end{subfigure}\hspace{0.1cm}
	\begin{subfigure}[]{0.325\textwidth}
		\includegraphics[width=1\linewidth]{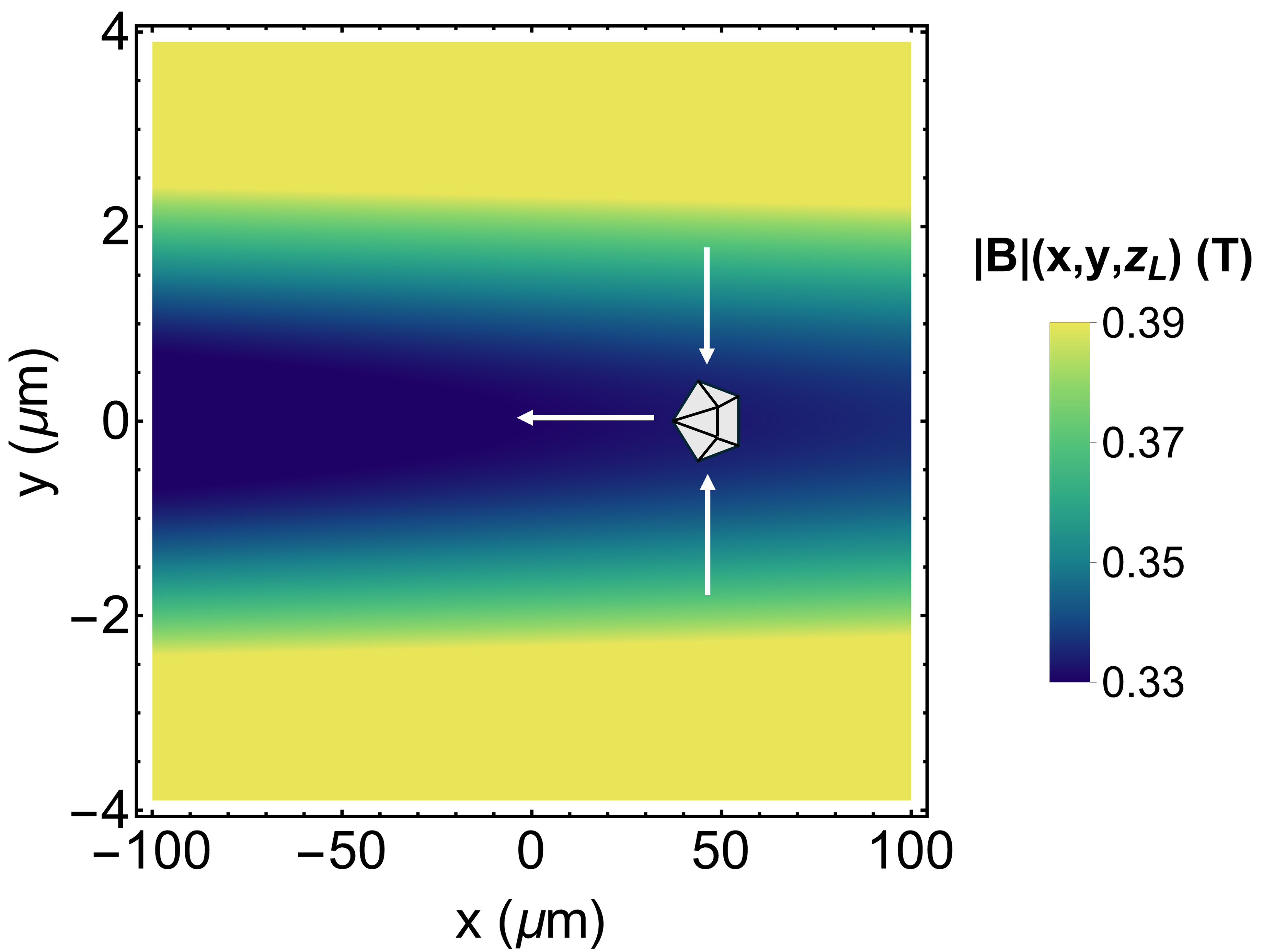}
		\subcaption*{(c)}
	\end{subfigure}
	\caption{
			Plots (a), (b), and (c) show the total magnetic field generated when the separation and levitation assemblies are operated simultaneously. Panel (a) shows a slice of the total magnetic field in the $x=0$ plane. The four blue squares represent the cross sections of the levitation wires and their corresponding current directions, with a current magnitude of $|I_\text{L}|=14~\mathrm{A}$. Their directions and magnitudes remain unchanged in time. Panels (b) and (c) show slices of the total magnetic field in the $z=z_\text{L}$ plane. Panel (b) corresponds to $|I_\text{S}|=6~\mathrm{A}$ and $B_0=0.33~\mathrm{T}$, for which the minimum of the magnetic field along the $x$ direction appears in the positive-$x$ region. Panel (c) corresponds to $I_\text{S}=-6~\mathrm{A}$, for which the magnetic-field minimum appears in the negative-$x$ region. In all three plots, the diamond particle is shown schematically, and the white arrows indicate the direction of the force acting on it. In panel (a), the illustrated spherical diameter is $1~\mu\mathrm{m}$, which corresponds to a mass of approximately $1.8\times10^{-15}~\mathrm{kg}$. For a mass of $10^{-17}~\mathrm{kg}$, the diameter of the particle is reduced by approximately a factor of five.
	}\label{sbpan}
\end{figure*}

Fig.~\ref{figure2} (a) shows the dependence of $\eta_\text{L}$ on the horizontal separation $2a$ for a fixed vertical separation $2b=14~\mu\mathrm{m}$ as well as the corresponding residual component at the levitation height, $|B_z(x,y,z_\text{L})|=|\eta_\text{L} z_\text{L}|$, as a function of $2a$. The solid and dashed curves correspond to $|I_\text{L}|=18$ and $|I_\text{L}|=14~\mathrm{A}$, respectively. The results show that reducing the separation between the levitation wires, or increasing the current density in the levitation wires, allows the levitated diamond particle to experience a larger magnetic-field gradient $\eta_\text{L}$ at the levitation point. This leads to a smaller residual field. The corresponding levitation height~\footnote{
		$F_z(x,y,z)$ refers to diamagnetic repulsion $F_z(x,y,z)= -\partial_z \frac{-\chi_\rho m}{2\mu_0}\mathbf{B}^2(x,y,z)$.
} is given in Fig.~\ref{figure2} where the geometric parameter has been set to $2a=18~\mu$m~\footnote{\label{FL}As a theoretical study, we have chosen smaller horizontal and vertical spacings along with a larger current, because such parameter settings can yield a significantly large magnetic field gradient $\eta_\text{L}$. This configuration can remarkably reduce the magnetic field's $z$-component experienced by the nanoparticle, as given by Eq. (\ref{Bz0}), thereby allowing an additional bias field $B_0$ to be applied to ensure that the NV spin remains predominantly aligned along the $x$-axis. }. As also discussed in Eq.~(\ref{Bz0}), to avoid state mixing caused by the residual field, one needs to reduce the residual field as much as possible. Therefore, a large current $|I_\text{L}|$ needs to be considered. For the two current cases shown in Fig.~\ref{figure2}, the corresponding current densities are $J=1.4\times 10^7~\mathrm{A/cm^2}$ and $J=1.8\times 10^7~\mathrm{A/cm^2}$, respectively. To sustain such high current densities, one needs to consider superconducting materials. For example, it has been reported that NbTiN-type superconductors can achieve current densities of $J_c=1.4\text{--}2\times 10^7~\mathrm{A/cm^2}$ \cite{herr2023scaling}. NbTiN is a type-II superconductor~\cite{rose2012introduction} with high upper critical field, $B_{c2}=18\text{--}22~\mathrm{T}$~\cite{klimov2016critical,zhang2024effect,lee2024penetration}. This current-carrying capability essentially covers the lower end of the current-density range corresponding to the examples shown in Fig.~\ref{figure2}. Other high-$J_c$ superconducting platforms, e.g., REBCO may also provide useful reference points for further material optimization \cite{majkic2018engineering}. In the subsequent discussion, we assume that the entire device is cooled to $2$ K~\cite{zhang2024effect}. And in the following discussion, we use the lower value $|I_L|=14~\mathrm{A}$ throughout. At this operating point, $\eta_\text{L}\approx 8.67\times10^4~\mathrm{T/m}$, and the levitation height is approximately $z_\text{L}\approx -0.265~\mu\mathrm{m}$, with a residual field of $|B_z|\approx 23~\mathrm{mT}$. Meanwhile, the levitation assembly forms stable harmonic traps in both the $y$ and $z$ directions, with trapping frequencies 
\begin{equation}
\omega_z\approx\omega_y\approx 6.1\times10^3~\mathrm{rad/s}
\end{equation}
\footnote{With the device parameters $I_\text{L}$, $2a$, and $2b$, the levitation
assembly generates a transverse magnetic-field gradient $\eta_\text{L}\approx 8.67\times 10^4~\mathrm{T/m}$. From Eqs.~(\ref{omegay}) and (\ref{omegaz}), the corresponding transverse trapping frequencies are $\omega_y\simeq\omega_z\simeq \eta_\text{L}\sqrt{-\chi_\rho/\mu_0} \simeq 6.1\times 10^3~\mathrm{rad/s}$, or $\omega_{y,z}/(2\pi)\simeq 9.7\times 10^2~\mathrm{Hz}$. The associated centre-of-mass ground-state widths $\sigma_{y,z}=\sqrt{\hbar/(2m\omega_{y,z})}$ are approximately $0.29,0.093,0.029~\mathrm{nm}$ for $m=10^{-19},10^{-18},10^{-17}~\mathrm{kg}$, respectively. For spherical diamond particles of density $\rho_\mathrm{d}\simeq 3500~\mathrm{kg/m^3}$, these widths correspond to about $1.6\%$, $0.23\%$, and $0.034\%$ of the corresponding classical particle radii. The transverse motion is therefore tightly confined, and the superposition dynamics can be treated as effectively one-dimensional along the $x$ direction.}

For the separation wires, since we need to change the combination of current directions to control the trajectory of the spatial superposition state, we consider using gold wires. This choice avoids the influence of vortex dynamics, which is common in type-II superconductors, on the current-carrying capability during current switching. One possible fabrication route is to electroplate the gold electrodes into trenches created by deep reactive ion etching in silicon, which can facilitate heat removal through the substrate, as demonstrated in related atom-chip implementations \cite{montoya2015resonant}. Fabrication-dependent effects such as surface roughness, current crowding, and microscopic current inhomogeneity may nevertheless influence the magnetic-field quality. Alternatively, high-quality evaporated or lithographically patterned gold conductors may also be employed to reduce surface roughness and microscopic current inhomogeneity \cite{groth2004atom,aigner2008long}. The chip architecture shown in Fig. \ref{spdevice} could be in principle fabricated by affixing a stack of two or more wafers after appropriate metal wire deposition takes place. 
For the separation wires, we mainly consider current configurations with $|I_{\mathrm{S}}|=6~\mathrm{A}$. For the gold wires considered here, with a width of $w_{\mathrm{S}}=32~\mu\mathrm{m}$, this corresponds to a current density of approximately $J\approx 5.9\times10^9~\mathrm{A/m^2}$ ~\footnote{The separation assembly is not superconducting and therefore produces finite Joule heating during operation. As an order-of-magnitude estimate, for four gold wires with $I_{\mathrm{S}}=6~\mathrm{A}$ and a cross section of $32\times32~\mu\mathrm{m}^{2}$, the total dissipated power is
	\begin{equation}
	P_{\mathrm{J}}
	=8I_{\mathrm{S}}^{2}\frac{\rho_{\mathrm{Au}}l}{w_{\mathrm{S}}^{2}}, 
	\end{equation}
	where $2l=300~\mu$m is the length of each separation wire.
	Taking $\rho_{\mathrm{Au}}\simeq2.4\times10^{-10}~\Omega\,\mathrm{m}$ at cryogenic temperatures \cite{bernat2007thermal} gives $P_{\mathrm{J}}\simeq 10.1~\mathrm{mW}$ for the four wires. Single-crystalline silicon provides efficient heat spreading and has been identified as a suitable substrate for dissipating heat from current-carrying gold microwires \cite{groth2004atom}. The resulting heat load of approximately $10~\mathrm{mW}$ should therefore be regarded as an engineering requirement on the cryogenic thermal design, requiring sufficient thermal anchoring of the gold conductors and silicon substrate to the cold stage.} \cite{groth2004atom}.
The separation assembly generates a magnetic-field gradient of approximately $106.7~\mathrm{T/m}$ with frequency $\omega_x\approx 7.5~$rad/s. The bias magnetic field aligned with the $x$ axis should be sufficiently large to avoid spin mixing induced by the residual field. The corresponding calculation is provided in the Appendix ~\ref{mix}. For the operating point $|I_\text{L}|$ considered here, we choose $B_0=330~\mathrm{mT}$. Therefore, combined with the levitation assembly, and in the region far from the wire ends, the diamond particle is tightly confined around the axis $y=0$ and $z_\text{L}=-0.265~\mu\mathrm{m}$, while allowing us to further manipulate the spatial-superposition c.o.m. trajectories along the flatter $x$ direction. The magnitude of the total magnetic field is shown in Fig.~\ref{sbpan}. 
It should be noted that, as shown in plot (a), the four superconducting wires are required to operate in a local magnetic field of approximately $0.8~\mathrm{T}$. Although this field is far below the upper critical field $B_{c2}$ of NbTiN, it can reduce the critical current density of the wires. For the present geometry, the required operating current density is already close to the estimated field-dependent critical-current-density limit. Furthermore, the practical implementation must account for additional challenges associated with fabrication conditions, current crowding, thermal stability, and vortex/flux-motion noise. The proposed configuration should therefore be regarded as a feasibility-level design with limited engineering margin. A detailed discussion of these limitations, together with an estimate of the field-dependent critical current density, is provided in Appendix~\ref{app:NbTiN_current}.



\begin{figure*}[t!]
	\begin{subfigure}[]{0.32\textwidth}
		\includegraphics[width=\textwidth]{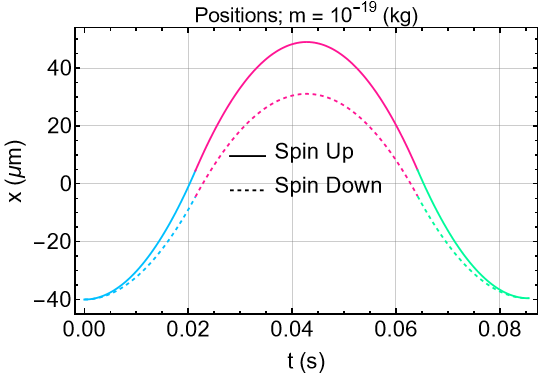}
		\subcaption*{(a)}
	\end{subfigure}\hspace{0.2cm}
	\begin{subfigure}[]{0.32\textwidth}
		\includegraphics[width=\textwidth]{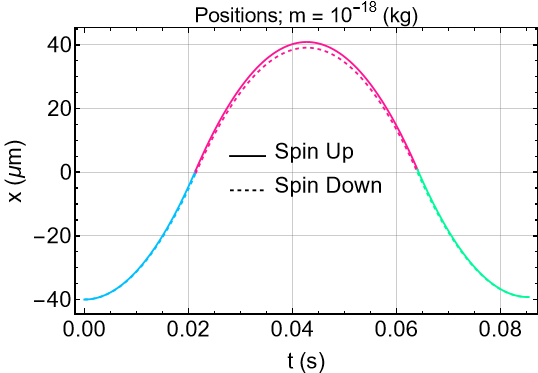}
		\subcaption*{(b)}
	\end{subfigure}\hspace{0.2cm}
	\begin{subfigure}[]{0.32\textwidth}
		\includegraphics[width=\textwidth]{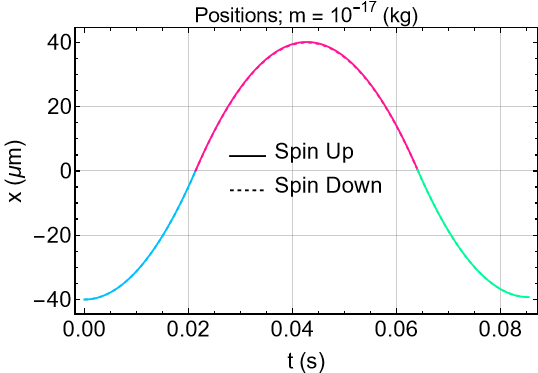}
		\subcaption*{(c)}
	\end{subfigure}\\
	\begin{subfigure}[]{0.32\textwidth}
		\includegraphics[width=\textwidth]{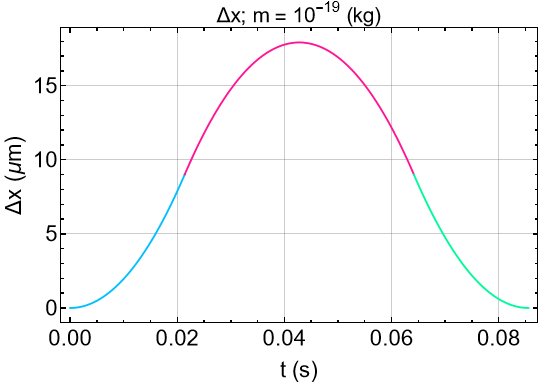}
		\subcaption*{(d)}
	\end{subfigure}\hspace{0.2cm}
	\begin{subfigure}[]{0.32\textwidth}
		\includegraphics[width=\textwidth]{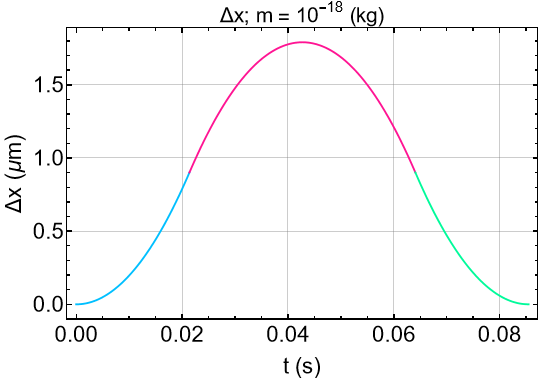}
		\subcaption*{(e)}
	\end{subfigure}\hspace{0.2cm}
	\begin{subfigure}[]{0.32\textwidth}
		\includegraphics[width=\textwidth]{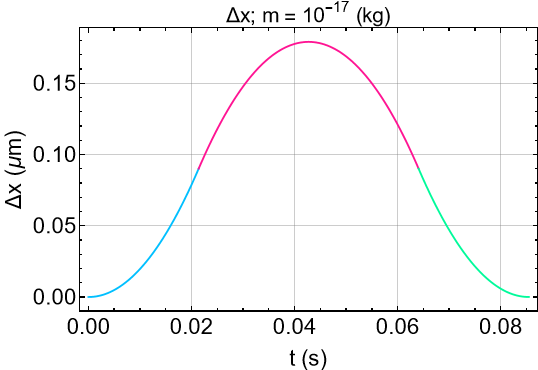}
		\subcaption*{(f)}
	\end{subfigure}
	\caption{Plots (a)--(f) show the superposition trajectories of diamond particles with three different masses, $m=10^{-19},~10^{-18},~10^{-17}~\mathrm{kg}$, including the spin-dependent trajectories in real space (a)--(c), where the spin-down trajectories are indicated by dashed lines, the superposition sizes (d)--(f). In all three examples, the particles are initially prepared at $x=-x_0=-40~\mu\mathrm{m}$. The three colors denote Stage 1 (blue), Stage 2 (red), and Stage 3 (green), respectively. The magnetic-field gradients in these three stages are $-\eta$, $\eta$, and $-\eta$, with the same magnitude $\eta\approx 106.7~\mathrm{T/m}$, corresponding to a current amplitude of $|I_\text{S}|=6~\mathrm{A}$ in the four separation wires.  }
	\label{example}
\end{figure*}



\section{Creating superposition}\label{seperation}

Here, we begin to discuss the specific details of generating the spatial superposition. First, recall that the function of the levitation assembly in Fig.~\ref{spdevice} is to create strong harmonic potential wells in the $y$- and $z$-directions, confining the diamagnetic nanoparticle along the axis at $y=0$, $z=z_\text{L}$. Along the $x$-direction, the separation assembly of the setup in Fig.~\ref{spdevice} is responsible for creating a weaker harmonic potential. By varying the direction of the currents $|I_\text{S}|$ in the separation assembly, one can adjust both the magnitude of the magnetic gradient $\eta_\text{S}$ and the direction of magnetic field along $x$ axis. Specifically, to achieve the creation of the spatial superposition and the closure of the two trajectories, we consider $\mathbf{B}_\text{S}$ of the following forms during different stages,
\begin{numcases}{\mathbf{B}_{\mathrm{S}} =}
	(B_0-\eta x)\mathbf{e}_x+ \eta y \mathbf{e}_y,& $0 \leq t \leq \tau_1$ \label{separation1}\\
	(B_0+\eta x)\mathbf{e}_x- \eta y \mathbf{e}_y,& $\tau_1\leq t\leq \tau_2$ \label{separation3} \\
	(B_0-\eta x)\mathbf{e}_x+ \eta y \mathbf{e}_y,& $\tau_2\leq t\leq \tau_3$\label{separation5}
\end{numcases}
where at different stages $\eta_\text{S}$ takes different values, $\pm \eta$, where $\eta>0$ is defined as a positive value. Such a magnetic-field configuration is similar to Refs.~\cite{Marshman:2021wyk,Zhou:2022epb}, where different field configurations are applied at different stages to control wave-packet separation and recombination.
Here, $\mathbf{B}_\text{S}$ in Eqs.~(\ref{separation1})--(\ref{separation5}) and the subsequent analytical expressions are introduced only to explain the protocol. 
The actual trajectories given below are calculated using the numerically simulated magnetic field produced by the separation wires for the specified geometric and current parameters. 
The numerical procedure used to obtain this magnetic field is described in Appendix~\ref{secsbpantaylor}.

For the magnetic fields in Eqs.~(\ref{separation1})--(\ref{separation5}), the parameters used here are $|I_\text{S}|=6~\mathrm{A}$ and a separation-wire spacing of $2L=300~\mu\mathrm{m}$. With this parameter setting, the magnetic-field gradient is $|\eta_\text{S}|=\eta\approx 106.7~\mathrm{T/m}$, and the corresponding angular frequency in each stage is $\omega_x\approx 7.5~\mathrm{rad/s}$. During the three stages, we reverse the current directions in the four separation wires to control the sign of $\eta_\text{S}$, thereby enabling the generation of the spatial superposition and ensuring the closure of the trajectories in both position and momentum space. The bias field used in the subsequent calculations is $B_0=0.33$ T. We will now perform an in-depth analysis of the protocol, as follows.

\begin{itemize}

\item{{\bf Preparation Stage}:
Note that, according to the equations of motion in the $y$- and $z$-directions, Eqs.~(\ref{ytrap}, \ref{zeom}), once the diamond is prepared near $(-x_0,0,z_\text{L})$ with a sufficiently cooled motional state, \footnote{Several magnetic approaches may be considered for loading and positioning the nanodiamond, e.g., magnetic-tweezer-type manipulation \cite{timonen2017tweezing}, diamagnetic trapping \cite{DUrso16_GM}, and current-carrying-wire configurations \cite{Elahi:2024dbb}.} the levitation assembly provides harmonic confinement around the axis $y=0$, $z=z_\text{L}$. The levitation assembly is therefore turned on before the interferometric sequence and remains active throughout the experiment. Meanwhile, the NV spin is prepared in the superposition state
 $(|+1\rangle_x+|-1\rangle_x)/\sqrt{2}$, with the NV axis predominantly aligned with the $x$-axis set by the bias magnetic field $B_0$. Under these conditions, the subsequent separation and recombination dynamics can be treated as motion along the $x$-direction, while the transverse degrees of freedom remain confined by the levitation potential.

The potential regarding diamond's motion in the $x$ direction can be written as,
\begin{equation}
	\begin{aligned}
		U(S_x)= -\frac{\chi_\rho m}{2\mu_0}\left(B_0+\eta_\text{S} x\right)^2+\hbar\gamma_\text{e}S_x\left(B_0+\eta_\text{S} x\right)\,,
	\end{aligned}\label{potential}
\end{equation}
where in different stages, $\eta_\text{S}$, takes different values $\pm\eta$ as shown in Eqs.~(\ref{separation1},\ref{separation3},\ref{separation5})}.


\item{{\bf Stage-1}:
 During $t\in [0,\tau_1]$, the magnetic field Eq.~(\ref{separation1}) is applied. The equation of motion for spin up and down trajectories can be written as,
\begin{equation}
	\begin{aligned}
	\frac{\text{d}^2x}{\text{d} t^2}&= -\frac{-\chi_\rho }{\mu_0}\eta^2 x -\left(\frac{\chi_\rho B_0}{\mu_0} -\frac{\hbar\gamma_\text{e}S_x}{m}\right)\eta \\
	&= -\omega^2 x-A_\pm.\\
	\end{aligned}\label{xeom1a} 
\end{equation}
where 
$\omega_x=\eta\sqrt{-\chi_\rho/\mu_0}$ and the constant $A_\pm= \left(\frac{\chi_\rho B_0}{\mu_0} -\frac{\hbar\gamma_\text{e}S_x}{m}\right)\eta.$ As can be seen from Eq.~(\ref{xeom1a}), if the spin $S_x$ were zero, the equation would reduce to a simple harmonic oscillator. It is due to the coupling between the spin superposition state and the magnetic field, the two c.o.m. experience different driving forces, leading to spatial separation during their motion. The solutions of Eq.~(\ref{xeom1a}) are,
\begin{align}
	x_1^\pm(t)&=-\left(x_0-\frac{A_\pm}{\omega_x^2}\right)\cos(\omega_x t)-
	\frac{A_\pm}{\omega_x^2}, \label{xsolution1a}\\
	\dot{x}^\pm_1(t)&=\omega_x\left(x_0-\frac{A_\pm}{\omega_x^2}\right)\sin(\omega_x t), \label{vsolution1a}
\end{align}
where $x=-x_0$ is the initialized position and $x^\pm$ correspond to the spin-up and spin-down trajectories, respectively. In Stage 1, the diamond is initialized at the negative position $x=-x_0$. For the magnetic field in Eq.~(\ref{separation1}), the minimum of the magnetic-field magnitude is located on the positive side of the $x$-axis, as shown in Fig.~\ref{sbpan}(b). Therefore, after release, both spin-dependent trajectories move toward the positive side of $x$-axis. During this motion, the Zeeman term produces different spin-dependent forces for the two branches, and the two c.o.m. trajectories gradually separate. This behavior is illustrated in Figs.~\ref{example} (a)--(c). The size of the separation $\Delta x (t)$ of this stage is then,
\begin{equation}
	\begin{aligned}
		\Delta x_1(t)&= x^+_1(t)- x^-_1(t)\\
		&=\frac{\Delta A}{\omega_x^2}\left[\cos(\omega_x t)-1\right],\label{deltax1a}
	\end{aligned}
\end{equation}
where,
\begin{equation}
	\begin{aligned}
		\Delta A=A_+-A_-=-\frac{2\hbar\gamma_e\eta}{m}, 
	\end{aligned}\label{deltak1a}
\end{equation}

where $\Delta A$ is a minus value since $\eta$ is defined positive. Therefore, for fixed frequency and magnetic-field gradient, the separation size is inversely proportional to the particle mass $m$, as shown in Figs.~\ref{example}(d)--(f). The choice of the ending moment $\tau_1$ of Stage-1 has some freedom. Here, our main consideration is to make sufficient use of the spatial region where the magnetic field generated by the separation assembly remains approximately linear. We therefore terminate Stage-1 when the two trajectories reach the vicinity of $x=0$, more precisely when their midpoint satisfies $\left[x_1^+(\tau_1)+x_1^-(\tau_1)\right]/2=0.$ Moment $\tau_1$ can be expressed analytically as
\begin{equation}
	\tau_1=\frac{1}{\omega_x}\arccos\left(\frac{B_0}{\eta x_0+B_0}\right).\label{tau1a}
\end{equation}
We will use $\tau_i$ and $T_i$ for moments and durations of different stages. The physical meaning of the symbols can be found in Tab.~\ref{tab1}.}
\end{itemize}

\begin{table}[h]
	\caption{\label{tab:symbols}Math symbols and physics interpretations.}
	\begin{ruledtabular}  
		\begin{tabular}{ll}
			\textrm{Symbol} & \textrm{Meaning} \\ 
			\colrule
			$x_i^\pm$    & Up and down trajectories of $i$-th stage. \\
			$\dot{x}_i^\pm$    & Up and down velocities of $i$-th stage. \\
			$X_i^\pm$    & Positions at the end of $i$-th stage. \\
			$\dot{X}^\pm_i$  & Velocities at the end of $i$-th stage. \\
			$T_i$        & Duration of $i$-th stage. \\
			$\tau_i$     & Moment at the end of $i$-th stage\tablenotemark[1]. \\
		\end{tabular}
	\end{ruledtabular}\label{tab1}
	\tablenotetext[1]{For example, at the end of the first stage the duration is $T_1$ and the moment is $\tau_1=T_1$, at the end of the second stage $\tau_2=T_1+T_2$.}
\end{table}

\begin{itemize}
\item{{\bf Stage-2}:
In the second stage, we map the magnetic field to the form given by Eq.~(\ref{separation3}), which corresponds to reversing the current direction in every wire of the separation assembly while keeping the current magnitude unchanged. In this stage, the equations of motion have the same harmonic form as Eq.~(\ref{xeom1a}), except that the magnetic-field gradient is reversed from $-\eta$ to $\eta$. As a result, the sign of the spin-dependent force term is reversed while the trap frequency $\omega_x$ remains unchanged. The solutions are,
\begin{equation}
	\begin{aligned}
		x_2^\pm(t)=&\frac{A_\pm}{\omega_x^2}
		+\left(X_1^\pm-\frac{A_\pm}{\omega_x^2}\right)
		\cos\left[\omega_x(t-\tau_1)\right]\\
		&\qquad+\frac{\dot{X}_1^\pm}{\omega_x}
		\sin\left[\omega_x(t-\tau_1)\right],
		\label{xsolution3a}
	\end{aligned}
\end{equation}
where $X_1^\pm=x^\pm(\tau_1)$ and $\dot{X}_1^\pm=\dot{x}^\pm|_{t=\tau_1}$ serve as the initial position and velocity in stage-2. As shown in Fig.~\ref{sbpan} (c), the minimum of the magnetic-field magnitude in this stage is located on the negative side of the $x$-axis. Consequently, the force acts opposite to the motion of the two branches, so that both trajectories are decelerated during this stage. The superposition size in this stage is then,
\begin{equation}
	\begin{aligned}
		\Delta x_2(t)=R\sin\left[\omega_x(t-\tau_1)+\alpha\right]+\frac{\Delta A}{\omega_x^2},
	\end{aligned}\label{deltax3}
\end{equation}
where, 
\begin{equation}
	\begin{aligned}
		R&=\sqrt{\left(\Delta X_1-\frac{\Delta A}{\omega_x^2}\right)^2+\left(\frac{\Delta \dot{X}_1}{\omega_x}\right)^2},\\
		\alpha&=\arctan\left[\frac{\Delta X_1-\dfrac{\Delta A}{\omega_x^2}}{\dfrac{\Delta \dot{X}_1}{\omega_x}}\right].
	\end{aligned}
\end{equation}
and $\Delta X_1=X_1^+-X_1^-$, $\Delta \dot{X}_1=\dot{X}_1^+-\dot{X}_1^-$. $\Delta A$ is defined in Eq. (\ref{deltak1a}).
At the moment,
\begin{equation}
	\begin{aligned}
		t_\text{max}=\tau_1+\frac{1}{\omega_x}\left(\frac{\pi}{2}-\alpha\right).
	\end{aligned}
\end{equation}
The size is maximized $\Delta x_\text{max}=R+\frac{\Delta A}{\omega_x^2}$. Written in more basic values, we have,
\begin{equation}
	\Delta x_{\max}=
	\frac{2\mu_0\hbar\gamma_e}{m(-\chi_\rho)\eta}\left[\sqrt{\frac{B_0+5\eta x_0}{B_0+\eta x_0}}-1\right].
	\label{Xmax}
\end{equation}
Recall that $x=-x_0$ is the initial position of the diamond in stage-1, which implies that $\Delta x_\text{max}$ is mainly determined by this parameter. The relation of $\Delta x_\text{max}$ and initial position parameter $x_0$ is given in Fig.~\ref{size}. As can be seen in the plot, releasing the diamond farther from the origin yields a larger superposition size. 

After this moment, the size begins to decrease. Here, we choose the ending moment of stage-2 as
\begin{equation}
	\begin{aligned}
		\tau_2= \tau_1+\frac{2}{\omega_x}\left(\frac{\pi}{2}-\alpha\right).\label{tau2a}
	\end{aligned}
\end{equation}
\begin{figure}[t]
	\includegraphics[width=.9\linewidth]{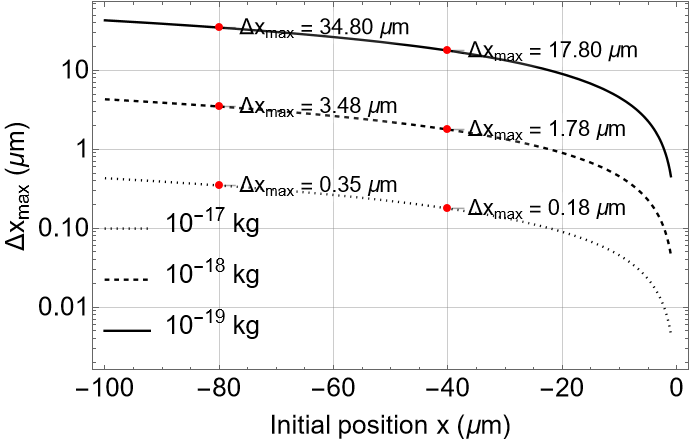}
	\caption{\centering 
		The achievable maximum superposition size vs. initial position is shown above. Initialising the diamond farther from the origin will take longer to complete a full loop. The magnitude of gradient presents here for stage-1 and -2 are $\eta\approx 106.7$ T/m, which correspond to $|I_\text{S}|=6$ A and $2L=300~\mu$m.}\label{size}
\end{figure}
indicating the stage-2 duration $T_2= \frac{2}{\omega_x}\left(\frac{\pi}{2}-\alpha\right)$. This duration is deliberately chosen such that, at the moment $\tau_2$, $\Delta x_2(\tau_2)$ is exactly equal to $\Delta X_1$. As can be seen from Figs.~\ref{example}(d)--(f), the red part of the curve is symmetric. In addition, both branches return to the vicinity of $x=0$, as shown in Figs.~\ref{example}(a)--(c), leaving a region with good magnetic-field linearity for closing their trajectories in Stage 3.

}

\item{\textbf{Stage-3}: 
At this stage, $\mathbf{B}_\text{S}$ is mapped to Eq.~(\ref{separation5}). Solutions at this stage are,
\begin{equation}
	\begin{aligned}
		x_3^\pm(t)=&-\frac{A_\pm}{\omega_x^2}+\left(X_2^\pm+\frac{A_\pm}{\omega_x^2}\right)\cos\left[\omega_x(t-\tau_2)\right]\\
		&\quad+\frac{\dot{X}_2^\pm}{\omega_x}\sin\left[\omega_x(t-\tau_2)\right],
	\end{aligned}
\end{equation}
where, now the final state of positions $X_2^\pm$ and velocities $\dot{X}_2^\pm$ from stage-2 become the initials conditions for stage-3. And note that $\Delta X_2=\Delta X_1$, $\Delta \dot{X}_2=-\Delta \dot{X}_1$, the separation in stage-3 can be written as,
\begin{equation}
	\Delta x_3(t)=R_\text{f}\sin\left[\omega_x(t-\tau_2)+\beta\right]-\frac{\Delta A}{\omega_x^2},\label{deltax3a}
\end{equation}
where, 
\begin{align}
	R_\text{f}&=\sqrt{\left(\Delta X_1+\frac{\Delta A}{\omega_x^2}\right)^2+\left(\frac{\Delta\dot{X}_1}{\omega_x}\right)^2},\label{deltax3aRf}\\
	&= -\frac{\Delta A}{\omega_x^2}\notag\\
	\beta&=\arctan\left[\frac{\Delta X_1+\dfrac{\Delta A}{\omega_x^2}}{\dfrac{-\Delta\dot{X}_1}{\omega_x}}\right]=\frac{\pi}{2}+\omega_x \tau_1.\label{deltax3abeta}
\end{align}
In Eqs.~(\ref{deltax3a})--(\ref{deltax3abeta}), we have substituted $\Delta X_1=\Delta x_1(\tau_1)$ and $\Delta \dot{X}_1=\Delta \dot{x}_1(\tau_1)$, and have simplified the expressions for $R_\text{f}$ and $\beta$.

Thus, one can find that at the moment
\begin{equation}
	\tau_3=\tau_2+\frac{1}{\omega_x}\left(\frac{3\pi}{2}-\beta\right),\label{tau3a}
\end{equation}
the two trajectories are closed in both position and momentum space, satisfying
\begin{equation}
	\Delta x_3(\tau_3)=0,\qquad\Delta \dot{x}_3(\tau_3)=0.\label{closure1}
\end{equation}}
\end{itemize}
In the above discussion of the protocol, we used the linear magnetic-field expressions in Eqs.~(\ref{separation1})--(\ref{separation5}) for convenience, and analytically derived the switching conditions for different stages. It should be noted that the results shown in Figs.~\ref{example}(a)--(f) are all obtained numerically. The spin-up and spin-down trajectories can be recombined to an indistinguishable degree without any fine-tuning of the current magnitude. The corresponding results regarding the numerical recombination are provided in Fig. \ref{example2}, Appendix~\ref{protocol}. In the example considered here, the only required operation is to reverse the current directions in the separation wires between different stages, as summarized in Tab.~\ref{tab}.
\begin{table}[htbp]
	\centering
	\renewcommand{\arraystretch}{1.4}
	\caption{Separation assembly currents in stage-1,2,3}
	\begin{tabular}{|c|c|c|c|}
		\hline
		\diagbox[width=3.2cm,height=1.3cm]{Wires}{Stages}
		&
		\begin{tabular}{c}
			Stage 1 \\
		\end{tabular}
		&
		\begin{tabular}{c}
			Stage 2 \\
		\end{tabular}
		&
		\begin{tabular}{c}
			Stage 3 \\
		\end{tabular}
		\\
		\hline
		$I_\text{S}$ in wire 1 & -6 A & 6 A & -6 A \\
		\hline
		$I_\text{S}$ in wire 2 & 6 A & -6 A & 6 A \\
		\hline
		$I_\text{S}$ in wire 3 & -6 A & 6 A & -6 A \\
		\hline
		$I_\text{S}$ in wire 4 & 6 A & -6 A & 6 A \\
		\hline
	\end{tabular}\tablenotetext[1]{Wires 1, 2, 3, and 4 refer to the wires in the first, second, third, and fourth quadrants of the separation assembly in Fig.~\ref{spdevice}, respectively. We define the current flowing along the positive $z$-direction as positive, and the current flowing along the negative $z$-direction as negative.}\label{tab}
\end{table}
The ending conditions for Stages 1, 2, and 3 are all calculated numerically. Specifically, in Stage 1, the midpoint satisfies $\left[x_1^+(\tau_1)+x_1^-(\tau_1)\right]/2=0$; in Stage 2, $\Delta x_2(\tau_2)$ is exactly equal to $\Delta X_1$; and in Stage 3, the ending moment $\tau_3$ is chosen such that $\Delta x_3(\tau_3)$ is minimized. This numerical treatment is required because the magnetic field generated by the separation-wire assembly is not perfectly linear. The difference between the analytical trajectories obtained from the linear-field model and the actual trajectories calculated from the simulated magnetic field is discussed in Appendix~\ref{protocol}, Fig. \ref{nuandan1}.

In addition, as indicated by Eq.~(\ref{Xmax}), the maximum superposition size achievable with this device can be tuned by adjusting the initial release position of the diamond particle, $x=-x_0$. Keeping the currents in Tab.~\ref{tab} fixed, the dependence of the maximum achievable superposition size $\Delta x_\text{max}$ on the initial position is shown in Fig.~\ref{size} for the three particle masses. For particles released from $x_0=40~\mu\mathrm{m}$, the maximum superposition sizes obtained for $m=10^{-19},~10^{-18},~10^{-17}~\mathrm{kg}$ are $\Delta x_\text{max}=17.8,~1.78,~0.18~\mu\mathrm{m}$, respectively. The corresponding full-loop durations are all below $100~\mathrm{ms}$, as can also be seen from Figs.~\ref{example}(d)--(f). If the NV spin coherence time can be further extended by improving the environmental vacuum, lowering the temperature, or other technical means, larger superposition sizes can be reached by increasing the initial release distance. For example, when the particle is released from $x_0=80~\mu\mathrm{m}$, the maximum superposition sizes for $m=10^{-19},~10^{-18},~10^{-17}~\mathrm{kg}$ increase to $\Delta x_\text{max}=34.8,~3.4,~0.35~\mu\mathrm{m}$, respectively.

Note that there is hardly any drift of the nanodiamond in the $y-z$ directions, as the traps are tight, and the frequencies along these directions are
$\omega_y\sim \omega_z\sim 6.1\times 10^{3}$~rad/s. 
If the diamond is cooled sufficiently in $y-z$ directions, the wave packet remains small, see footnote-6.

\subsection{Magnetic-field-noise-induced dephasing}
The interferometric signal is determined by the relative phase accumulated between the two spin-dependent c.o.m. trajectories. Fluctuations of the magnetic field introduce a stochastic contribution to this relative phase, such that repeated realizations of the interferometric sequence acquire different phase shifts. After averaging over these fluctuations, the phase coherence and hence the interferometric contrast are reduced.

For the three-stage protocol considered here, we restrict the analysis to the dominant one-dimensional motion along the $x$ direction and evaluate the phase fluctuations along the unperturbed trajectories $x_i^\pm(t)$, with $i=1,2,3$. A small fluctuation of the magnetic field modifies the Lagrangian along the two interferometric branches and therefore changes their relative action. To first order in the perturbation, the corresponding stochastic phase can be written as~\cite{Wu:2024tcr}
\begin{equation}
	\delta\phi=\frac{1}{\hbar}\sum_{i=1}^{3}\int_{\tau_{i-1}}^{\tau_i}\left[\delta L_i^+(t)-\delta L_i^-(t)\right]\mathrm{d}t,
	\label{phase1}
\end{equation}
where $\tau_0=0$ and $\tau_3$ denotes the final moment of the full-loop protocol. Equation~\eqref{phase1} provides the general starting point for the dephasing analysis below. We consider two relevant sources of magnetic-field fluctuations: fluctuations of the separation gradient induced by the separation-wire currents, and fluctuations of the externally applied bias field $B_0$.

\subsubsection{Separation-current fluctuations}

The separation gradient used for the spin-dependent splitting is generated by the current flowing through the separation wires. For the wire geometry considered in Fig.~\ref{spdevice}, the magnetic-field gradient is proportional to the separation current,
\begin{equation}
	\eta_\mathrm{S}
	=
	\frac{\mu_0}{\pi}
	\frac{I_\mathrm{S}}{L^2}.
\end{equation}

For a fixed wire geometry, a fractional current fluctuation therefore produces the same fractional fluctuation in the separation gradient. In Stage $i$, we define
\begin{equation}
	\xi_i(t)\equiv\frac{\delta\eta_{\mathrm S,i}(t)}{\eta_i}=\frac{\delta I_{\mathrm S,i}(t)}{I_i},
	\label{deltaeta}
\end{equation}
where $\eta_i$ and $I_i$ denote the nominal gradient and current in Stage $i$, respectively. The actual instantaneous gradient in this stage is therefore written as
\begin{equation}
	\eta_{\mathrm S,i}(t)
	=
	\eta_i\bigl[1+\xi_i(t)\bigr].
	\label{actualgradient}
\end{equation}

For the three-stage protocol considered here, the nominal gradients are $\eta_1=-\eta$, $\eta_2=+\eta$, and $\eta_3=-\eta$, corresponding to the field configurations introduced in Eqs.~\eqref{separation1}, \eqref{separation3}, and \eqref{separation5}, respectively. The same nominal current magnitude is used in all three stages, while the sign of the gradient is reversed by reversing the current directions in the separation wires. The quantity $\xi_i(t)$ denotes the fractional deviation from the corresponding stage-dependent set point, whereas the sign of the gradient is contained in $\eta_i$. Consequently, a positive value of $\xi_i(t)$ corresponds to an increase in the gradient magnitude, irrespective of the sign of $\eta_i$.

For the four outer wires in Fig.~\ref{spdevice}, such current fluctuations may arise from technical noise in the current source as well as Johnson--Nyquist noise \cite{johnson1928thermal} associated with the finite resistance of the wires. Johnson--Nyquist noise has also been discussed in the context of QGEM, e.g. see Ref.~\cite{marshman2020mesoscopic}. Using the general phase-response expression in Eq.~\eqref{phase1}, we evaluate the resulting stochastic phase along the unperturbed trajectories $x_i^\pm(t)$.

To first order in $\delta\eta_{\mathrm S,i}(t)$, the Zeeman part of the Lagrangian fluctuation in Stage $i$ is
\begin{equation}
	\delta L_{\mathrm Z,i}^\pm(t) = -\hbar\gamma_e S_x x_i^\pm(t) \delta\eta_{\mathrm S,i}(t).
	\label{zeelag}
\end{equation}

Here, $S_x=+1$ and $S_x=-1$ correspond to the spin-up and spin-down trajectories $x_i^+(t)$ and $x_i^-(t)$, respectively. The diamagnetic contribution follows from the term proportional to $\mathbf{B}^2$ in the Lagrangian. Retaining the part affected by the separation gradient gives
\begin{equation}
	\delta L_{\mathrm{dia},i}^\pm(t) = \frac{\chi_\rho m}{\mu_0}
	\left[ B_0x_i^\pm(t) + \eta_i x_{i,\pm}^2(t) \right] \delta\eta_{\mathrm S,i}(t).
	\label{dialag}
\end{equation}

In Eqs.~\eqref{zeelag} and \eqref{dialag}, $\delta\eta_{\mathrm S,i}(t)$ denotes the effective gradient fluctuation induced by the current fluctuation in Stage $i$, as defined in Eq.~\eqref{deltaeta}. For clarity, $x_{i,\pm}^2(t)$ denotes the square of the corresponding trajectory, namely $x_{i,\pm}^2(t)=[x_i^\pm(t)]^2$. Substituting Eqs.~\eqref{zeelag} and \eqref{dialag} into Eq.~\eqref{phase1}, and using $\delta\eta_{\mathrm S,i}(t)=\eta_i\xi_i(t)$, gives
\begin{equation}
	\begin{aligned}
		\delta\phi_I ={}& \frac{1}{\hbar} \sum_{i=1}^{3} \int_{\tau_{i-1}}^{\tau_i} \eta_i
		\Bigg\{ \frac{\chi_\rho m}{\mu_0} \Bigl[ B_0\bigl(x_i^+(t)-x_i^-(t)\bigr)
		\\
		&\qquad\qquad +\eta_i \bigl(x_{i,+}^2(t)-x_{i,-}^2(t)\bigr) \Bigr]
		\\
		&\qquad -\hbar\gamma_e \bigl(x_i^+(t)+x_i^-(t)\bigr) \Bigg\} \xi_i(t) \mathrm{d}t .
		\label{phase2}
	\end{aligned}
\end{equation}

It is convenient to introduce the stage-dependent response function
\begin{equation}
	\begin{aligned}
		\mathcal{F}_i(t) ={}& \frac{\eta_i}{\hbar} \Bigg\{ \frac{\chi_\rho m}{\mu_0} \Bigl[
		B_0\bigl(x_i^+(t)-x_i^-(t)\bigr) \\
		&\qquad +\eta_i \bigl(x_{i,+}^2(t)-x_{i,-}^2(t)\bigr) \Bigr]
		\\
		&\qquad -\hbar\gamma_e \bigl(x_i^+(t)+x_i^-(t)\bigr) \Bigg\},
	\end{aligned}
	\label{calF}
\end{equation}
so that Eq.~\eqref{phase2} can be expressed compactly as
\begin{equation}
	\delta\phi_I = \sum_{i=1}^{3} \int_{\tau_{i-1}}^{\tau_i} \mathcal{F}_i(t)\xi_i(t) \mathrm{d}t.
	\label{phasecompact}
\end{equation}

In the present dephasing estimate, $\xi_i(t)$ is treated as a zero-mean time-dependent stochastic process. For a white fractional current-noise model, we have \cite{gardiner1181stochastic}
\begin{equation}
	\left\langle
	\xi_i(t)\xi_j(t')
	\right\rangle
	=
	S_{\delta I/I}
	\delta(t-t'),
	\label{whitenoisecorrelation}
\end{equation}
where $S_{\delta I/I}$ is the white-noise spectral density of the fractional current fluctuation. Since the three stage intervals do not overlap, Eq.~\eqref{whitenoisecorrelation}, together with Eq.~\eqref{phasecompact}, gives
\begin{equation}
	\begin{aligned}
		\left\langle
		\delta\phi_I^2
		\right\rangle
		&=\sum_{i,j}\int_{\tau_{i-1}}^{\tau_i}\mathrm{d}t
		\int_{\tau_{j-1}}^{\tau_j}\mathcal{F}_i(t)\mathcal{F}_j(t^\prime)\langle\xi_i(t)\xi_j(t^\prime)\rangle\mathrm{d}t^\prime\\
		&=S_{\delta I/I}\sum_{i,j}\int_{\tau_{i-1}}^{\tau_i}\mathrm{d}t\int_{\tau_{j-1}}^{\tau_j}\mathcal{F}_i(t)\mathcal{F}_j(t^\prime)\delta(t-t^\prime)\mathrm{d}t^\prime\\
		&=S_{\delta I/I}\sum_i \int_{\tau_{i-1}}^{\tau_i}\mathcal{F}_i^2(t)\mathrm{d}t.
		\label{phasevariance}
	\end{aligned}
\end{equation}

\begin{table}[t] 
	\centering 
	\caption{Effective fractional and absolute separation-current fluctuations.}
	\label{dephasingcurrent} 
	\begin{tabular}{c@{\hspace{28pt}}c@{\hspace{28pt}}c} 
		\hline\hline 
		\noalign{\vskip 2pt} 
		$x_0~(\mu\mathrm{m})$ 
		& $\xi_{\mathrm{eff}}$ 
		& $\delta I_{\mathrm{eff}}$ \\ 
		\hline 
		\noalign{\vskip 2pt} 
		$10$ & $1.2\times10^{-8}$ & $72~\mathrm{nA}$ \\ 
		$40$ & $1.5\times10^{-9}$ & $9.0~\mathrm{nA}$ \\ 
		$70$ & $0.7\times10^{-9}$ & $4.2~\mathrm{nA}$ \\ 
		\hline\hline 
	\end{tabular} 
	
	\vspace{2mm} 
	\parbox{0.92\linewidth}{ 
		\footnotesize 
		Here, $\xi_{\mathrm{eff}}$ denotes the effective fractional fluctuation of the separation current corresponding to $C_{\phi,I}=0.9$. The absolute current fluctuation is obtained from $\delta I_{\mathrm{eff}}=|I_{\mathrm S}|\xi_{\mathrm{eff}}$, using the nominal separation current $|I_{\mathrm S}|=6~\mathrm{A}$. The values therefore provide an estimate of the separation-current stability required to limit dephasing to the chosen reference contrast level.
	} 
\end{table}
Following the convention adopted in Ref.~\cite{narasimha2025magnetic}, we define the effective dephasing rate as
\begin{equation}
	\Gamma_{\phi,I}=\frac{1}{\tau_3}\left\langle\delta\phi_I^2\right\rangle.
	\label{currentdephasing}
\end{equation}
Using the same convention, the corresponding phase-coherence contrast is estimated as
\begin{equation}
	C_{\phi,I} \simeq \exp\left(-\Gamma_{\phi,I}~\tau_3\right).
	\label{currentcontrast}
\end{equation}

Here, $\Gamma_{\phi,I}\tau_3$ is the dimensionless dephasing exponent. It is worth noting that Eq.~\eqref{calF} is independent of the particle mass, since in each stage the difference between the spin-up and spin-down trajectories is proportional to $1/m$; see Appendix~\ref{protocol}.

\begin{figure}[t]
	\centering
	\includegraphics[width=.9\linewidth]{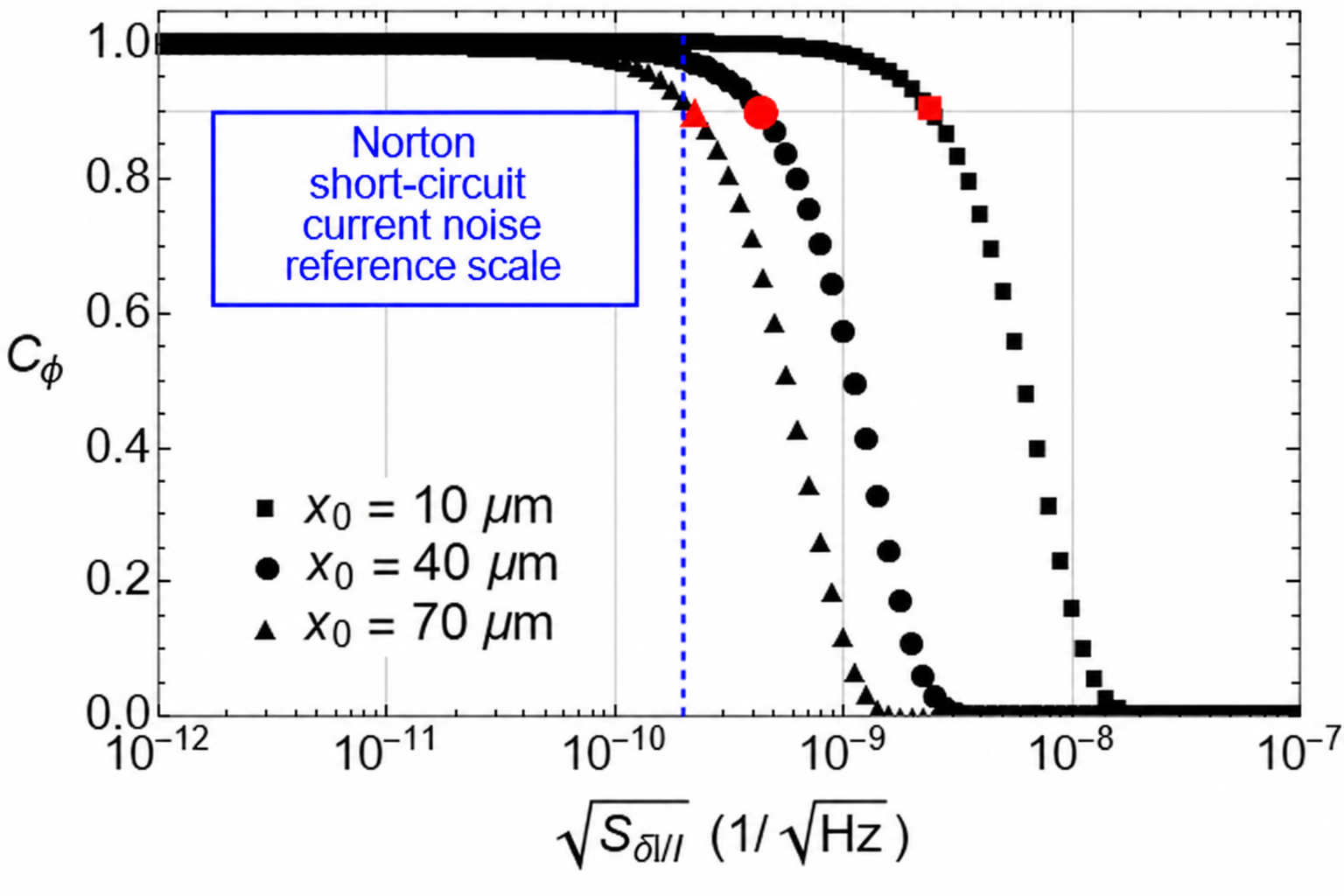}
	\caption{
		\centering
			We show the contrast $C_{\phi,I}$ as a function of the fractional current-noise amplitude spectral density $\sqrt{S_{\delta I/I}}$ for the full-loop trajectories with fixed particle mass $m=10^{-19}~\mathrm{kg}$. The three curves correspond to the initial release positions $x_0=10$, $40$, and $70~\mu\mathrm{m}$, respectively. The red markers indicate the points at which the contrast reaches $C_{\phi,I}=0.9$. The corresponding threshold values are $\sqrt{S_{\delta I/I}}\simeq2.39\times10^{-9}$, $4.35\times10^{-10}$, and $2.23\times10^{-10}~\mathrm{Hz}^{-1/2}$ for $x_0=10$, $40$, and $70~\mu\mathrm{m}$, respectively. The blue dashed line indicates the Norton short-circuit current-noise reference scale, $\sqrt{S_{\delta I/I}^{\mathrm{JN}}}\approx2\times10^{-10}~\mathrm{Hz}^{-1/2}$, estimated from Eq.~\eqref{jnfraction} for the gold separation-wire resistance at $T=2~\mathrm{K}$.
		}
	\label{Cphi}
\end{figure}

To express the required current stability in terms of an effective fractional fluctuation, we introduce the root-mean-square quantity $\xi_{\mathrm{eff}}$ using the characteristic bandwidth $\Delta f=1/\tau_3$. Since $S_{\delta I/I}$ is taken to be frequency independent within the white-noise model, we have
\begin{equation}
	\begin{aligned}
		\xi_{\mathrm{eff}}&=
		\sqrt{\int_{0}^{\Delta f}S_{\delta I/I}\mathrm{d}f}\\
		&=\sqrt{S_{\delta I/I}\Delta f}.
	\end{aligned}
	\label{xieff}
\end{equation}
As shown in Fig.~\ref{Cphi}, the three curves correspond to $x_0=10$, $40$, and $70~\mu\mathrm{m}$, respectively, and the red markers indicate the points at which the contrast reaches $C_{\phi,I}=0.9$. For $|I_{\mathrm S}|=6~\mathrm{A}$, the corresponding effective fractional current fluctuations are $\xi_{\mathrm{eff}}\simeq1.2\times10^{-8}$, $1.5\times10^{-9}$, and $0.7\times10^{-9}$, respectively. For a more direct experimental interpretation, these fractional fluctuations can be converted into effective absolute current fluctuations using $\delta I_{\mathrm{eff}}=|I_{\mathrm S}|\xi_{\mathrm{eff}}$. The corresponding values are summarized in Tab.~\ref{dephasingcurrent}.

For comparison with the required current stability, we consider the Johnson--Nyquist noise of the gold separation wires. In an actual drive circuit, the contribution of the wire's own Johnson noise to the current fluctuation is filtered by circuit impedance and can be written as~\cite{qu2019johnson}
\begin{equation}
	S_I^{\mathrm{JN}}(\omega)=\frac{4k_{\mathrm B}TR_{\mathrm{Au}}}{|R_{\mathrm{Au}}+Z_{\mathrm{ext}}(\omega)|^2},
	\label{jncircuit}
\end{equation}
where $Z_{\mathrm{ext}}(\omega)$ denotes the impedance presented by the external drive circuit and is in general frequency dependent. The total current noise generally also contains contributions from the current source and other circuit elements, and therefore cannot be determined from the gold-wire resistance $R_{\mathrm{Au}}$ alone.

Since a specific drive circuit is not considered in the present work, we use the Norton short-circuit limit, $Z_{\mathrm{ext}}(\omega)\rightarrow0$, to obtain an order-of-magnitude reference scale for comparison with the required current stability. By neglecting the suppression associated with a finite external circuit impedance, this limit then gives a larger estimate of this contribution. In this limit, Eq.~\eqref{jncircuit} reduces to
\begin{equation}
	S_I^{\mathrm{JN}}=\frac{4k_{\mathrm B}T}{R_{\mathrm{Au}}}.\label{jnshort}
\end{equation}
For the assumed operating temperature $T=2~\mathrm{K}$, effective wire length $2l=300~\mu\mathrm{m}$, and square cross section with $w_{\mathrm S}=32~\mu\mathrm{m}$, the resistance of a gold separation wire is given by $R_{\mathrm{Au}}=2l\rho_{\mathrm{Au}}/w_{\mathrm S}^2$, where $\rho_{\mathrm{Au}}\simeq2.4\times10^{-10}~\Omega\mathrm{m}$ at cryogenic temperatures \cite{bernat2007thermal}. Assuming white Johnson--Nyquist noise over the effective bandwidth $\Delta f=1/\tau_3$, the corresponding root-mean-square current fluctuation is \cite{marshman2020mesoscopic}
\begin{equation}
	\delta I_{\mathrm{JN}}
	=
	\sqrt{S_I^{\mathrm{JN}}\Delta f}
	=
	\sqrt{\frac{4k_{\mathrm B}T\Delta f}{R_{\mathrm{Au}}}},
	\label{jncurrent}
\end{equation}
where $k_{\mathrm B}=1.380649\times10^{-23}~\mathrm{J/K}$ is the Boltzmann constant \cite{mohr2025codata}. Combining Eq.~\eqref{jncurrent} with the definition of the fractional current fluctuation in Eq.~\eqref{deltaeta}, the corresponding fractional current-noise amplitude spectral density is\footnote{Using the short-circuit Johnson--Nyquist spectrum in Eq.~\eqref{jnshort},
	\begin{equation}
		\frac{\delta I_{\mathrm{JN}}}{|I_{\mathrm S}|}
		=
		\frac{1}{|I_{\mathrm S}|}\sqrt{S_I^{\mathrm{JN}}\Delta f}
		=
		\sqrt{S_{\delta I/I}^{\mathrm{JN}}\Delta f},
		\notag
	\end{equation}
	where the Johnson--Nyquist noise is treated in the classical white-noise limit and $S_{\delta I/I}^{\mathrm{JN}}=S_I^{\mathrm{JN}}/I_{\mathrm S}^{2}$.}
\begin{equation}
	\sqrt{S_{\delta I/I}^{\mathrm{JN}}}
	=
	\frac{1}{|I_{\mathrm S}|}
	\sqrt{\frac{4k_{\mathrm B}T}{R_{\mathrm{Au}}}},
	\label{jnfraction}
\end{equation}
which gives $\sqrt{S_{\delta I/I}^{\mathrm{JN}}}\approx2\times10^{-10}~\mathrm{Hz}^{-1/2}$ for $|I_{\mathrm S}|=6~\mathrm{A}$, as indicated by the blue dashed line in Fig.~\ref{Cphi}.

In the figure, the Norton short-circuit current-noise reference scale lies slightly below the fractional current-noise level corresponding to $C_{\phi,I}=0.9$ for the most demanding case, $x_0=70~\mu\mathrm{m}$. As discussed above, the short-circuit limit neglects the suppression associated with a finite external circuit impedance and therefore gives a slightly conservative estimate of the Johnson-noise contribution from the gold wire itself. From Fig.~\ref{size}, $x_0\simeq70~\mu\mathrm{m}$ corresponds to maximum spatial separations of approximately $\Delta x_{\mathrm{max}}\simeq30$, $3$, and $0.3~\mu\mathrm{m}$ for particle masses $m=10^{-19}$, $10^{-18}$, and $10^{-17}~\mathrm{kg}$, respectively. These results indicate that maintaining a high interferometric contrast requires a rather high stability of the separation current. The Johnson--Nyquist contribution estimated here remains below this stability threshold, while the corresponding short-circuit value should be regarded only as an order-of-magnitude reference rather than as an unavoidable current-noise floor. In addition, one should note that in the above estimate the gold separation wires are treated as resistive circuit elements. The resulting Johnson--Nyquist noise therefore corresponds to lumped-circuit transport-current noise, rather than to local magnetic near-field noise generated by microscopic thermal current-density fluctuations within the conductor.

\subsubsection{Bias-field fluctuations}

We now consider fluctuations of the externally applied bias field $B_0$, while keeping the separation gradients at their nominal values. In the present work, the bias field is supplied by an external device that is not included in the chip design, and its actual stability therefore depends on the specific configuration and noise characteristics of the corresponding drive circuit. Rather than modelling a particular bias-field source, we estimate here the absolute field stability required to preserve the interferometric coherence. We write the instantaneous bias field as
\begin{equation}
	B_0(t)=B_0+\delta B_0(t),
	\label{biasfluctuation}
\end{equation}
where $\delta B_0(t)$ denotes a small stochastic fluctuation around the nominal bias field.

To evaluate the bias-field-induced phase fluctuation using Eq.~\eqref{phase1}, we first determine the corresponding first-order variation of the Lagrangian. To first order in $\delta B_0(t)$, the Zeeman contribution in Stage $i$ is
\begin{equation}
	\delta L_{\mathrm ZB,i}^{\pm}(t)=-\hbar\gamma_e S_x\delta B_0(t).\label{biaszeelag}
\end{equation}
This term gives the direct Larmor-dephasing contribution, since the two interferometric branches correspond to $S_x=\pm1$, a fluctuation $\delta B_0(t)$ shifts their Zeeman energies in opposite directions and therefore produces a differential energy fluctuation of $2\hbar\gamma_e\delta B_0(t)$.

In addition to the Zeeman contribution, fluctuations of the bias field also modify the diamagnetic energy through the term proportional to $\mathbf{B}^2$. The corresponding diamagnetic contribution is
\begin{equation}
	\delta L_{\mathrm{diaB},i}^{\pm}(t)
	=
	\frac{\chi_\rho m}{\mu_0}
	\left[B_0+\eta_i x_i^\pm(t)\right]
	\delta B_0(t).
	\label{biasdialag}
\end{equation}

Substituting Eqs.~\eqref{biaszeelag} and \eqref{biasdialag} into Eq.~\eqref{phase1}, the common terms proportional to $B_0$ cancel between the two interferometric branches. Using $S_x=+1$ and $S_x=-1$ for the spin-up and spin-down branches, respectively, the resulting stochastic phase can be written as
\begin{equation}
	\begin{aligned}
		\delta\phi_B={}&\frac{1}{\hbar}\sum_{i=1}^{3}\int_{\tau_{i-1}}^{\tau_i}
		\Bigg[
		\frac{\chi_\rho m\eta_i}{\mu_0}
		\bigl(x_i^+(t)-x_i^-(t)\bigr)
		\\
		&\qquad\qquad-2\hbar\gamma_e\Bigg]\delta B_0(t)\mathrm{d}t.
	\end{aligned}
	\label{biasphase}
\end{equation}
Likewise, we define the stage-dependent response function,
\begin{equation}
	\mathcal{G}_i(t)=\frac{\chi_\rho m\eta_i}{\hbar\mu_0}\bigl(x_i^+(t)-x_i^-(t)\bigr)-2\gamma_e,\label{biasresponse}
\end{equation}
Since the trajectory separation $x_i^+(t)-x_i^-(t)$ scales as $1/m$ in the present protocol, the factor $m[x_i^+(t)-x_i^-(t)]$ is independent of the particle mass. Consequently, the response function $\mathcal{G}_i(t)$ in Eq.~\eqref{biasresponse}, and hence the bias-field-induced dephasing, is also independent of $m$. In Eq.~\eqref{biasresponse}, the first term represents the diamagnetic contribution, denoted by $\mathcal{G}_{\mathrm{dia}}(t)$, which is proportional to the instantaneous spatial separation between the two interferometric branches. The second term represents the direct differential Larmor contribution, denoted by $\mathcal{G}_{\mathrm{Z}}=-2\gamma_e$, which is independent of both position and time.

\begin{figure}[t]
	\centering
	\includegraphics[width=.9\linewidth]{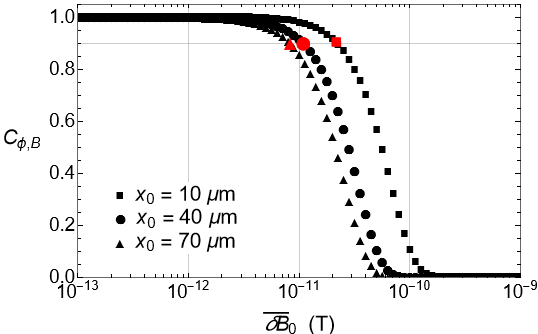}
	\caption{
		\centering
			We show the contrast $C_{\phi,B}$ as a function of the effective absolute bias-field fluctuation $\overline{\delta B_0}$ for the full-loop trajectories. The three curves correspond to the initial release positions $x_0=10$, $40$, and $70~\mu\mathrm{m}$, respectively. The quantity $\overline{\delta B_0}$ is obtained by integrating the white bias-field noise over the effective bandwidth $\Delta f=1/\tau_3$. The red markers indicate the points at which the contrast reaches $C_{\phi,B}=0.9$. The corresponding threshold values are $\overline{\delta B_0}=21.55$, $10.85$, and $8.26~\mathrm{pT}$ for $x_0=10$, $40$, and $70~\mu\mathrm{m}$, respectively.
	}
	\label{CphiB}
\end{figure}
To illustrate the relative importance of these two contributions, we estimate their maximum relative magnitude by taking the trajectory separation to be its maximum value $\Delta x_{\mathrm{max}}$. This gives
\begin{equation}
	\left|\frac{\mathcal{G}_{\mathrm{dia}}}{\mathcal{G}_{\mathrm{Z}}}\right|
	=
	\frac{|\chi_\rho|m|\eta|\Delta x_{\mathrm{max}}}{2\hbar\mu_0\gamma_e}.
	\label{biasratio}
\end{equation}
For the representative trajectory with $x_0=40~\mu\mathrm{m}$, where $\Delta x_{\mathrm{max}}\simeq17.8~\mu\mathrm{m}$ for $m=10^{-19}~\mathrm{kg}$, this ratio is approximately $2.5\times10^{-2}$. The direct Larmor response therefore dominates over the diamagnetic contribution. In the following calculations, however, the full time-dependent trajectory separation $x_i^+(t)-x_i^-(t)$ is retained in $\mathcal{G}_i(t)$.

Substituting Eq.~\eqref{biasresponse} into Eq.~\eqref{biasphase}, the resulting phase difference is
\begin{equation}
	\delta\phi_B=\sum_{i=1}^{3}\int_{\tau_{i-1}}^{\tau_i}\mathcal{G}_i(t)\delta B_0(t)\mathrm{d}t.\label{biasphasecompact}
\end{equation}

It is also worth noting that the stochastic perturbation in Eq.~\eqref{biasphasecompact} enters directly through the absolute field fluctuation $\delta B_0(t)$, rather than through the fractional fluctuation $\delta B_0(t)/B_0$. This does not imply that the resulting dephasing is completely independent of the nominal bias field $B_0$, since $B_0$ enters the unperturbed trajectories $x_i^\pm(t)$ and therefore affects the diamagnetic part of $\mathcal{G}_i(t)$. The direct Larmor contribution, $-2\gamma_e$, is independent of $B_0$ and, as shown above, dominates the total bias-field response for the parameters considered here. For a given nominal bias field and the corresponding interferometric trajectories, we therefore characterize the bias-field noise by its absolute power spectral density $S_B$ and model $\delta B_0(t)$ as a zero-mean white-noise process,
\begin{equation}
	\left\langle\delta B_0(t)\delta B_0(t')\right\rangle=S_{B}\delta(t-t'),
	\label{biascorrelation}
\end{equation}
\begin{table}[t] 
	\centering 
	\caption{Effective bias-field fluctuations.} 
	\label{dephasingbias} 
	\begin{tabular}{c@{\hspace{40pt}}c} 
		\hline\hline 
		\noalign{\vskip 2pt} 
		$x_0~(\mu\mathrm{m})$ 
		& $\overline{\delta B_0}~(\mathrm{pT})$ \\ 
		\hline 
		\noalign{\vskip 2pt} 
		$10$ & $21.55$ \\ 
		$40$ & $10.85$ \\ 
		$70$ & $8.26$ \\ 
		\hline\hline 
	\end{tabular} 
	
	\vspace{2mm} 
	\parbox{0.92\linewidth}{ 
		\footnotesize 
		Here, $\overline{\delta B_0}$ denotes the estimated effective absolute fluctuation of the externally applied bias magnetic field corresponding to $C_{\phi,B}=0.9$.
	} 
\end{table}
Using Eq.~\eqref{biasphasecompact} and \eqref{biascorrelation}, the corresponding phase variance is
\begin{equation}
	\begin{aligned}
		\left\langle\delta\phi_B^2\right\rangle
		&=\sum_{i,j}\int_{\tau_{i-1}}^{\tau_i}\mathrm{d}t
		\int_{\tau_{j-1}}^{\tau_j}\mathcal{G}_i(t)\mathcal{G}_j(t^\prime)\left\langle\delta B_0(t)\delta B_0(t^\prime)\right\rangle\mathrm{d}t^\prime\\
		&=S_B\sum_{i,j}\int_{\tau_{i-1}}^{\tau_i}\mathrm{d}t\int_{\tau_{j-1}}^{\tau_j}\mathcal{G}_i(t)\mathcal{G}_j(t^\prime)\delta(t-t^\prime)\mathrm{d}t^\prime\\
		&=S_B\sum_i\int_{\tau_{i-1}}^{\tau_i}\mathcal{G}_i^2(t)\mathrm{d}t.\label{biasvariance}
	\end{aligned}
\end{equation}

Following the same convention as in the separation-current-noise analysis, we have
\begin{equation}
	\Gamma_{\phi,B}	=\frac{1}{\tau_3}\left\langle\delta\phi_B^2\right\rangle,\label{biasdephasing}
\end{equation}
and estimate the corresponding phase-coherence contrast as
\begin{equation}
	C_{\phi,B}\simeq\exp\left(-\Gamma_{\phi,B}\tau_3\right).\label{biascontrast}
\end{equation}
To express the bias-field-noise requirement in terms of a more experimentally intuitive absolute field-stability scale, we introduce the effective root-mean-square fluctuation, denoted by $\overline{\delta B_0}$. For consistency with the separation-current-noise analysis, we take the characteristic bandwidth to be $\Delta f=1/\tau_3$. Within the white-noise model, where $S_B$ is frequency independent, this quantity is,
\begin{equation}
	\begin{aligned}
		\overline{\delta B_0}&=\sqrt{\int_0^{\Delta f}S_B\mathrm{d}f}\\
		&=\sqrt{S_B\Delta f}.
	\end{aligned}\label{effectivebiasnoise}
\end{equation}
Here, $\overline{\delta B_0}$ represents the effective absolute root-mean-square fluctuation of the bias field within the bandwidth relevant to one full interferometric sequence. Combining Eq.~\eqref{effectivebiasnoise} with Eq.~\eqref{biascontrast}, the dependence of the contrast $C_{\phi,B}$ on $\overline{\delta B_0}$ is shown in Fig.~\ref{CphiB}. As the absolute bias-field fluctuation increases, the contrast decreases rapidly, and the threshold $C_{\phi,B}=0.9$ is reached at $\overline{\delta B_0}=21.55$, $10.85$, and $8.26~\mathrm{pT}$ for $x_0=10$, $40$, and $70~\mu\mathrm{m}$, respectively. These results show that, for the present protocol, maintaining $C_{\phi,B}\geq0.9$ requires the effective absolute bias-field fluctuation to be controlled at the order of $10~\mathrm{pT}$. Since the bias-field response is dominated by the differential Larmor contribution, this represents a stringent absolute-field-stability requirement for the present protocol in the absence of additional refocusing control. Spin-echo or dynamical-decoupling techniques can in principle suppress low-frequency bias-field fluctuations by reversing the sign of the accumulated Larmor phase. In the present form of the protocol, however, they cannot be incorporated without modifying the demonstrated motional sequence, since a reversal of $S_x$ also reverses the Stern--Gerlach force and hence changes the interferometric trajectories. Their implementation would therefore require a modified control sequence.

\subsection{Finite switching and recombination contrast}

In the three-stage protocol discussed above, the magnetic-field gradient generated by the separation assembly is assumed to change instantaneously between $-\eta$ and $+\eta$. In practice, the reversal of the separation-wire currents requires a finite switching time. Recall that the moments $\tau_1$, $\tau_2$, and $\tau_3$ are determined by the functional conditions of the three interferometric stages, as discussed in detail in Appendix~\ref{protocol}. In Stage 1, the two spin-dependent trajectories are allowed to evolve until their midpoint reaches the vicinity of the origin, with $\tau_1$ defined by $\left[x_1^+(\tau_1)+x_1^-(\tau_1)\right]/2=0$; see Appendix~\ref{separation}. Stage 2 reverses the relative motion of the two branches and initiates their recombination. Its ending time $\tau_2$ is chosen such that the spatial separation returns to its value at the end of Stage 1, $\Delta x_2(\tau_2)=\Delta x_1(\tau_1)=\Delta X_1$; see Appendix~\ref{closing}. Finally, Stage 3 is used to recombine the two spin-conditioned motional states in phase space. Exact phase-space closure corresponds to vanishing residual differences in both position and velocity. In the numerical treatment, $\tau_3$ is chosen such that the residual mismatches $\left|\Delta x_3(\tau_3)\right|$ and $\left|\Delta\dot{x}_3(\tau_3)\right|$ are minimized,  see Appendix~\ref{recombination}.

We retain these functional definitions for $\tau_{1,~2,~3}$ and introduce a finite switching interval $T_{\rm sw}$ only after the corresponding condition of the preceding stage has been reached. Since the magnetic-field gradient is proportional to the separation-wire current, we model each current reversal by a linear variation of $\eta_S$ over the duration $T_{\rm sw}$. The resulting gradient sequence is
\begin{numcases}
	{\eta_S(t)=}
	\eta_1, & $0\leq t\leq\tau_1$, \label{etaswitcha}\\
	\eta_1+\dfrac{\eta_2-\eta_1}{T_{\rm sw}}(t-\tau_1), & $\tau_1<t\leq\tau_1+T_{\rm sw}$, \label{etaswitchb}\\
	\eta_2, & $\tau_1+T_{\rm sw}<t\leq\tau_2$, \label{etaswitchc}\\
	\eta_2+\dfrac{\eta_3-\eta_2}{T_{\rm sw}}(t-\tau_2), & $\tau_2<t\leq\tau_2+T_{\rm sw}$, \label{etaswitchd}\\
	\eta_3, & $\tau_2+T_{\rm sw}<t\leq\tau_3$. \label{etaswitche}
\end{numcases}

Here, $\eta_1$, $\eta_2$, and $\eta_3$ denote the magnetic-field gradients in the three constant-gradient stages. For the nominal protocol considered in this subsection, they are given by $\eta_1=-\eta$, $\eta_2=+\eta$, and $\eta_3=-\eta$, reproducing the original three-stage sequence. This notation is also convenient for introducing independent stage-dependent deviations below without changing the structure of the switching protocol. The finite-switching profile in Eqs.~(\ref{etaswitcha})--(\ref{etaswitche}) is defined for $T_{\rm sw}>0$, since the slopes of the two linear ramps contain the factor $1/T_{\rm sw}$. However, as $T_{\rm sw}\rightarrow0^+$, the corresponding ramp intervals also shrink to zero duration, with $t-\tau_1\rightarrow0$ and $t-\tau_2\rightarrow0$ within the respective switching intervals. The finite ramps therefore reduce continuously to instantaneous changes of the gradient, recovering the piecewise-constant protocol in Eqs.~(\ref{separation1})--(\ref{separation5}). For finite $T_{\rm sw}$, the spin-dependent trajectories are calculated numerically using the time-dependent gradient in Eqs.~(\ref{etaswitcha})--(\ref{etaswitche}), while retaining the original conditions that determine the end of each stage. In the instantaneous-switching limit, the analytical solution of the nominal piecewise-constant protocol gives exact phase-space closure at $t=\tau_3$, with $\Delta X_3(\tau_3)=0$ and $\Delta\dot{X}_3(\tau_3)=0$. For finite $T_{\rm sw}$, the modified trajectories generally no longer satisfy these closure conditions exactly, resulting in finite residual mismatches $\Delta X_3(\tau_3)$ and $\Delta\dot{X}_3(\tau_3)$. 

In this subsection, we examine two complementary control effects on the phase-space recombination of the two spin-conditioned motional states. Throughout the analysis, the nominal constant-stage gradient magnitude is fixed at $|\eta_i^{(0)}|=106.7~\mathrm{T/m}$ and the initial position at $x_0=40~\mu\mathrm{m}$, corresponding to the trajectory configuration shown in Fig.~\ref{example}. We first study the effect of a finite switching duration $T_{\rm sw}$ while keeping the three constant-stage gradients at their nominal values $\eta_1=-\eta$, $\eta_2=+\eta$, and $\eta_3=-\eta$. We then fix the switching duration and examine the effect of independent current deviations $\overline{\delta I}_i$ in the three constant-gradient stages on the recombination contrast. In this analysis, $\overline{\delta I}_i$ are assumed to remain constant within each stage but to vary independently between stages.

\subsubsection{Recombination contrast at finite motional occupation}\label{4211}

In the preceding analysis, the longitudinal c.o.m. motion was taken to be initially prepared in the harmonic ground state, providing a convenient reference for examining the phase-space closure of the two motional states. This assumption separates the dynamical mismatch generated during the interferometric sequence from the additional phase-space broadening associated with finite motional occupation. We now relax this assumption and consider an initial thermal Gaussian state with mean longitudinal occupation $\bar{n}_x$. At the onset of the interferometric sequence, the preparation of the internal spin superposition is assumed not to modify the c.o.m. motional state. Accordingly, the two spin-conditioned motional states are taken to have the same initial phase-space centre and the same initial covariance matrix. During the interferometric sequence, the spin dependence enters only through the linear force term, while the quadratic part of the motional Hamiltonian is identical for the two spin states. Their covariance matrices therefore undergo the same evolution, whereas their phase-space centres follow different spin-dependent trajectories. At the end of Stage 3, the remaining phase-space displacement between the two motional states is characterized by the residual position and momentum differences $\Delta X_3$ and $\Delta P_3$. The recombination contrast at $t=\tau_3$ can then be written as \cite{Roura:2014vwa}
\begin{align}
	C(\bar n_x,T_{\rm sw})&=\exp\left[-(2\bar n_x+1)\Lambda(T_{\rm sw})\right].\label{thermalcontrast}
\end{align}
where
\begin{align}
	\Lambda(T_{\rm sw})&=\frac{1}{2\hbar^2}\Big[\sigma_{p,0}^2(\tau_3)\Delta X_3^2+\sigma_{x,0}^2(\tau_3)\Delta P_3^2\notag\\
	&\qquad-2C_{xp,0}(\tau_3)\Delta X_3\Delta P_3\Big].\label{closureexponent}
\end{align}
Here $\Delta X_3$ and $\Delta P_3$ denote the residual position and momentum differences between the centres of the two spin-conditioned motional states at the end of Stage 3. The quantities $\sigma_{x,0}^2(\tau_3)$ and $\sigma_{p,0}^2(\tau_3)$ denote the position and momentum variances of the evolved zero-temperature reference state, respectively, while $C_{xp,0}(\tau_3)$ is the corresponding position--momentum covariance. All quantities entering $\Lambda$ depend on $T_{\rm sw}$. 
Equation~(\ref{thermalcontrast}) also shows that finite motional occupation enhances the reduction in recombination contrast caused by a given residual phase-space mismatch.
For a fixed target contrast $C$, increasing $\bar n_x$ requires a smaller value of $\Lambda$, thereby imposing a more stringent requirement on the residual phase-space mismatch at recombination. In the limit $T_{\rm sw}\rightarrow0$, the analytical solution guarantees phase-space closure at $t=\tau_3$, such that $\Delta X_3=\Delta P_3=0$, while the covariance remains diagonal and hence $C_{xp,0}(\tau_3)=0$. For finite $T_{\rm sw}$, the switching intervals modify both the spin-dependent c.o.m. trajectories and the evolution of the motional covariance. As a result, finite residual values of $\Delta X_3$ and $\Delta P_3$ generally remain at the end of Stage 3, while a nonzero $C_{xp,0}(\tau_3)$ can also generate an additional position--momentum cross-covariance contribution to $\Lambda$. In the following calculations, the c.o.m. trajectories and the covariance evolution are obtained numerically for each $T_{\rm sw}$. Details of the covariance evolution and the derivation of Eqs.~(\ref{thermalcontrast}) and (\ref{closureexponent}) are given in Appendix~\ref{app:thermalcontrast}. 
\begin{figure}[t]
	\centering
	\includegraphics[width=.9\linewidth]{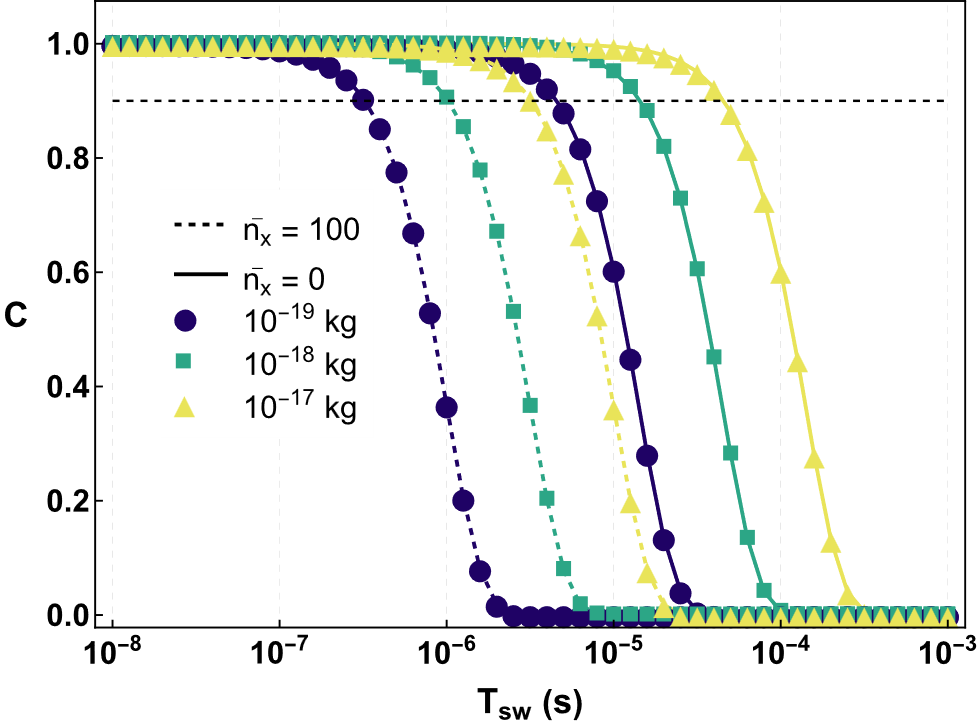}
	\caption{Recombination contrast $C$ as a function of the finite switching time $T_{\rm sw}$ for three nanodiamond masses, $m=10^{-19}$, $10^{-18}$, and $10^{-17}\,\mathrm{kg}$, all released from the same initial position $x_0=40\,\mu\mathrm{m}$, corresponding to the representative trajectories discussed in Fig.~\ref{example}. The circular, square, and triangular markers correspond to the three masses, respectively. Solid curves denote the zero-temperature reference case with $\bar n_x=0$, whereas dashed curves correspond to $\bar n_x=100$. The horizontal dashed line at $C=0.9$ serves as a reference for the switching-time tolerance, and the corresponding values of $T_{\rm sw}$ at which the contrast reaches this reference level are summarized in Table~\ref{tabswitchingtime}. For sufficiently short $T_{\rm sw}$, the recombination contrast remains close to unity, while increasing $T_{\rm sw}$ produces a rapid contrast reduction due to imperfect phase-space recombination. Finite motional occupation shifts this reduction toward shorter switching times, indicating a more stringent switching-time requirement. In contrast, increasing the nanodiamond mass shifts the curves toward longer $T_{\rm sw}$. This behavior follows from the mass dependence of the spin-dependent relative acceleration, $\Delta A\propto1/m$, and the motional phase-space widths, which together yield the approximate scaling $\Lambda\propto1/m$ for a fixed switching protocol (see footnote~\ref{massscaling}).}
	\label{CreM}
\end{figure}

The corresponding dependence of the recombination contrast on the switching time is shown in Fig.~\ref{CreM} for three representative masses, $m=10^{-19}$, $10^{-18}$, and $10^{-17}\,\mathrm{kg}$, all released from $x_0=40\,\mu\mathrm{m}$. The cases $\bar n_x=0$ and $\bar n_x=100$ are shown by solid and dashed curves, respectively, and $C=0.9$ is included as a reference level for comparison. For sufficiently short $T_{\rm sw}$, the recombination contrast remains close to unity, whereas increasing $T_{\rm sw}$ leads to a rapid loss of contrast due to the increasing residual phase-space mismatch. Finite motional occupation shifts the contrast reduction toward shorter switching durations, consistent with the enhanced contrast loss predicted by Eq.~(\ref{thermalcontrast}). Furthermore, as the nanodiamond mass increases, the contrast curves shift toward longer switching durations, indicating that heavier particles can tolerate longer switching times before reaching the same reference contrast. This mass dependence can be understood from the approximate scaling $\Lambda\propto1/m$ for a fixed switching protocol.\footnote{\label{massscaling}
		From Eq.~(\ref{deltak1a}), the spin-dependent relative acceleration scales as $1/m$, while the trapping frequency $\omega_x=\eta\sqrt{-\chi_\rho/\mu_0}$ is independent of $m$. Consequently, for the same gradient waveform and switching duration, the residual relative position and velocity satisfy $\Delta X_3\propto1/m$ and $\Delta\dot{X}_3\propto1/m$, whereas the residual momentum difference $\Delta P_3=m\Delta\dot{X}_3$ is independent of $m$. Meanwhile, Eqs.~(\ref{thermalinitial1}) and (\ref{thermalinitial2}) give $\sigma_{x,0}^2\propto1/m$ and $\sigma_{p,0}^2\propto m$, while the position--momentum covariance also remains independent of $m$ under the corresponding linear evolution. Substituting these relations into Eq.~(\ref{closureexponent}) shows that all three contributions to the contrast-loss exponent scale as $1/m$, yielding $\Lambda\propto1/m$. For fixed interferometric parameters and switching protocol, lighter particles achieve a larger maximum spatial superposition size, $\Delta x_{\max}\propto1/m$, and consequently exhibit larger residual position and velocity mismatches under the same switching process.}

\begin{table}[t]
	\centering
	\caption{Finite switching durations corresponding to $C=0.9$}
	\label{tabswitchingtime}
	\begin{tabular}{c@{\hspace{28pt}}c@{\hspace{28pt}}c}
		\hline\hline
		\noalign{\vskip 2pt}
		$m~(\mathrm{kg})$
		& $T_{\rm sw}~(\mu\mathrm{s}),~\bar n_x=0$
		& $T_{\rm sw}~(\mu\mathrm{s}),~\bar n_x=100$ \\
		\hline
		\noalign{\vskip 2pt}
		$10^{-19}$ & $4.5$ & $0.32$ \\
		$10^{-18}$ & $14$ & $1.0$ \\
		$10^{-17}$ & $45$ & $3.2$ \\
		\hline\hline
	\end{tabular}
	
	\vspace{2mm}
	\parbox{0.92\linewidth}{
		\footnotesize
		The listed values denote approximate finite switching durations at which the recombination contrast reaches the reference level $C=0.9$. Increasing the motional occupation reduces the allowable switching duration.
	}
\end{table}

\begin{figure*}[t!]
	\begin{subfigure}[]{0.41\textwidth}
		\includegraphics[width=\textwidth]{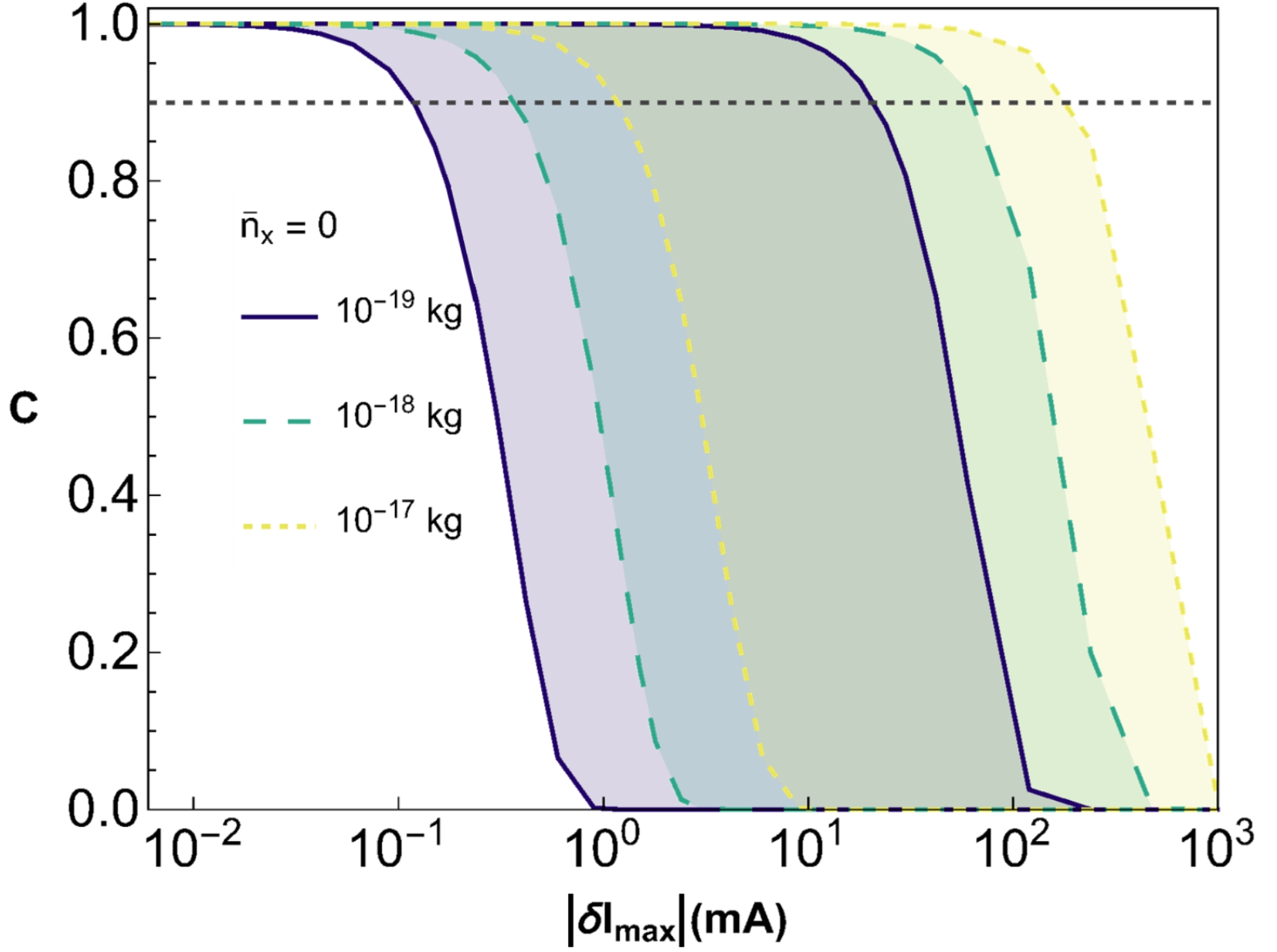}
		\subcaption*{(a)}
	\end{subfigure}\hspace{0.2cm}
	\begin{subfigure}[]{0.41\textwidth}
		\includegraphics[width=\textwidth]{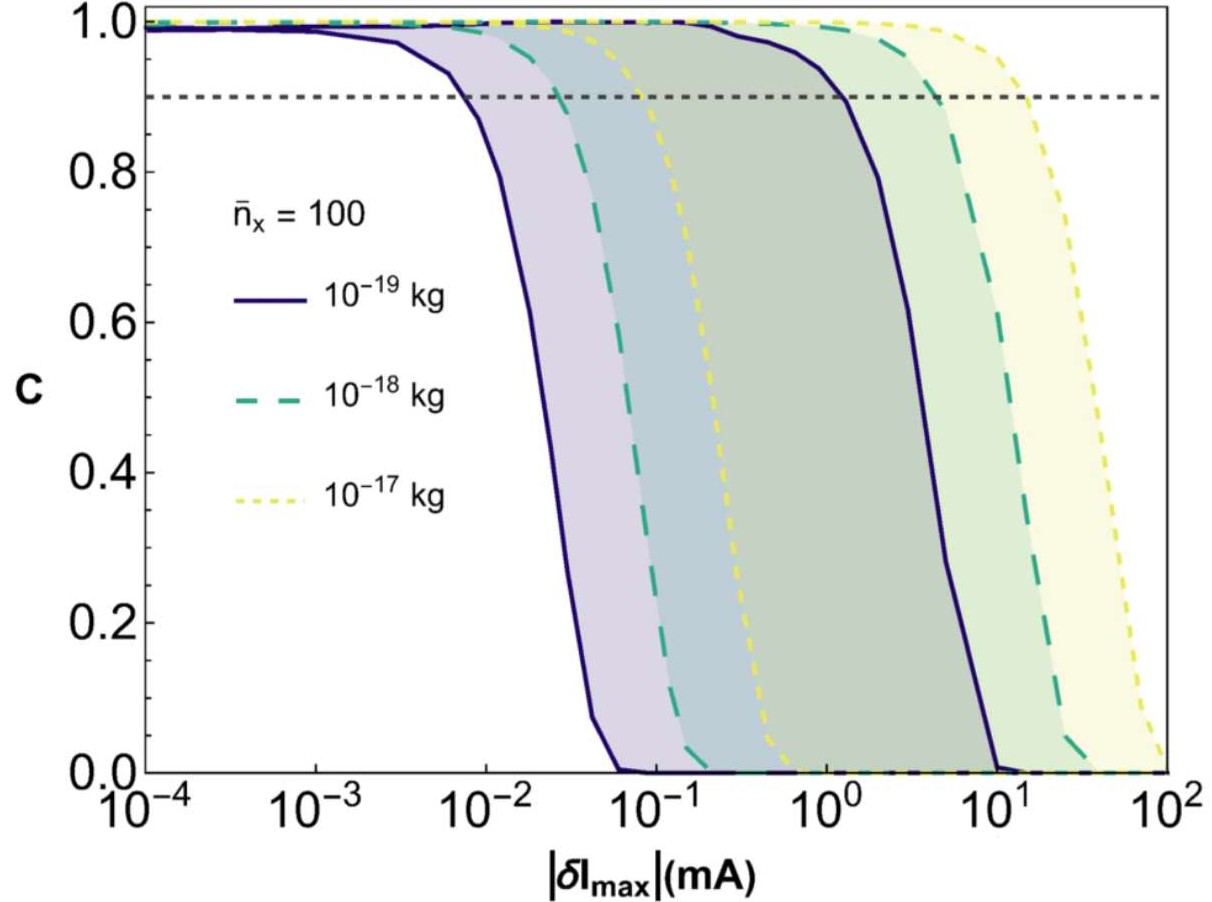}
		\subcaption*{(b)}
	\end{subfigure}\hspace{0.2cm}
	\caption{ Effect of independent current deviations in the three constant-gradient stages on the recombination contrast $C$ for (a) $\bar{n}_x=0$ and (b) $\bar{n}_x=100$. In both panels, the switching duration is fixed at $T_{\rm sw}=0.1~\mu\mathrm{s}$, the nominal gradient magnitude is $|\eta_i^{(0)}|=106.7~\mathrm{T/m}$, the nominal separation-current magnitude is $|I_{\mathrm S}|=6~\mathrm{A}$, and the initial position is $x_0=40~\mu\mathrm{m}$. These parameters correspond to the trajectory configuration shown in Fig.~\ref{example}. The gradients in the three constant-gradient stages are modified by independent current deviations $\overline{\delta I}_i$, which remain constant within each stage and satisfy $|\overline{\delta I}_i|\leq|\delta I_{\max}|$, while the nominal timing parameters remain fixed for all sampled realizations. The solid, long-dashed, and short-dashed curves correspond to particle masses $m=10^{-19}$, $10^{-18}$, and $10^{-17}~\mathrm{kg}$, respectively. A set of $500$ independent random triplets is generated once and reused for all values of $|\delta I_{\max}|$ and all three particle masses. The lower and upper boundaries of each shaded region correspond to the minimum sampled contrast $C_{\rm low}$ and the maximum sampled contrast, respectively, enclosing all $500$ sampled contrast values. The horizontal black dashed line indicates the reference contrast $C=0.9$. The intersections of $C_{\rm low}$ with this reference line determine the estimated maximum absolute current-deviation amplitudes $|\delta I_{\max}|$, which are summarized in Tab.~\ref{currentdev}. Comparison of panels (a) and (b) shows that this reference level is reached at smaller current deviations for the larger thermal occupation.} 
	\label{robustness}
\end{figure*}

\subsubsection{Effect of independent current deviations on the recombination contrast}\label{422}

We further examine the effect of independent current deviations in the three constant-gradient stages defined in Eqs.~(\ref{etaswitcha}), (\ref{etaswitchc}), and (\ref{etaswitche}), where the index $i=1,2,3$ labels these stages.

We assume that the current deviations remain constant within each stage but are independent between different stages. We denote the corresponding constant current-amplitude deviation in the $i$-th stage by $\overline{\delta I}_i$, which may take either positive or negative values relative to the nominal separation-current magnitude $|I_{\mathrm S}|=6~\mathrm{A}$. Denoting the nominal gradients by $\eta_i^{(0)}$, with $\eta_1^{(0)}=-\eta$, $\eta_2^{(0)}=+\eta$, and $\eta_3^{(0)}=-\eta$, and using the proportionality between the magnetic-field gradient and the separation current established in Eq.~(\ref{deltaeta}), the modified gradients are written as
\begin{equation}
	\begin{aligned}
		\eta_i=\eta_i^{(0)}\left(1+\frac{\overline{\delta I}_i}{|I_{\mathrm S}|}\right), \qquad i=1,2,3.
	\end{aligned}
	\label{etaerror}
\end{equation}
Here, $\overline{\delta I}_i$ represents the signed deviation of the current magnitude from its nominal value, where positive and negative values indicate an increase and a decrease in the current magnitude, respectively, rather than the current direction itself. The current direction in each stage is already accounted for by the sign of $\eta_i^{(0)}$. The corresponding constant fractional deviation is therefore given by $\bar{\xi}_i=\overline{\delta I}_i/|I_{\mathrm S}|$, allowing the current stability to be characterized directly in units of amperes.

For a given set of deviations $(\overline{\delta I}_1,\overline{\delta I}_2,\overline{\delta I}_3)$, the three constant-gradient stages in Eqs.~(\ref{etaswitcha}), (\ref{etaswitchc}), and (\ref{etaswitche}) take the modified values $\eta_1$, $\eta_2$, and $\eta_3$, respectively. During the two finite switching intervals defined in Eqs.~(\ref{etaswitchb}) and (\ref{etaswitchd}), the magnetic-field gradient varies linearly from $\eta_1$ to $\eta_2$ and from $\eta_2$ to $\eta_3$, respectively. Throughout this analysis, the switching duration is fixed at $T_{\rm sw}=0.1~\mu\mathrm{s}$, independent of the current deviations. Consequently, the current deviations modify both the constant-gradient values and the endpoints and slopes of the two linear ramps, without introducing additional independent deviations during the switching intervals.

The timing parameters $\tau_1$, $\tau_2$, and $\tau_3$ are determined from the nominal sequence with $\overline{\delta I}_1=\overline{\delta I}_2=\overline{\delta I}_3=0$ and are subsequently kept fixed for all sampled current deviations. No recalculation or re-optimization of these parameters is performed for individual realizations. Consequently, the effect of independent current deviations on the recombination contrast can be evaluated without compensating for the resulting trajectory mismatch through adjustments to the timing parameters.

For this analysis, we retain the nominal constant-stage gradient magnitude $|\eta_i^{(0)}|=106.7~\mathrm{T/m}$ used for the trajectories discussed above, with the finite switching duration fixed at $T_{\rm sw}=0.1~\mu\mathrm{s}$. As indicated by Tab.~\ref{tabswitchingtime}, even in the absence of current deviations, this finite switching duration already produces a recombination contrast below unity. For $x_0=40~\mu\mathrm{m}$, the corresponding nominal timing parameters are $\tau_1=21.348~\mathrm{ms}$, $\tau_2=62.985~\mathrm{ms}$, and $\tau_3=84.333~\mathrm{ms}$. These timing parameters are common to the three masses considered here and remain fixed for all sampled current deviations. Both the ground-state case with $\bar{n}_x=0$ and the thermal case with $\bar{n}_x=100$ are considered.

To investigate the effect of independent current deviations, we consider $500$ sets of stage-dependent deviations $(\overline{\delta I}_1^{(k)},\overline{\delta I}_2^{(k)},\overline{\delta I}_3^{(k)})$, where $k=1,2,\ldots,500$ labels the individual realizations. For each realization, the corresponding deviations modify the three stage gradients through Eq.~(\ref{etaerror}) and are incorporated into the piecewise switching sequence defined by Eqs.~(\ref{etaswitcha})--(\ref{etaswitche}). The resulting evolution is then used to evaluate the recombination contrast from Eq.~(\ref{thermalcontrast}), yielding
\begin{equation}
	\begin{aligned}
		C^{(k)}\equiv C\left(\overline{\delta I}_1^{(k)},\overline{\delta I}_2^{(k)},\overline{\delta I}_3^{(k)}\right), \quad k=1,2,\ldots,500
	\end{aligned}
	\label{csample}
\end{equation}
The current deviations for each realization are generated according to
\begin{equation}
	\begin{aligned}
		\overline{\delta I}_i^{(k)}=|\delta I_{\max}|u_i^{(k)}, \qquad i=1,2,3,
	\end{aligned}
	\label{xirandom}
\end{equation}
where $|\delta I_{\max}|$ specifies the maximum magnitude of the current deviation, and $u_i^{(k)}$ are normalized random variables~\footnote{The normalized random variables $u_i^{(k)}$ are independently sampled from a uniform distribution over $[-1,1]$, with a constant probability density $p(u)=1/2$ within this interval. For any subinterval $[a,b]$, where $-1<a<b<1$, the probability of obtaining a value within this interval is $P(a\leq u\leq b)=\int_a^b p(u)\,du=\int_a^b \frac{1}{2}\,du=\frac{b-a}{2}$.} independently and uniformly sampled over $[-1,1]$. We generate $500$ normalized triplets $(u_1^{(k)},u_2^{(k)},u_3^{(k)})$ and use the same set for all values of $|\delta I_{\max}|$, with their amplitudes rescaled according to Eq.~(\ref{xirandom}). The same normalized triplets are also used for the three particle masses considered here, allowing the dependence on $|\delta I_{\max}|$ and the particle mass to be compared under identical deviation patterns. The sampling ranges of $|\delta I_{\max}|$ are selected separately for $\bar{n}_x=0$ and $\bar{n}_x=100$ to resolve their respective contrast variations.

For each fixed $|\delta I_{\max}|$, particle mass, and thermal occupation, the lower boundary is defined as the minimum contrast among the $500$ sampled realizations, $C_{\rm low}$, while the upper boundary is given by the maximum sampled contrast as shown in Fig.~\ref{robustness}. These two boundaries therefore enclose all $500$ sampled contrast values at each value of $|\delta I_{\max}|$.

The resulting recombination contrasts are shown in Fig.~\ref{robustness}, with panels (a) and (b) corresponding to $\bar{n}_x=0$ and $\bar{n}_x=100$, respectively. The shaded regions represent the ranges between the minimum and maximum sampled contrasts for $m=10^{-19}$, $10^{-18}$, and $10^{-17}~\mathrm{kg}$. In both cases, the contrast remains close to its zero-deviation value for sufficiently small $|\delta I_{\max}|$ and generally decreases as the current deviations increase. Increasing the particle mass shifts the contrast reduction toward larger values of $|\delta I_{\max}|$, indicating that heavier nanodiamonds can tolerate larger current deviations while maintaining the same recombination contrast under the considered interferometric protocol.

The reference level $C=0.9$, indicated by the horizontal black dashed line in Fig.~\ref{robustness}, is used to determine the maximum absolute current-deviation amplitudes corresponding to $C_{\rm low}=0.9$. The intersections of the lower boundaries with this reference level for the three particle masses and two thermal occupations are summarized in Tab.~\ref{currentdev}. At these thresholds, all $500$ sampled realizations maintain $C\geq0.9$, providing a numerical estimate of the allowable current-deviation amplitude within the adopted sampling procedure.

Comparing Figs.~\ref{robustness}(a) and (b), the same reference contrast is reached at smaller current deviations when the thermal occupation is increased. This behavior follows from the thermal factor $(2\bar{n}_x+1)$ in the recombination exponent, which enhances the contrast reduction caused by a given residual phase-space displacement. Conversely, increasing the particle mass allows larger current deviations for the same reference contrast, consistent with the mass dependence of the phase-space closure discussed in Fig.~\ref{CreM}.

\begin{table}[t]
	\centering
	\caption{Maximum absolute separation-current deviation amplitudes $|\delta I_{\max}|$ corresponding to the intersections of the minimum sampled contrast $C_{\rm low}$ with the reference level $C=0.9$ for the two motional occupations considered.}
	\label{currentdev}
	\begin{ruledtabular}
		\begin{tabular}{c c c}
			Mass (kg) &$|\delta I_\text{max}|$, $\bar n_x=0$ &$|\delta I_\text{max}|$, $\bar n_x=100$ \\
			\colrule\noalign{\vskip 2pt}
			$10^{-19}$ & $120~\mu\mathrm{A}$ & $7.4~\mu\mathrm{A}$ \\
			$10^{-18}$ & $365~\mu\mathrm{A}$ & $26.6~\mu\mathrm{A}$ \\
			$10^{-17}$ & $1.2~\mathrm{mA}$ & $81.9~\mu\mathrm{A}$ \\
		\end{tabular}
	\end{ruledtabular}
\end{table}


\color{black}
\section{Conclusion and Discussion}

In this work, we have proposed a chip-based apparatus for generating spatial quantum superpositions of a levitated nanodiamond. The setup comprises two functional components: a levitation assembly and a separation assembly, each comprising four current-carrying parallel wires. 
The levitation assembly consists of superconducting wires carrying a constant current, whereas the gold separation wires provide a switchable magnetic-field gradient by reversing their current directions.
The superimposed magnetic fields create approximately harmonic potentials in all three spatial dimensions. This configuration provides strong confinement in the $y$ and $z$ directions, while enabling the generation of a mesoscopic spatial superposition along the $x$ direction by applying suitable currents in the separation wires. 
A bias magnetic field applied along the $x$ direction keeps the NV spin predominantly quantized along this axis and reduces mixing induced by the residual transverse magnetic field. Furthermore, the levitation and separation assemblies generate magnetic fields that serve distinct purposes and have only weak mutual influence. This separation of roles leads to nearly decoupled motional degrees of freedom and offers precise and flexible control of the spatial superposition trajectories.

For the representative device parameters considered here, the levitation wires carry a current $|I_{\rm L}|=14~\mathrm{A}$ and generate a transverse magnetic-field gradient $\eta_{\rm L}\simeq8.67\times10^{4}~\mathrm{T/m}$, resulting in tight confinement with $\omega_y\simeq\omega_z\simeq6.1\times10^{3}~\mathrm{rad/s}$. The separation wires carry a current magnitude $|I_{\rm S}|=6~\mathrm{A}$ and generate a nominal gradient magnitude $|\eta_{\rm S}|\simeq106.7~\mathrm{T/m}$. In the nominal three-stage full-loop protocol, the current magnitude is kept fixed and only its direction is reversed between consecutive stages. Numerical simulations show that the two spin-conditioned motional states can be recombined in both position and momentum without requiring independent tuning of the current magnitude in the three stages.

For particles released from $x=-40~\mu\mathrm{m}$, the maximum spatial separations are approximately $17.8~\mu\mathrm{m}$, $1.78~\mu\mathrm{m}$, and $0.18~\mu\mathrm{m}$ for masses $m=10^{-19}$, $10^{-18}$, and $10^{-17}~\mathrm{kg}$, respectively, with full-loop durations below $0.1~\mathrm{s}$. The achievable separation can be further increased by increasing the initial release distance.

We have further examined the phase-space recombination beyond the assumptions of instantaneous gradient reversal and ground-state motional preparation. Finite current-reversal times are described by linear switching ramps, while the longitudinal motional state is treated as a thermal Gaussian state with mean occupation $\bar n_x$. The resulting recombination contrast depends on both the residual phase-space displacement and the evolved motional covariance. Taking $C=0.9$ as a reference, the corresponding switching durations for $m=10^{-19}$, $10^{-18}$, and $10^{-17}~\mathrm{kg}$ are $4.5~\mu\mathrm{s}$, $14~\mu\mathrm{s}$, and $45~\mu\mathrm{s}$ for $\bar n_x=0$, and $0.32~\mu\mathrm{s}$, $1.0~\mu\mathrm{s}$, and $3.2~\mu\mathrm{s}$ for the representative finite-occupation case $\bar n_x=100$. Finite motional occupation therefore causes the same reference contrast to be reached at shorter switching durations. We emphasize that $T_{\rm sw}$ represents a prescribed finite switching duration, which is distinct from stochastic timing jitter (see for example Ref.~\cite{narasimha2025magnetic}) associated with fluctuations in the switching moments.

The effect of independent current deviations in the three constant-gradient stages was also examined while keeping the nominal timing sequence fixed. For $T_{\rm sw}=0.1~\mu\mathrm{s}$ and $x_0=40~\mu\mathrm{m}$, a set of $500$ independent random triplets was generated once and reused for different values of $|\delta I_{\max}|$ and the three particle masses. Taking the minimum sampled contrast $C_{\rm low}=0.9$ as a reference, the corresponding maximum absolute current-deviation amplitudes are $120~\mu\mathrm{A}$, $365~\mu\mathrm{A}$, and $1.2~\mathrm{mA}$ for $\bar n_x=0$, and $7.4~\mu\mathrm{A}$, $26.6~\mu\mathrm{A}$, and $81.9~\mu\mathrm{A}$ for $\bar n_x=100$, for masses $m=10^{-19}$, $10^{-18}$, and $10^{-17}~\mathrm{kg}$, respectively.

The stability requirements associated with interferometric dephasing are more stringent. For the representative trajectory with $x_0=40~\mu\mathrm{m}$, the reference contrast $C_{\phi,I}=0.9$ corresponds to an effective fractional separation-current fluctuation $\xi_{\rm eff}\simeq1.5\times10^{-9}$, or an effective absolute current fluctuation of approximately $9~\mathrm{nA}$ for $|I_{\rm S}|=6~\mathrm{A}$. The corresponding bias-field requirement for $C_{\phi,B}=0.9$ is an effective absolute fluctuation of approximately $11~\mathrm{pT}$. Notably, even within the instantaneous-switching approximation adopted in the dephasing analysis, achieving a high phase-coherence contrast already requires stringent current stability.

The open geometry of the device along the $z$ and $x$ directions provides flexibility for the initial loading and positioning of the nanodiamond, as well as optical or microwave access for subsequent spin-state preparation and readout. These auxiliary procedures are not tied to a specific implementation and may therefore be incorporated independently of the interferometric sequence considered here.

Several experimental challenges nevertheless remain before the proposed scheme can be implemented. We assume throughout this work that the nanodiamond contains a single addressable NV centre. Realizing a low-defect-density nanodiamond while retaining the required particle size and spin properties represents an additional materials and fabrication challenge. Recent progress in the fabrication and selection of single-NV nanodiamonds for massive-particle matter-wave interferometry provides important guidance for future experimental developments in this direction \cite{Givon:2025wyb}. The separation-current drive and the externally supplied bias field require substantial stabilization, while the superconducting levitation wires must sustain the required operating current density in the presence of the local magnetic field. Their practical performance will ultimately depend on the fabrication quality, thermal operating margin, current distribution, and vortex dynamics of the superconducting conductors, as well as on experimental characterization of the fabricated device.

Finally, the recombination contrast evaluated here should be understood as a longitudinal Gaussian phase-space closure estimate rather than a complete prediction of the experimentally observable visibility. A more complete experimental model would additionally need to incorporate circuit-specific switching dynamics, transverse and rotational motion, residual anharmonicities, material-dependent thermal effects, local magnetic near-field noise, and the detailed loading, cooling, spin-control, and readout procedures. These effects will need to be addressed in a complete experimental realization. Nevertheless, within the assumptions and parameter ranges considered here, the present analysis suggests that a single chip-based magnetic architecture may provide a possible route toward combining stable magnetic levitation, controlled generation of spatial superposition, and phase-space recombination. The effects of finite switching duration, finite motional occupation, independent current deviations, and magnetic-field-induced dephasing have been examined within the present theoretical framework, while a more comprehensive assessment of experimental feasibility remains necessary.

\color{black}

\section*{acknowledgments} 

QX thanks Run Zhou, Meng-zhi Wu, and Tian Zhou for extensive discussions and acknowledges the support from the China Scholarship Council (CSC). AG is supported in part by NSF grants PHY-2409472 and PHY-2111544, the Heising-Simons Foundation, the W.M. Keck Foundation, the John Templeton Foundation, DARPA, and ONR Grant N00014-18-1-2370. SB’s work is supported by an 
EPSRC grant EP/X009467/1. AG, SB, and AM's research is funded by the Gordon and Betty Moore Foundation through Grant GBMF12328, DOI 10.37807/GBMF12328. This material is based on work supported by the Alfred P. Sloan Foundation under Grant No. G-2023-21130. AM warmly acknowledges support from an NSERC discovery grant (RGPIN-2025-06182).

\section{Appendices}
\begin{appendices}
\section{Magnetic field}\label{sbfromwidthtothin}

Considering that the four wires of the levitation assembly in Fig.~\ref{spdevice} are in very close proximity to the diamond particle, when numerically calculating the magnetic field they generate, we account for both their width and thickness, i.e., their extension in the $z$- and $y$-direction, respectively. In the main text, we assume these wires carry a current magnitude of $14$ A and have identical width and thickness, both of $w_\text{L}=10~\mu$m. It should be noted that this relates to a very high current density, which relies on advanced superconducting material. As a theoretical study, we use this configuration to achieve a smaller $B_z$, ensuring that the NV-centre primarily couples with $B_x$. The calculation of the magnetic field generated by a wire with a quadrilateral cross-section is relatively straightforward, see for example, \cite{panofsky2012classical}. Here we provide a brief description.
\begin{table}[h]
	\caption{\label{levtab}Positions and currents of the levitation wires.}
	\begin{ruledtabular}
		\begin{tabular}{lll}\label{tab2}  
			$k$ & \textrm{($y_k,z_k$) position}& current \\ 
			\colrule
			$k=1$ & ($a,b$)&   $I_{\text{L},1}=-|I_{\text{L}}|$\\
			$k=2$ & ($-a,b$)&  $I_{\text{L},2}=|I_{\text{L}}|$\\
			$k=3$ & ($-a,-b$)& $I_{\text{L},3}=-|I_{\text{L}}|$\\
			$k=4$ & ($a,-b$)&  $I_{\text{L},4}=|I_{\text{L}}|$
		\end{tabular}
	\end{ruledtabular}
\end{table}

$y$-component of levitation assembly:
\begin{align}
	\mathbf{B}_\text{L}\cdot\mathbf{e}_y&=\sum_{k=1}^{k=4}\int_{z_{k}-\frac{w}{2}}^{z_{k}+\frac{w}{2}} \int_{y_{k}-\frac{w}{2}}^{y_{k}+\frac{w}{2}} f_k(y_{s},z_{s})~ \text{d} y_{s}\text{d} z_{s} \label{finite1int}\\ 
	f_k&=-\frac{\mu_0 J_{k}}{2\pi}\frac{(z-z_s)}{(y-y_s)^2+(z-z_s)^2}\label{finite1}
\end{align}

$z$-component of levitation assembly:
\begin{align}		  			   \mathbf{B}_\text{L}\cdot\mathbf{e}_z&=\sum_{k=1}^{k=4}\int_{z_{k}-\frac{w}{2}}^{z_{k}+\frac{w}{2}} \int_{y_{k}-\frac{w}{2}}^{y_{k}+\frac{w}{2}}g_k(y_{s},z_{s}) ~ \text{d} y_{s}\text{d} z_{s} \label{finite2int}\\ 
	g_k&=\frac{\mu_0 J_{k}}{2\pi}\frac{(y-y_s)}{(y-y_s)^2+(z-z_s)^2}\label{finite2}
\end{align}
In the expressions above, Eqs.~(\ref{finite1}, \ref{finite2}) represent the simplest expressions for the $y$- and $z$-components of the magnetic field generated by an infinitely thin wire, with the original current replaced by the current density $J = I_L/w^2$. Here, $(y,z)$ denotes the field point, $(y_s,z_s)$ denotes arbitrary source point on the wire cross-section where we define the centre point of $k^\text{th}$ wire to be $(y_k,z_k)$, and the subscript $k$ indicates the specific wire number as listed in Tab.~\ref{tab2}. Therefore, after summing up contributions from all four wires, $y$- and $z$-components of the levitation assembly are derived in Eqs.~(\ref{finite1int}, \ref{finite2int}).

For the four wires of the separation assembly, recall that their mutual separation is $2L=300~\mu\text{m}$ as shown in Fig.~\ref{spdevice}. Recall that, considering the current-carrying capability of the gold wires, we assume their width and thickness to be $w_{\mathrm S}=32~\mu\mathrm{m}$. However, since the separation wires are sufficiently far from the nanodiamond, their finite cross-sectional dimensions introduce only a minor correction to the magnetic field at the particle position. Therefore, we directly adopt the simplest infinitely thin wire model to calculate their magnetic field components, which are given as follows:
\begin{table}[h]
	\caption{\label{splittab}Positions and currents of the separation wires.}
	\begin{ruledtabular}
		\begin{tabular}{lll}\label{table3}
			$k$ & \textrm{($x_k,y_k$) position}& current \\ 
			\colrule
			$k=1$ & ($L,L$)&   $I_{\text{S},1}=-|I_{\text{S}}|$\\
			$k=2$ & ($-L,L$)&  $I_{\text{S},2}=|I_{\text{S}}|$\\
			$k=3$ & ($-L,-L$)& $I_{\text{S},3}=-|I_{\text{S}}|$\\
			$k=4$ & ($L,-L$)&  $I_{\text{S},4}=|I_{\text{S}}|$
		\end{tabular}
	\end{ruledtabular}
\end{table}

$x$-component:
\begin{align}
	&\mathbf{B_\text{S}\cdot\mathbf{e}_x}=\\
	&-\frac{\mu_0 I_{\text{S},k}}{4\pi}\frac{y-y_k}{R^2_{\text{S},k}}\left[\frac{z+z_s}{\sqrt{(z+z_s)^2+R^2_{\text{S},k}}}\right]_{z_s=-l}^{z_s=l}\label{finite3}
\end{align}

$y$-component:
\begin{align}
	&\mathbf{B_\text{S}\cdot\mathbf{e}_y}=\\
	&\frac{\mu_0 I_{\text{S},k}}{4\pi}\frac{x-x_k}{R^2_{\text{S},k}}\left[\frac{z+z_s}{\sqrt{(z+z_s)^2+R^2_{\text{S},k}}}\right]_{z_s=-l}^{z_s=l}\label{finite4}
\end{align}
Similarly, $I_{k}$ denotes the current magnitude in each of these four wires, where $k=1, 2, 3, 4$ serves as a dummy index. The coordinates ($x_k$, $y_k$) represent both the position of the wire in the $x$-$y$ plane and the location of the source point. $R_{\text{S},k}=\sqrt{(x-x_k)^2+(y-y_k)^2}$ corresponds to the shortest distance from the field point ($x$, $y$) to the wire positioned at ($x_k$, $y_k$). Position of these four wires in the ($x,y$) plane is given in 
Tab.~\ref{table3}. Regarding the lengths of these wires, denoted as $2l$, since all eight wires can extend freely along their longitudinal direction, this value can be set arbitrarily. In our numerical calculations, we have selected $2l=300~\mu$m.


\subsection{Magnetic levitation \& confinement}\label{secsbpantaylor}

Considering the magnetic field 
Eqs.~(\ref{Bfield1}),(\ref{Bfield2}) and (\ref{Bfield3}), we aim to create the magnitude of the field, which can be written as, 
\begin{equation}
    \begin{aligned}
          &B(x,y,z):=|\mathbf{B}| =\sqrt{(\mathbf{B}_\text{L}+\mathbf{B}_\text{S})^2}\\
          &= \sqrt{(\eta_\text{S}x+B_0)^2+(\eta_\text{L}^2 z^2)+(\eta_\text{L}+\eta_\text{S})^2y^2},
    \end{aligned}\label{Bfield4}
\end{equation}
Substituting this magnetic field into the diamagnetic term of Eq.~(\ref{3dHamiltonian}) reveals that this form of linear magnetic field creates mutually uncoupled harmonic oscillator potentials along the $x$, $y$, and $z$ directions. For $y$ and $z$ direction, the diamagnetic forces are,

    \begin{align}
          F_\text{L}&=-\partial_z\frac{-\chi_\rho m}{2\mu_0}\mathbf{B}^2= \frac{\chi_\rho m}{\mu_0}\eta_\text{L}^2 z=-m\omega^2_zz\label{fz}\\
          F_y&=-\partial_y\frac{-\chi_\rho m}{2\mu_0}\mathbf{B}^2= \frac{\chi_\rho m}{\mu_0}(\eta_\text{L}+\eta_\text{S})^2 y=-m\omega^2_yy\label{fy}
    \end{align}

Here, $F_\text{L}$ represents the levitation force along the positive $z$-direction that balances gravity, while $F_y$ denotes the diamagnetic restoring force in the $y$-direction. The frequencies $\omega_y$ and $\omega_z$ correspond to those defined in Eqs.~(\ref{ytrap}) and (\ref{zeom}), respectively. It is evident that, in the absence of gravity, both forces act to confine the particle along the central axis $(x, 0, 0)$. When $z<0$, $F_\text{L}$ offers a positive force against gravity as shown in Fig.~\ref{figure2} (b).

For the target magnetic field Eqs.~(\ref{Bfield1}, \ref{Bfield2}, \ref{Bfield3}), we plan to employ four wires to generate it, taking the levitation assembly as an example. For the four infinitely long straight wires parallel to the $x$-axis that we consider, they form a rectangle in the $y$-$z$ plane (see 
Fig.~\ref{spdevice}), with any two adjacent wires carrying currents in opposite directions. In this configuration, in the vicinity of the geometric central axis of the levitation assembly, the four wires can only produce magnetic field components along the $y$ and $z$ direction,
	\begin{align}
		B_{\text{L},z}&=\frac{ \mu_0}{2\pi}\sum_{k=1}^4\frac{(y-y_k)I_{\text{L},k}}{(y-y_k)^2+(z-z_k)^2}\label{blz}\\
		B_{\text{L},y}&=\frac{\mu_0}{2\pi}\sum_{k=1}^4\frac{-(z-z_k)I_{\text{L},k}}{(y-y_k)^2+(z-z_k)^2}\label{bly}
	\end{align}
where $k$ indicates four wires placed at different places in the $y-z$ plane with current $I_{\text{L},k}$. 

The magnetic field of the levitation assembly is the superposition of the eight components mentioned above (note that, for ease of discussion, we present only the results for infinitely long straight wires here). Therefore, we can perform a Taylor expansion of Eqs.~(\ref{blz}, \ref{bly}) around $(x, 0, 0)$, summing up the contribution from each wire to obtain the gradient expressions,
\begin{equation}
	\begin{aligned}
	\left.\frac{\partial B_{\text{L},z}}{\partial z}\right|_{(0,0)}&=-\frac{ \mu_0}{\pi}\sum_{k=1}^4\frac{y_kz_k}{(y_k^2+z_k^2)^2}I_{\text{L},k}\\
	&=\frac{4\mu_0}{\pi}\frac{ab}{(a^2+b^2)^2}I_\text{L}
	\end{aligned}\label{sbpanexpantion}
\end{equation}
Taking the partial derivative of Eq.~(\ref{bly}) with respect to $y$ yields a gradient identical to Eq.~(\ref{sbpanexpantion}), differing only in sign.

Similarly, one can estimate the magnetic field gradient of the separation assembly in the vicinity of the geometric centre of the setup using the infinitely long wire model. For our specific parameters, the magnetic field components of the separation assembly can be expressed in the following form,
	\begin{align}
	B_{\text{S},x}&=\frac{ \mu_0}{2\pi}\sum_{k=1}^4\frac{-(y-y_k)I_{\text{S},k}}{(x-x_k)^2+(y-y_k)^2}\label{bsx}\\
	B_{\text{S},y}&=\frac{\mu_0}{2\pi}\sum_{k=1}^4\frac{(x-x_k)I_{\text{S},k}}{(x-x_k)^2+(y-y_k)^2}\label{bxy}
\end{align}
where $k$ indicates four wires placed at different places in the $x-y$ plane with current $I_{k}$.  

Similarly, one can perform a Taylor expansion around the geometric centre of the setup at $(0,0,z)$ to obtain the expression for the magnetic field gradient in the $x$-$y$ plane. 
\begin{equation}
	\begin{aligned}
		\left.\frac{\partial B_{\text{S},x}}{\partial x}\right|_{(0,0)}&=-\frac{ \mu_0}{\pi}\sum_{k=1}^4\frac{x_ky_k}{(x_k^2+y_k^2)^2}I_{\text{S},k}\\
	\end{aligned}\label{sbpanexpantion1}
\end{equation}
Note that this result should be used for estimation purposes only. 

\subsection{Field-dependent critical current density of the NbTiN wires}
\label{app:NbTiN_current}
	
The transport current carried by a type-II superconducting wire is limited by the motion of magnetic vortices in the mixed state. To estimate the current-carrying capability of the NbTiN levitation wires in the presence of the local magnetic field, we use the semi-empirical scaling relation \cite{tsui2012critical}
\begin{equation}
	\begin{aligned}
	J_c(B,T,\varepsilon)={}&A(\varepsilon)\left[T_c(\varepsilon)\left(1-\Lambda^2\right)
	\right]^2B_{c2}(T,\varepsilon)^{n-3}\\
	&\times\left[\frac{B}{B_{c2}(T,\varepsilon)}\right]^{p-1}\left[1-\frac{B}{B_{c2}(T,\varepsilon)}\right]^q ,
	\end{aligned}\label{eq:Jc_full_scaling}
\end{equation}
where
\begin{equation}
	\Lambda=\frac{T}{T_c(\varepsilon)}.\label{eq:reduced_parameters}
\end{equation}
Here, $T$ is the operating temperature, $\varepsilon$ is the mechanical strain, and
$\Lambda=T/T_c(\varepsilon)$ is the reduced temperature. The quantities
$T_c(\varepsilon)$ and $B_{c2}(T,\varepsilon)$ denote the strain-dependent transition temperature and the upper critical field, respectively. The coefficient $A(\varepsilon)$ sets the overall scale of $J_c$, while $n$ characterizes the scaling of the pinning-force amplitude with $B_{c2}$. The exponents $p$ and $q$ describe the reduced-field dependence through $B/B_{c2}(T,\varepsilon)$ and are determined by the dominant vortex-pinning mechanism.
	
For the present estimate, the operating temperature and mechanical strain are taken to be fixed. Therefore, $T_c(\varepsilon)$ and $B_{c2}(T,\varepsilon)$ are treated as constants, and all field-independent factors in Eq.~\eqref{eq:Jc_full_scaling} can be absorbed into
\begin{equation}
    J_0=A(\varepsilon)\left[T_c(\varepsilon)(1-\Lambda^2)\right]^2B_{c2}^{\,n-3}.\label{eq:J0_definition}
\end{equation}
The critical current density then reduces to \cite{talantsev2022new}
\begin{equation}
		J_c(B)=J_0\left(\frac{B}{B_{c2}}\right)^{p-1}
		\left(1-\frac{B}{B_{c2}}\right)^q.\label{eq:Jc_reduced}
\end{equation}
	
The reduced-field dependence in Eq.~\eqref{eq:Jc_reduced} is consistent with the Kramer--Dew-Hughes scaling law for the bulk vortex-pinning force \cite{DewHughes1974},
\begin{equation}
	F_p(B)=F_0\left(\frac{B}{B_{c2}}\right)^p\left(1-\frac{B}{B_{c2}}\right)^q,\label{eq:Fp_scaling}
\end{equation}
where $F_0$ is the pinning-force scale at fixed temperature and strain. At the transport critical current, the Lorentz-force density acting on the vortex system is balanced by the maximum pinning-force density. For a current perpendicular to the local magnetic field, this gives $F_p(B)=J_c(B)B$. Dividing Eq.~\eqref{eq:Fp_scaling} by $B$ therefore yields the same reduced-field dependence as Eq.~\eqref{eq:Jc_reduced}, with the corresponding amplitude identified as $J_0=F_0/B_{c2}$. Hence, the semi-empirical expression in Eq.~\eqref{eq:Jc_full_scaling} can be regarded as an extension of the Kramer--Dew-Hughes scaling law that additionally incorporates the temperature- and strain-dependent variations of $T_c$, $B_{c2}$, and the pinning-force amplitude.
	
For an order-of-magnitude estimate, we assume normal-core point pinning to be dominant. In the Dew--Hughes classification, this pinning mechanism corresponds to $p=1, q=2,$ and hence
\begin{equation}
	J_c(B)=J_0\left(1-\frac{B}{B_{c2}}\right)^2.\label{eq:Jc_point_pinning}
\end{equation}
The reported low-temperature critical current densities for NbTiN are approximately $J_0\simeq1.5\text{--}1.92\times10^{11}~\mathrm{A/m^{2}},$ at approximately $2.2~\mathrm{K}$~\cite{klimov2016critical,zhang2024effect}. Representative values of the upper critical field are $B_{c2}\simeq18\text{--}22~\mathrm{T}$ at low temperature~\cite{lee2024penetration}.
	
For the superconducting wires shown in Fig.~\ref{spdevice}, the local magnetic-field environment includes the bias field, the field generated by the gold wires, and the self-fields of the superconducting wires. The resulting total field is approximately $B\approx0.8~\mathrm{T}$, as shown in Fig.~\ref{sbpan}(a). Although the bias field, $B_0=330~\mathrm{mT}$, is parallel to the transport-current direction and therefore does not directly contribute to the Lorentz force, we conservatively include it in the effective perpendicular field when estimating the critical current density. This represents a worst-case estimate of the magnetic-field-induced reduction in the current-carrying capability. Equation~\eqref{eq:Jc_point_pinning} then gives
\begin{equation}
	J_c(0.8~\mathrm{T})\simeq1.37\text{--}1.78\times10^{11}~\mathrm{A/m^{2}},\label{eq:Jc_NbTiN_estimate}
\end{equation}
This range is comparable to the current-density scale required for the levitation-wire geometry considered in this work.

Nevertheless, the estimated current density lies close to the experimentally reported critical-current-density scale of thin NbTiN \cite{klimov2016critical,zhang2024effect}. The applicability of such current-density values to micron-scale superconducting wires is expected to depend on the conductor geometry and fabrication conditions. A complete device assessment would need to account for the fabricated-wire thickness and microstructure, as well as nonuniform and edge-enhanced current distributions. The thermal operating margin and possible current crowding at conductor edges will further affect the practically accessible current density. In addition, because NbTiN is a type-II superconductor, vortex motion and flux dynamics may introduce additional magnetic-field noise. A more detailed quantitative assessment of these material- and fabrication-dependent effects would require a dedicated device-level analysis.

\section{Forces induced by nearby surfaces}
\subsection{Estimate of the Casimir--Polder force}\label{cpforce}
To obtain a conservative upper-bound estimate, we approximate the two superconducting wires on one side of the diamond as ideal conductors and further replace them by an infinite perfectly conducting plane. The perpendicular distance from the centre of the
diamond to the inner surface of the superconducting wires is
\begin{equation}
	d=a-\frac{w_{\mathrm L}}{2},
\end{equation}
where $a$ is the distance from the central axis of the device to the centre of each superconducting wire, and $w_{\mathrm L}$ is the wire width. In the retarded regime, the Casimir--Polder potential between a spherical dielectric particle and an infinite perfectly conducting plane is
	\cite{Canaguier-Durand:2011pvh}
\begin{equation}
	U_{\mathrm{CP}}(d)=-\frac{3\hbar cR^3}{8\pi d^4}\frac{\epsilon_r-1}{\epsilon_r+2},\label{UCP}
\end{equation}
where $R$ is the particle radius and $\epsilon_r\approx5.7$ is the static relative permittivity of diamond. The corresponding attractive force is
\begin{equation}
	F_{\mathrm{CP}}(d)=-\frac{\partial U_{\mathrm{CP}}}{\partial d}=-\frac{3\hbar cR^3}{2\pi d^5}\frac{\epsilon_r-1}{\epsilon_r+2}.\label{FCP}
\end{equation}
Since the actual superconducting wires have finite cross sections and do not form a continuous plane, this ideal-plane approximation is expected to overestimate the Casimir--Polder attraction and therefore provides an upper-limit estimate.

For the three particle masses considered in this work, $m=10^{-19}$, $10^{-18}$, and $10^{-17}~\mathrm{kg}$, the corresponding particle radii are obtained from
\begin{equation}
	R=\left(\frac{3m}{4\pi\rho_{\mathrm d}}\right)^{1/3},
\end{equation}
where the density of diamond is taken as $\rho_{\mathrm d}=3500~\mathrm{kg/m^3}$. For the device parameters $2a=18~\mu\mathrm{m}$ and $w_{\mathrm L}=10~\mu\mathrm{m}$, the particle--surface distance is $d=a-\frac{w_{\mathrm L}}{2}=4~\mu\mathrm{m}.$ Using $\epsilon_r=5.7$~\cite{greenwood2012chemistry}, the corresponding Casimir--Polder forces are listed in Tab.~\ref{tab:CPforce}.
\begin{table}[ht]
	\centering
	\caption{Upper-limit estimates of the
			Casimir--Polder force at a particle--surface distance
			$d=4~\mu\mathrm{m}$ for the three particle masses.}
	\label{tab:CPforce}
\begin{tabular}{c c c c}
		\hline\hline
		\noalign{\vskip 2pt}
		$m~(\mathrm{kg})$
		& $R~(\mathrm{nm})$
		& $|F_{\mathrm{CP}}|~(\mathrm{N})$
		& $|F_{\mathrm{CP}}|/|F_L|$ \\
		\hline
		\noalign{\vskip 2pt}
		$10^{-19}$ & $19.0$ & $6.1\times10^{-23}$ & $6.2\times10^{-5}$ \\
		$10^{-18}$ & $40.9$ & $6.1\times10^{-22}$ & $6.2\times10^{-5}$ \\
		$10^{-17}$ & $88.0$ & $6.1\times10^{-21}$ & $6.2\times10^{-5}$ \\
		\hline\hline
\end{tabular}\label{tabCP}
	
\vspace{2mm}
\parbox{0.92\linewidth}{
		\footnotesize
		\textit{} Here, $F_L$ denotes the diamagnetic levitation force given in footnote \ref{FL}. At the equilibrium position, $|F_L|=mg$, and therefore $|F_{\mathrm{CP}}|/|F_L|=|F_{\mathrm{CP}}|/(mg)$. At the equilibrium position, the net forces along the transverse directions vanish because of the symmetry of the trapping configuration, whereas the vertical diamagnetic force balances gravity.}
\end{table}
The ratio $|F_{\mathrm{CP}}|/(mg)$ is independent of the particle mass because $R^3\propto m$, and hence $|F_{\mathrm{CP}}|\propto m$. Even under this conservative ideal-plane approximation, the Casimir--Polder force at $d=4~\mu\mathrm{m}$ is only of order $10^{-5}$ of the gravitational force and is therefore negligible for the present setup. The above estimate accounts only for the attractive force exerted by the superconducting structure on one side of the particle. In the symmetric configuration considered here, an identical superconducting structure is located on the opposite side at the same perpendicular distance. Therefore, when the particle is positioned on the central axis (which is ensured by the trap given by the levitation assembly), the two Casimir--Polder forces are equal in magnitude and opposite in direction; therefore, the leading symmetric contribution cancels at the central axis. 
\subsection{Estimate of the patch potential}\label{patch}
The nanodiamond is assumed to be free of residual net charge and to possess no permanent electric dipole moment. Nevertheless, an external electric field can polarize the particle and induce an electric dipole moment,
\begin{equation}
	\mathbf{p}_{\mathrm{ind}}=\alpha\mathbf{E},
\end{equation}
where $\alpha$ is the static electric polarizability of the nanodiamond. Treating the particle as a homogeneous dielectric sphere, its polarizability is \cite{BohrenHuffman}
\begin{equation}
	\alpha=4\pi\varepsilon_0R^3\frac{\varepsilon_r-1}{\varepsilon_r+2},
	\label{alpha}
\end{equation}
where $R$ is the particle radius and $\varepsilon_r$ is the relative permittivity of diamond.

One possible source of such an electric field is the spatially and temporally fluctuating potential generated by microscopic inhomogeneities on the nearby conductor surfaces. To estimate the resulting electrostatic force, the two nearby conductor surfaces are conservatively approximated as a single infinite planar surface at the minimum particle--surface separation $d=4~\mu\mathrm{m}$. The microscopic surface patches are assumed to be statistically independent and randomly distributed over the surface. The electric-field noise power spectral density at a distance $d$ from the surface is given by \cite{Dubessy:2008krl},
\begin{equation}
	S_E(\omega,d)=2\frac{N\zeta^2}{d}S_V(\omega)\int_0^\infty\frac{k^3e^{-2k}}{\left(d^2+\zeta^2k^2\right)^{3/2}}\,dk,
	\label{patchSE}
\end{equation}
where $S_E(\omega,d)$ is the two-sided power spectral density of the electric-field fluctuations at the particle position, with units of $\mathrm{V^2\,m^{-2}\,s}$. Here, $S_V(\omega)$ is the voltage-noise power spectral density associated with an individual surface patch, $N$ is the effective number of statistically independent patches, and $\zeta$ is the spatial correlation length of the surface potential. The integration variable $k$ is dimensionless and accounts for the contributions from different spatial Fourier components of the surface potential.

For the present geometry, the patch correlation length is assumed to be much smaller than the particle--surface separation, i.e., $d\gg\zeta$.

Under this assumption, Eq.~\eqref{patchSE} reduces to
\begin{equation}
	S_E(\omega,d)\simeq\frac{3N\zeta^2}{4d^4}S_V(\omega).
	\label{patchSE1}
\end{equation}

The mean-square electric field is obtained by integrating the electric-field power spectral density over all temporal frequencies,
\begin{equation}
	\left\langle E^2(d)\right\rangle=\int_{-\infty}^{\infty}\frac{d\omega}{2\pi}S_E(\omega,d)=\frac{3V_{\mathrm{rms}}^2\zeta^2}{4d^4}.\label{E2}
\end{equation}
where the effective root-mean-square surface-potential variation is defined as
\begin{equation}
	V_{\mathrm{rms}}^2\equiv N\int_{-\infty}^{\infty}\frac{d\omega}{2\pi}S_V(\omega).
	\label{eq:Vrms_definition}
\end{equation}
The statistically independent contributions of the surface patches have therefore been absorbed into $V_{\mathrm{rms}}$. Physically, $V_{\mathrm{rms}}$ represents the typical magnitude of the local voltage variations produced by microscopic inhomogeneities on the conductor surface.

The interaction energy of the induced electric dipole with the fluctuating patch field is
\begin{equation}
	\left\langle U_{\mathrm{patch}}\right\rangle=-\frac{\alpha}{2}\left\langle E^2(d)\right\rangle.\label{patchpotential}
\end{equation}
Substitute Eqs. (\ref{alpha},~\ref{E2}) into Eq. (\ref{patchpotential}) the average patch force is
\begin{align}
	\left\langle F_{\mathrm{patch}}\right\rangle&=-\frac{\partial\left\langle U_{\mathrm{patch}}\right\rangle}{\partial d}\nonumber\\
	&=\frac{\alpha}{2}\frac{\partial\left\langle E^2(d)\right\rangle}{\partial d}\nonumber\\
	&\simeq-6\pi\varepsilon_0R^3\frac{\varepsilon_r-1}{\varepsilon_r+2}\frac{V_{\mathrm{rms}}^2\zeta^2}{d^5}.
	\label{Fpatch}
\end{align}
It is worth noting that, since diamond has a positive electric polarizability, the induced dipole is driven toward regions of stronger electric field. Although the patch potentials have random signs and may give a vanishing mean field, $\langle\mathbf{E}\rangle=0$, the mean-square field remains positive, $\langle E^2\rangle>0$, and increases toward the surface. The resulting force is therefore attractive and directed toward the planar surface. Moreover, for a statistically homogeneous and isotropic distribution of surface patches, there is no preferred direction parallel to the plane, so the mean tangential force vanishes.

\begin{table}[ht]
	\centering
	\caption{Estimated patch-potential force relative to the magnetic levitation force for nanodiamonds of different masses. The distance between diamond and wire surface is $d=4~\mu\mathrm{m}$, and the characteristic patch size is assumed to be $\zeta=100~\mathrm{nm}$ ~\cite{bimonte2021measurement,zhuang2022effective}, $\varepsilon_r=5.7$, and $\rho_\text{d}=3500~\mathrm{kg\,m^{-3}}$.}
	\label{tab:patch_force}
	\begin{tabular}{c c c c}
		\hline\hline
		\noalign{\vskip 2pt}
		$m~(\mathrm{kg})$
		& $R~(\mathrm{nm})$
		& \multicolumn{2}{c}{$|\langle F_{\mathrm{patch}}\rangle|/|F_L|$} \\
		\cline{3-4}
		& &
		$V_{\mathrm{rms}}=10~\mathrm{mV}$
		& $V_{\mathrm{rms}}=30~\mathrm{mV}$ \\
		\hline
		\noalign{\vskip 2pt}
		$10^{-19}$ & $19.0$ & $6.92\times10^{-7}$ & $6.23\times10^{-6}$ \\
		$10^{-18}$ & $40.9$ & $6.92\times10^{-7}$ & $6.23\times10^{-6}$ \\
		$10^{-17}$ & $88.0$ & $6.92\times10^{-7}$ & $6.23\times10^{-6}$ \\
		\hline\hline
	\end{tabular}\label{tabpatch}
	\vspace{2mm}
	\parbox{0.92\linewidth}{
		\footnotesize
		\textit{} The levitation assembly also provides a comparably tight trap along the $y$ direction, with $\omega_y\simeq\omega_z$. For a common force-scale comparison, we normalize the estimated patch-potential force by the magnitude of the magnetic levitation force $F_{\mathrm L}$ at the equilibrium position, where $|F_{\mathrm L}|=mg$.  }
\end{table}

For the estimate of the patch-potential force in Eq.~\eqref{Fpatch}, we consider two representative values, $V_{\mathrm{rms}}\approx10$ and $30~\mathrm{mV}$, and take the characteristic patch size to be $\zeta=100~\mathrm{nm}$~\cite{bimonte2021measurement,zhuang2022effective}. The estimated patch-potential forces for the three nanodiamond masses are presented in Tab.~\ref{tabpatch}. For an electrically neutral nanodiamond, the electrostatic force induced by surface-potential variations associated with conductor-surface inhomogeneities is approximately one to two orders of magnitude smaller than the corresponding Casimir--Polder force, as shown in Tab.~\ref{tabCP}. As in the Casimir--Polder estimate, only the contribution from the conductor surfaces on one side of the nanodiamond is included here. The conductors on the opposite side generate a similar patch-potential interaction. Since the patch distributions on the two sides are statistically independent, their forces do not cancel exactly. Nevertheless, the symmetric conductor arrangement is expected to reduce the resulting net patch-potential force compared with the contribution from a single side.

Overall, both the Casimir--Polder and patch-potential forces are small compared with the magnetic forces governing the nanodiamond dynamics. Moreover, the values reported in Tabs.~\ref{tabCP} and~\ref{tabpatch} correspond to the contributions from the conductors on only one side of the particle. For the Casimir--Polder interaction, the symmetric conductor geometry leads to a strong cancellation of the forces from the two opposing sides. For the patch-potential interaction, the surface-potential distributions on the two sides are generally independent and therefore do not cancel exactly; nevertheless, the symmetric geometry is expected to reduce the resulting systematic force imbalance compared with a single-sided configuration. For the parameters considered in Sec.~\ref{seperation}, namely $\eta_{\mathrm S}\approx106.7~\mathrm{T\,m^{-1}}$ and $B_0=0.33~\mathrm{T}$, the Casimir--Polder force is approximately two to three orders of magnitude weaker than the magnetic force acting along the superposition axis, while the estimated patch-potential force is approximately three to five orders of magnitude weaker. Therefore, within the scope of the present theoretical study, the dephasing and recombination-contrast analyses primarily focus on perturbations arising from current fluctuations.

\subsection{state mixing}\label{mix}
Recall that in Eq.~(\ref{Bfield2}), a static bias magnetic field $B_0$ is applied along the laboratory $x$-direction, which is taken to be aligned with the NV axis. This bias field establishes a predominantly $x$-directed spin-quantization axis. However, the residual $z$-component of the magnetic field at the levitation point acts as a transverse perturbation and induces mixing between different spin states. Therefore, $B_0$ must be chosen sufficiently large to suppress this transverse-field-induced state mixing.

To quantify this effect, we isolate and approximate the spin-dependent part of the Hamiltonian in Eq.~(\ref{3dHamiltonian}) as, see Ref. \cite{perdriat2021spin} for example,
\begin{align}
	H_{\mathrm{spin}}
	&=\hbar D \hat{S}_x^2+\hbar \gamma_e B_x \hat{S}_x+\hbar \gamma_e B_z \hat{S}_z\label{spindynamics1}\\
	&\approx\hbar\begin{pmatrix}
		D + \gamma_e B_0 & \gamma_e B_z / \sqrt{2} & 0 \\
		\gamma_e B_z / \sqrt{2} & 0 & \gamma_e B_z / \sqrt{2} \\
		0 & \gamma_e B_z / \sqrt{2} & D - \gamma_e B_0
	\end{pmatrix}.
	\label{spindynamics2}
\end{align}
Equation~(\ref{spindynamics1}) is generally time dependent, since the local longitudinal field $B_x$ experienced by the particle depends on its trajectory. However, the displacement of the particle along the $x$ direction is of order $100~\mu\mathrm{m}$, and we consider a small gradient $\eta_S$. The spatially varying contribution to the longitudinal Zeeman term is therefore negligible, i.e. $\hbar \gamma_e\left(B_0 + \eta_S x\right)\hat{S}_x\approx\hbar \gamma_e B_0 \hat{S}_x.$ Accordingly, we approximate Eq.~(\ref{spindynamics1}) by Eq.~(\ref{spindynamics2}). Here the matrix is written in the common eigen basis of $\hat{S}_x$ and $\hat{S}^2_x$,
\begin{align}
	\{|+1\rangle_x,~|0\rangle_x,~|-1\rangle_x\}.
	\label{eigenbasis}
\end{align}
In this basis, the bias field $B_0$ shifts the energies of the states
$\lvert \pm 1 \rangle_x$, whereas a nonzero residual transverse component
$B_z$ introduces off-diagonal couplings between $\lvert 0 \rangle_x$ and
$\lvert \pm 1 \rangle_x$. For the resulting three-level
Hamiltonian in Eq.~(\ref{spindynamics2}), the eigenvalue equation is
\begin{equation}
	H_{\mathrm{spin}}
	\lvert \phi_n \rangle
	=\hbar \lambda_n\lvert \phi_n \rangle,\qquad n = 1,2,3.
	\label{spineigensystem}
\end{equation}
Here, $\lvert \phi_n \rangle$ and $\hbar \lambda_n$ are the associated eigenstate
and the eigenenergy, respectively.\footnote{$\lvert \phi_n(B_0,B_z) \rangle$ and $\lambda_n(B_0,B_z)$ in (\ref{spineigensystem}) are all functions of $B_0$ and $B_z$, where the eigenstates form a complete
orthonormal basis,
\begin{equation}
	\langle\phi_m(B_0,B_z)\vert\phi_n(B_0,B_z)\rangle=\delta_{mn},
\end{equation}
and
\begin{equation}
	\sum_{n=1}^{3}
	\lvert \phi_n(B_0,B_z) \rangle
	\langle \phi_n(B_0,B_z) \rvert
	=
	\mathbb{I}_3.
	\label{spincompleteness}
\end{equation}
}
Considering the initial spin state
\begin{equation}
	\lvert \psi(0) \rangle
	=
	\frac{1}{\sqrt{2}}
	\left(
	\lvert +1 \rangle_x
	+
	\lvert -1 \rangle_x
	\right),
	\label{initialspinstate}
\end{equation}
Using the completeness of the eigenstates introduced in Eq.~(\ref{spineigensystem}), the time-evolved spin state can be written as \cite{doherty2012theory}
\begin{equation}
	\begin{aligned}
		\lvert \psi(t) \rangle
		&=
		e^{-i H_{\mathrm{spin}} t / \hbar}
		\lvert \psi(0) \rangle
		\\
		&=
		\sum_{n=1}^{3}
		e^{-i \lambda_n t}
		\lvert \phi_n \rangle
		\langle \phi_n \vert \psi(0) \rangle.
	\end{aligned}
	\label{spintimeevolution}
\end{equation}
The probability of leakage into the state $\lvert 0 \rangle_x$ is
\begin{equation}
	\left|c_0(t)\right|^2=\left|\sum_{n=1}^{3}e^{-i \lambda_n t}a_n\right|^2\leq\left(\sum_{n=1}^{3}\left|a_n\right|\right)^2,
	\label{instantaneousleakage}
\end{equation}
where
\begin{equation}
	a_n	={}_x\langle 0 \vert \phi_n \rangle\langle \phi_n \vert \psi(0) \rangle.\label{spectralamplitude}
\end{equation}
The inequality in Eq. (\ref{instantaneousleakage}) is guaranteed by the triangle inequality. We therefore define an upper bound probability for the leakage,
\begin{equation}
	P_0=\left(\sum_{n=1}^{3}\left|a_n\right|\right)^2.
	\label{conservativeleakage}
\end{equation}
\begin{figure*}[t!]
	\begin{subfigure}[]{0.45\textwidth}
		\includegraphics[width=\textwidth]{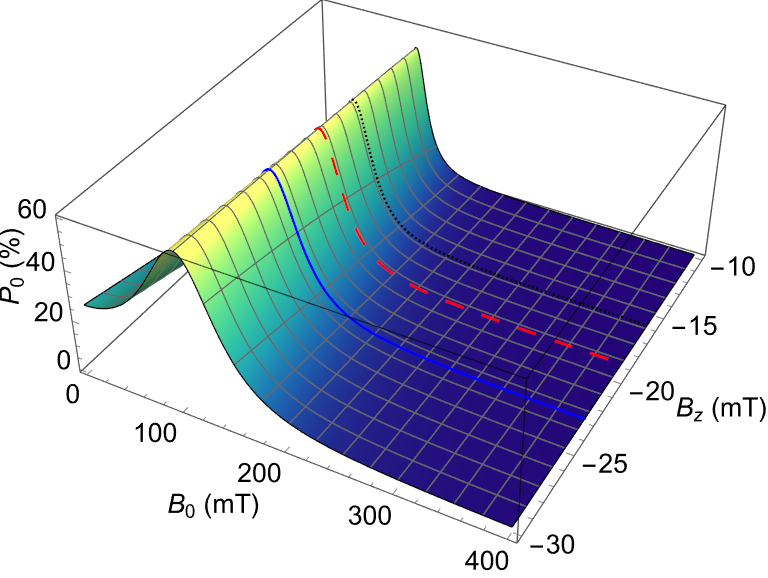}
		\subcaption*{(a)}
	\end{subfigure}
	\hspace{0.01\textwidth}
	\begin{subfigure}[]{0.45\textwidth}
		\includegraphics[width=\textwidth]{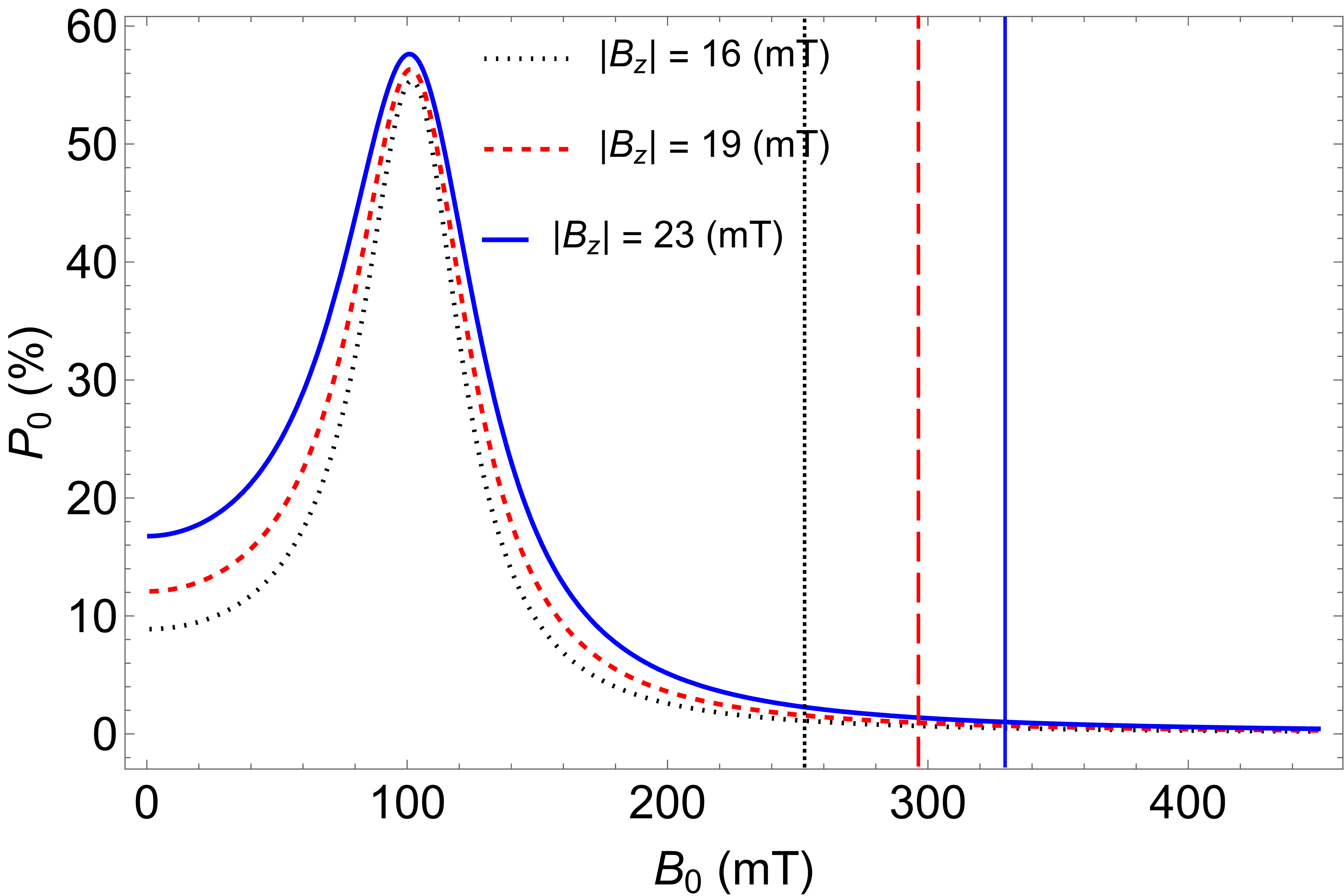}
		\subcaption*{(b)}
	\end{subfigure}
	\caption{We show the upper bound $P_0$ on the state-mixing probability induced by the residual transverse magnetic field. Panel (a) shows the dependence of $P_0$ on the bias field $B_0$ and the residual transverse component $B_z$, while panel (b) shows one-dimensional cuts at $|B_z|=16$, $19$, and $23~\mathrm{mT}$, corresponding to the black, red, and blue curves, respectively. The results show that a sufficiently large bias field is required to suppress transverse-field-induced state mixing. For the three values of $B_z$ considered here, reducing the upper bound below $1\%$ requires $B_0 \gtrsim 255~\mathrm{mT}$, $298~\mathrm{mT}$, and $330~\mathrm{mT}$, respectively.
	}\label{1d3d}
\end{figure*}

This quantity is obtained by diagonalizing the original three-level Hamiltonian in Eq.~(\ref{spindynamics2}). The resulting upper bound depends on both the bias field $B_0$ and the residual transverse component $B_z$. Its dependence on these two field components is shown in Fig.~\ref{1d3d}. In panel (a), we present a three-dimensional plot of $P_0(B_0, B_z)$. Three representative cases are highlighted: $B_z = 16~\mathrm{mT}$ (black dotted line), $19~\mathrm{mT}$ (red dashed line), and $23~\mathrm{mT}$ (blue solid line). These cases are shown in greater detail in panel (b). The results show that, as the bias field $B_0$ increases, the conservative upper bound on the probability for the spin superposition state to mix into the spin-zero state first increases, reaches a maximum near $B_0 \approx 100~\mathrm{mT}$, and then decreases continuously. The maximum occurs because, as can be seen from Eq. (\ref{spindynamics2}), $D - \gamma_e B_0$ approaches zero in this regime, such that the spin-down state becomes nearly degenerate with the spin-zero state. These results indicate that the transition from the spin states to the spin-zero state can be suppressed by increasing the ratio $B_0/B_z$.

\color{black}
\section{Protocol}\label{protocol}
\subsection{Stage-1}\label{separation}
In the separation stage, the magnetic field is given by Eq.~(\ref{separation1}), the potential regarding diamond's $x$-motion can be written as,
\begin{equation}
    \begin{aligned}
          U(x,z_\text{L})= -\frac{\chi_\rho m}{2\mu_0}\mathbf{B}^2(x,z_\text{L})+\hbar \gamma_\text{e}S_x B_x(x,z_\text{L})\,.
    \end{aligned}\label{potential1}
\end{equation}
The equation of motion is then,
\begin{equation}
	\begin{aligned}
		\frac{\text{d}^2x}{\text{d} t^2}&= -\frac{-\chi_\rho }{\mu_0}\eta^2 x -\left(\frac{\chi_\rho B_0}{\mu_0} -\frac{\hbar\gamma_\text{e}S_x}{m}\right)\eta \\
		&= -\omega^2 x-A_\pm.\\
	\end{aligned}\label{xeom1}
\end{equation}
where we define 
\begin{equation}
	\begin{aligned}
		A_\pm=\left(\frac{\chi_\rho B_0}{\mu_0}-\frac{\hbar\gamma_e S_x}{m}\right)\eta
	\end{aligned}\label{apm}
\end{equation}
The solution can be solved as,  
    \begin{align}
          x_1^\pm(t)&=-\left(x_0-\frac{A_\pm}{\omega_x^2}\right)\cos(\omega_x t)-
          \frac{A_\pm}{\omega_x^2}, \label{xsolution1pm}\\
          x_1(t)&=-\left(x_0-\frac{A_0}{\omega_x^2}\right)\cos(\omega_x t)-
          \frac{A_0}{\omega_x^2},\label{xsolution10}\\
          \dot{x}^\pm_1(t)&=\omega_x\left(x_0-\frac{A_\pm}{\omega_x^2}\right)\sin(\omega_x t),\label{vsolution1pm}\\
          \dot{x}_1(t)&=\omega_x\left(x_0-\frac{A_0}{\omega_x^2}\right)\sin(\omega_x t),\label{vsolution10}
    \end{align}
where $x^\pm$ corresponds to spin-up and spin-down, respectively. We also give spin-zero solution $x_1$ for the convenience of discussion. $-x_0<0$ is the initial position of stage-1 which suggests the wave packets initially move toward the positive side of the $x$ axis. The separation in this stage can be written as,
\begin{equation}
    \begin{aligned}
          \Delta x_1(t)&= x^+_1(t)- x^-_1(t)\\
          &=\frac{\Delta A}{\omega_x^2}\left[\cos(\omega_x t)-1\right],
    \end{aligned}\label{deltax1}
\end{equation}
where we define, 
\begin{equation}
    \begin{aligned}
          \Delta A=A_+-A_-=-\frac{2\hbar\gamma_e\eta}{m}.
    \end{aligned}\label{deltak1}
\end{equation}
From Eqs. (\ref{xsolution1pm}-\ref{deltax1}), it is not difficult to find the relation,
\begin{equation}
	\begin{aligned}
		x_1^\pm(t)&=x_1(t)\pm\frac{1}{2}\Delta x_1(t)\\
		\dot{x}_1^\pm(t)&=\dot{x}_1(t)\pm\frac{1}{2}\Delta \dot{x}_1(t).\\
	\end{aligned}
\end{equation}

The duration $\tau_1$ of stage 1 can be chosen freely. Here, our main consideration is that the linearity of the gradient field generated by the separation wires is maintained only over a limited spatial range. Therefore, in order to make maximal use of the region where the field exhibits good linearity, we choose to terminate this stage at the moment $T_1=\tau_1$ when $x_1(\tau_1)=0$.
\begin{equation}
	\begin{aligned}
		\tau_1= \frac{1}{\omega_x}\arccos\left(\frac{B_0}{\eta x_0+B_0}\right)\,.
	\end{aligned}\label{tau1}
\end{equation}

At this moment,
\begin{equation}
	\begin{aligned}
		&X_1^\pm:=x_1^\pm(T_1)=\mp\frac{\Delta A\,x_0}{2\left(\omega_x^2x_0-A_0\right)},\\
		&\dot{X}_1^\pm:=\dot{x}_1^\pm(T_1)=\\
		&\qquad\frac{\omega_x^2x_0-A_0\mp\frac{1}{2}\Delta A}{\omega_x}\sqrt{1-\left(\frac{A_0}{\omega_x^2x_0-A_0}\right)^2},\\
		&\Delta X_1:=\Delta x_1(T_1)=-\frac{\Delta A\,x_0}{\omega_x^2x_0-A_0},\\
		&\Delta \dot{X}_1:=\Delta \dot{x}_1(T_1)=-\frac{\Delta A}{\omega_x}
		\sqrt{1-\left(\frac{A_0}{\omega_x^2x_0-A_0}\right)^2
		}.
	\end{aligned}
\end{equation}


\subsection{Stage-2}\label{closing}
In the second stage, we map the magnetic field from Eq.~(\ref{separation1}) to Eq.~(\ref{separation3}). The only difference between the magnetic field in this stage and that in Stage-1 is that the gradient of the magnetic field is reversed ($-\eta \rightarrow \eta$). This operation corresponds to reversing the current $I_\text{S}$ in the separation device.
The potential and dynamics are essentially the same as 
Eqs.~(\ref{potential1}, \ref{xeom1}) but with different gradient and initial conditions.
Trajectories in the stage-2 can be written as
\begin{align}
	x_2^\pm(t)=&\frac{A_\pm}{\omega_x^2}
	+\left(X_1^\pm-\frac{A_\pm}{\omega_x^2}\right)
	\cos\left[\omega_x(t-T_1)\right]\notag\\
	&\qquad+\frac{\dot{X}_1^\pm}{\omega_x}
	\sin\left[\omega_x(t-T_1)\right],
	\label{xsolution2pm}\\
	x_2(t)=&\frac{A_0}{\omega_x^2}\left\{
	1-\cos\left[\omega_x(t-T_1)\right]\right\}\notag\\
	&\qquad+\frac{\dot{X}_1^++\dot{X}_1^-}{2\omega_x}
	\sin\left[\omega_x(t-T_1)\right],
	\label{xsolution2}\\
	\dot{x}_2^\pm(t)=&
	-\omega_x\left(X_1^\pm-\frac{A_\pm}{\omega_x^2}\right)
	\sin\left[\omega_x(t-T_1)\right]\notag\\
	&\qquad+\dot{X}_1^\pm
	\cos\left[\omega_x(t-T_1)\right].
	\label{vsolution2pm}\\
	\dot{x}_2(t)=&\frac{A_0}{\omega_x}\sin\left[\omega_x(t-T_1)\right]\notag\\
	&\qquad+\frac{\dot{X}_1^++\dot{X}_1^-}{2}
	\cos\left[\omega_x(t-T_1)\right].\label{vsolution2}
\end{align}
\begin{figure}[h]
	\centering
	\includegraphics[width=.75\linewidth]{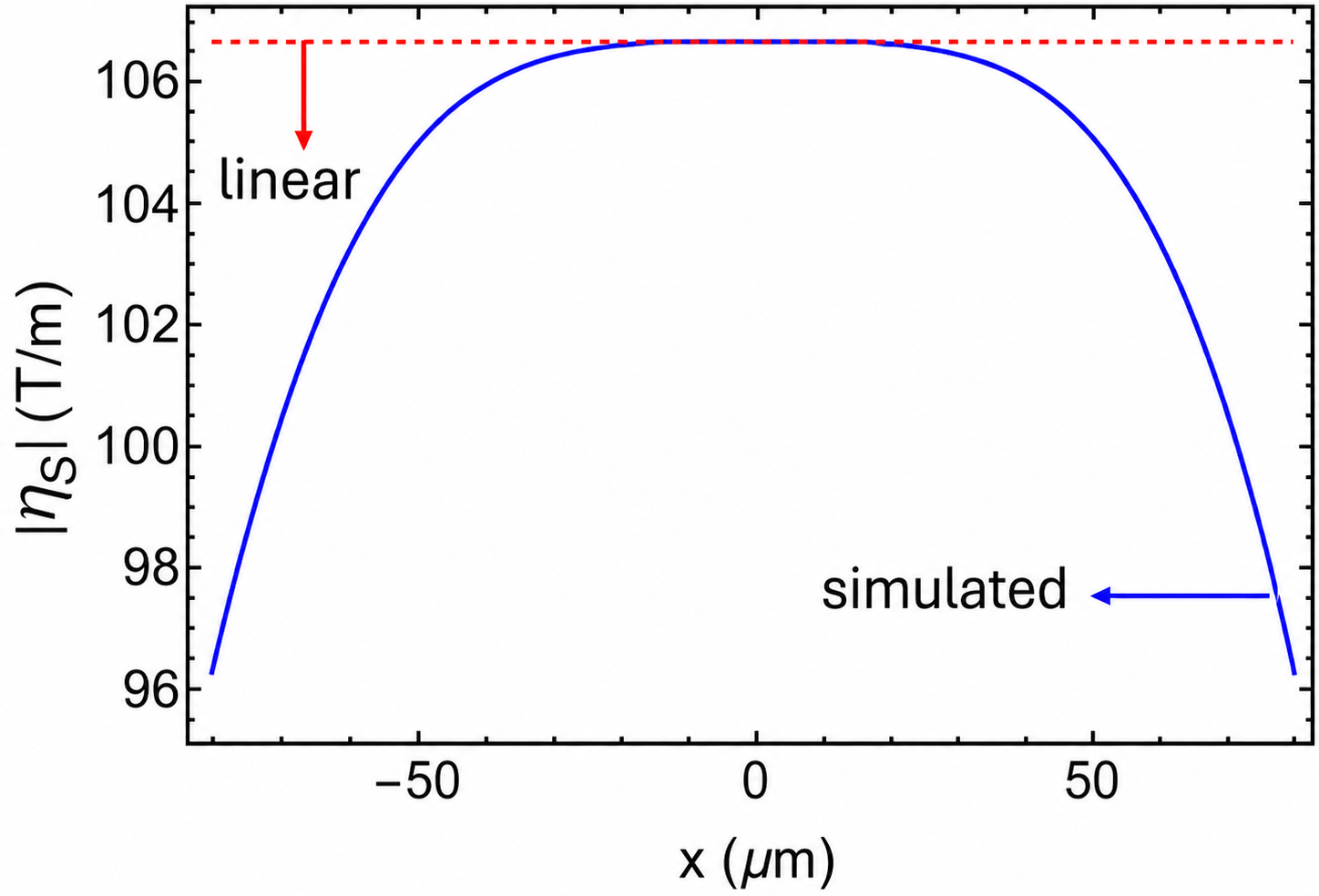}
	\includegraphics[width=.75\linewidth]{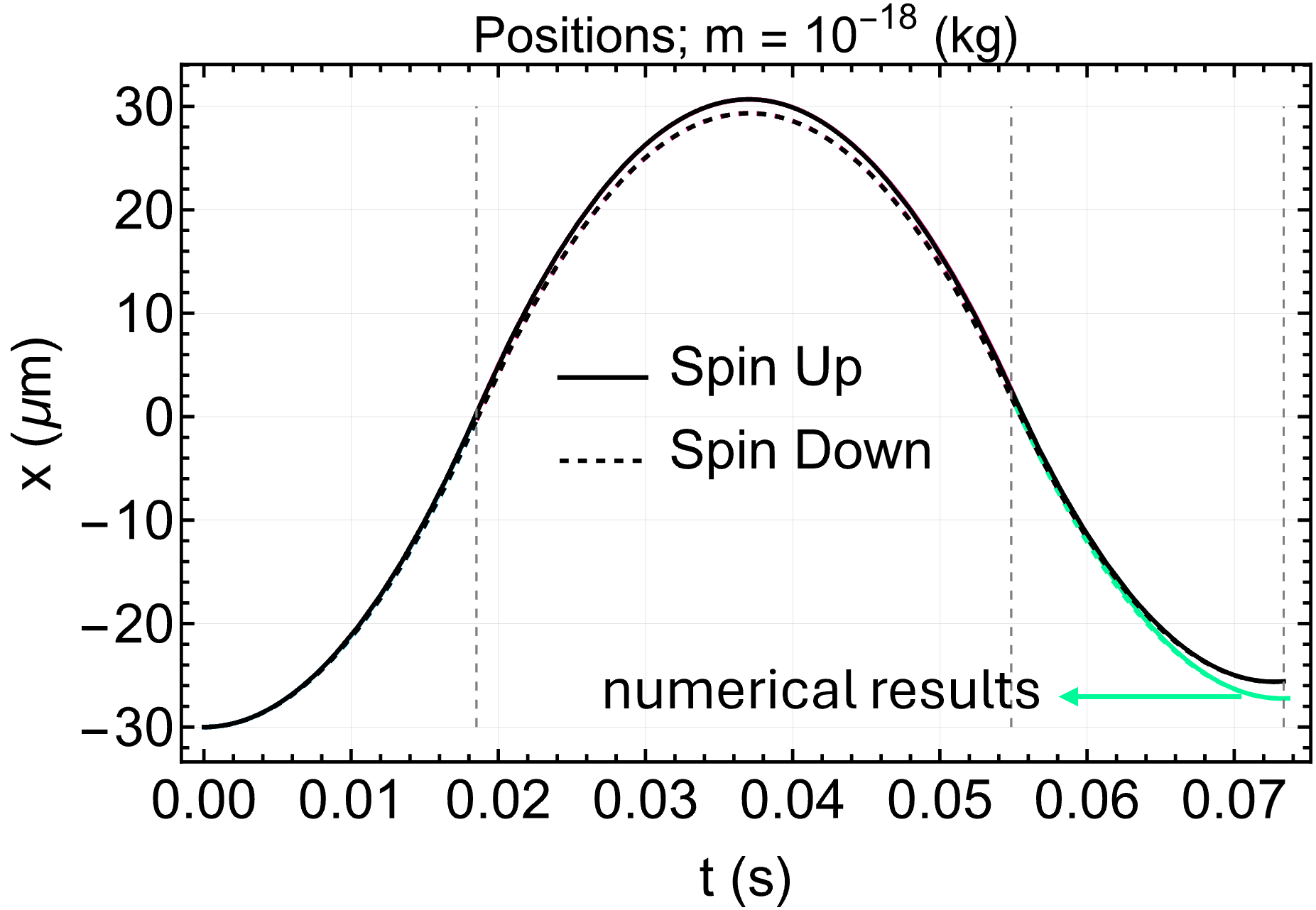}
	\caption{\centering The upper panel compares the ideal linear magnetic field gradient \eqref{Bfield3} used in the analytical treatment (red dashed line) and the simulated magnetic field gradient (blue solid curve) generated from Appendix \ref{secsbpantaylor}. The parameters for simulated gradient are $|I_{\mathrm S}|=6~\mathrm{A}$ and $2L=300~\mu\mathrm{m}$. The lower panel shows the difference between the numerical and analytical trajectories. In this example, the particle has a mass of $m=10^{-18}~\mathrm{kg}$ and is initialized at $x=-x_0=-30~\mu\mathrm{m}$. For the analytical trajectories, the switching times $\tau_1$, $\tau_2$, and $\tau_3$ are calculated from Eqs.~(\ref{tau1a}), (\ref{tau2a}), and (\ref{tau3a}), respectively. For the numerical trajectories, the corresponding switching times are determined numerically according to the criteria introduced in Sec.~\ref{seperation}.}
	\label{nuandan1}
\end{figure}
\begin{figure*}[t]
	\begin{subfigure}[]{0.32\textwidth}
		\includegraphics[width=\textwidth]{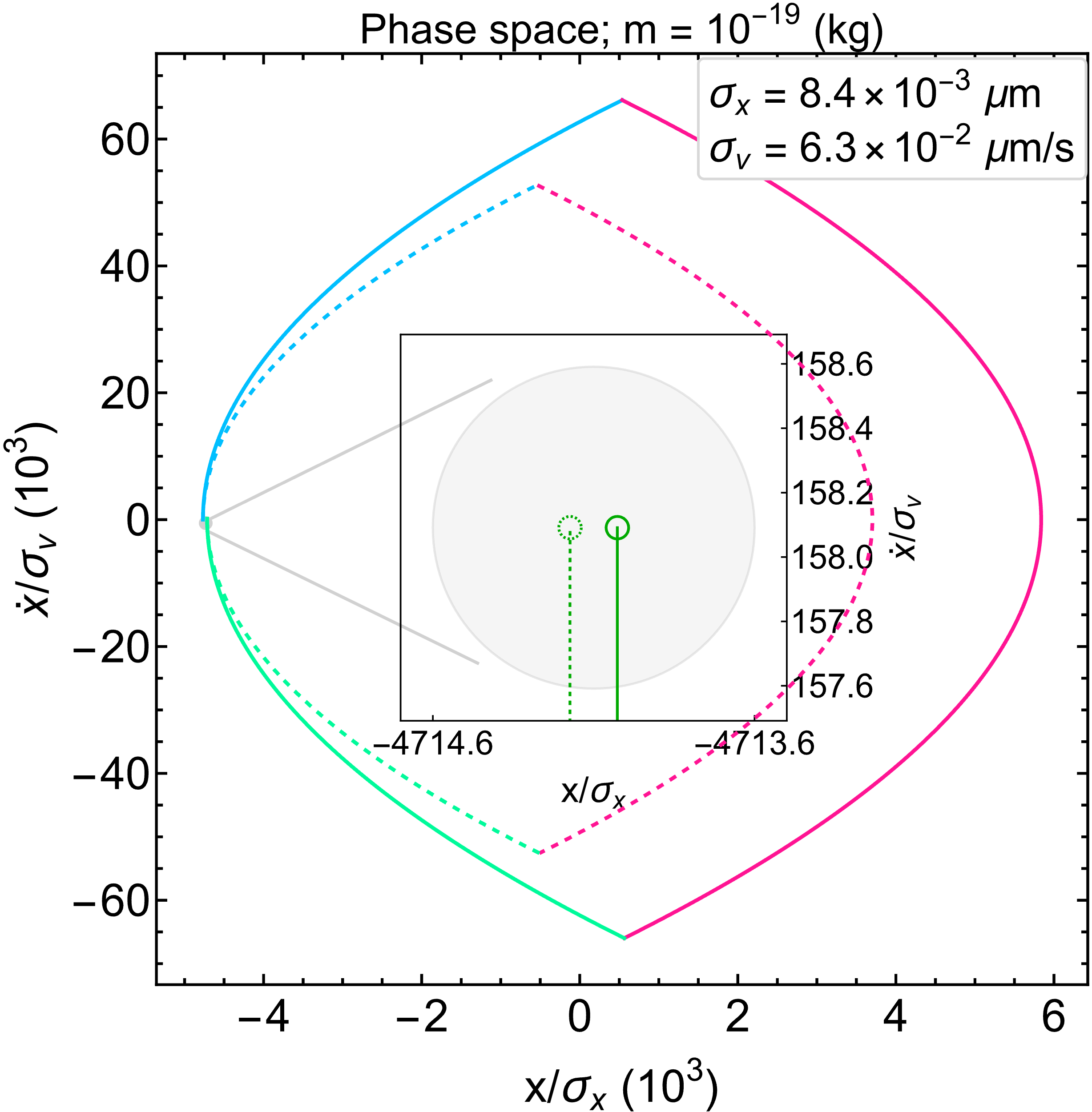}
		\subcaption*{(a)}
	\end{subfigure}\hspace{0.1cm}
	\begin{subfigure}[]{0.32\textwidth}
		\includegraphics[width=\textwidth]{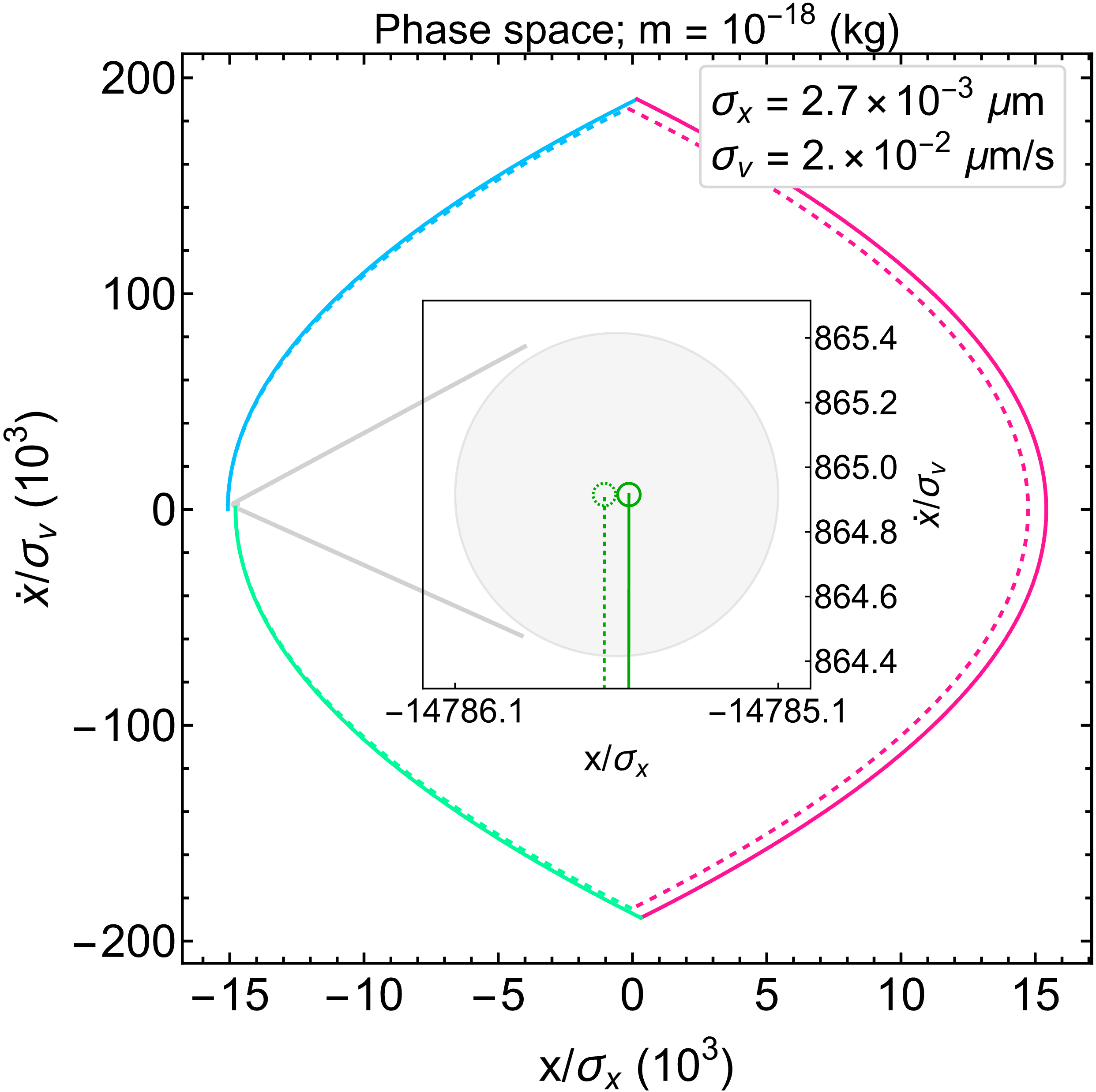}
		\subcaption*{(b)}
	\end{subfigure}\hspace{0.1cm}
	\begin{subfigure}[]{0.336\textwidth}
		\includegraphics[width=\textwidth]{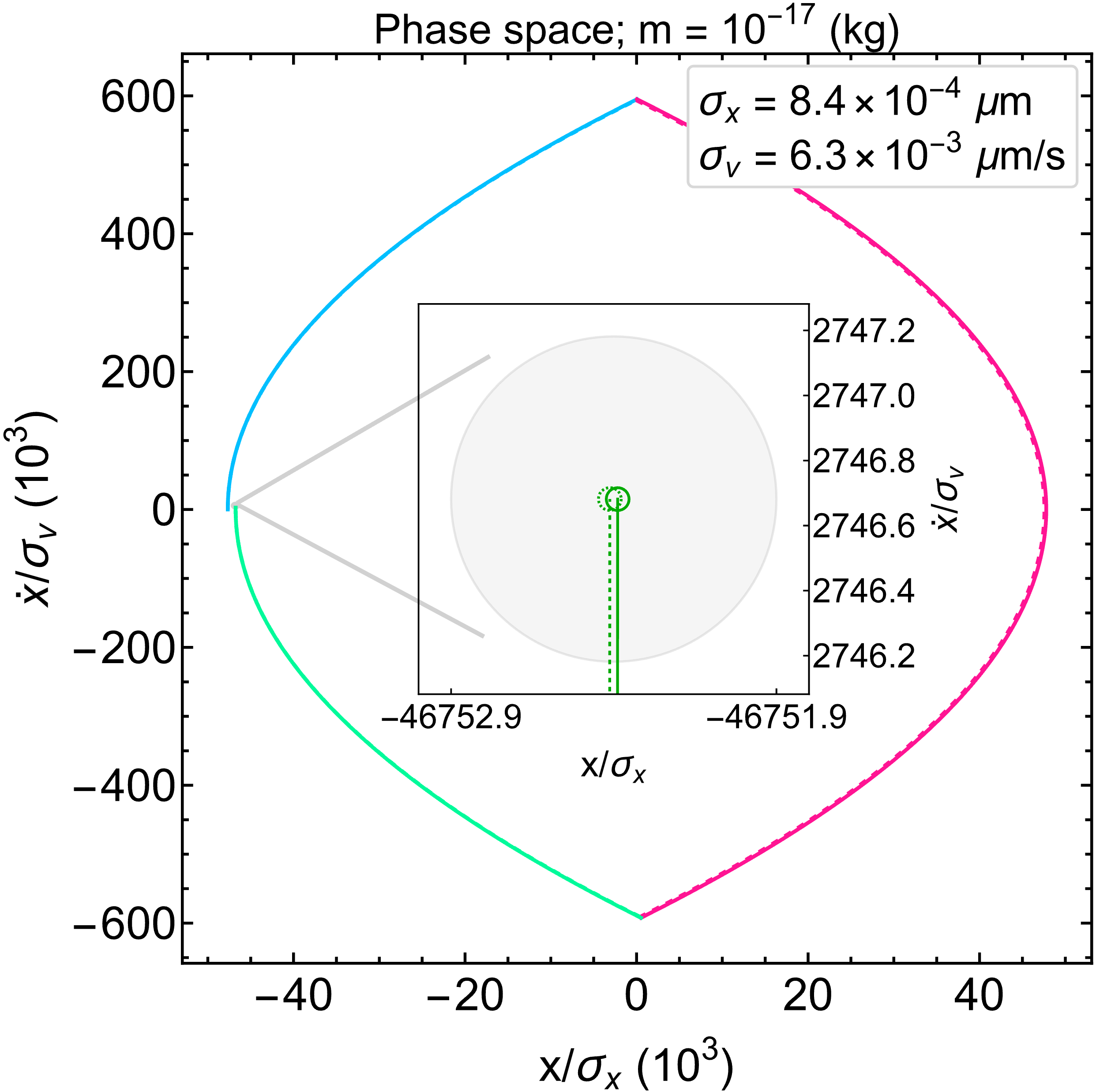}
		\subcaption*{(c)}
	\end{subfigure}\\
	\caption{We show in panels (a)--(c), the position and velocity in phase space scaled by the ground-state position uncertainty $\sigma_x=\sqrt{\hbar/(2m\omega_x)}$ and ground-state velocity uncertainty $\sigma_v=\sqrt{\hbar\omega_x/(2m)}$ of the c.o.m., respectively, with $\omega_x\approx 7.5~\mathrm{rad/s}$. In the three insets, the grey circle represents the region $\sqrt{(\Delta x/\sigma_x)^2+(\Delta \dot{x}/\sigma_v)^2}\leq 1$, indicating that, at the final moment, the classical trajectories return to within one wave-packet scale in phase space, and hence become indistinguishable within the resolution set by the c.o.m. wave packet. }
	\label{example2}
\end{figure*}
The size of the superposition in this stage is then,
\begin{equation}
    \begin{aligned}
         \Delta x_2(t)=R\sin\left[\omega_x(t-T_1)+\alpha\right]+\frac{\Delta A}{\omega_x^2},\label{deltaxposition2}
    \end{aligned}
\end{equation}
where,
\begin{equation}
    \begin{aligned}
         R&=\sqrt{\left(\Delta X_1-\frac{\Delta A}{\omega_x^2}\right)^2+\left(\frac{\Delta \dot{X}_1}{\omega_x}\right)^2},\\
         \alpha&=\arctan\left[\frac{\Delta X_1-\dfrac{\Delta A}{\omega_x^2}}{\dfrac{\Delta \dot{X}_1}{\omega_x}}\right].
    \end{aligned}
\end{equation}

Similar to the trajectories, the superposition size Eq.~(\ref{deltaxposition2}) in this stage continues to increase, at the moment,

\begin{equation}
    \begin{aligned}
        t_\text{max}=T_1+\frac{1}{\omega_x}\left(\frac{\pi}{2}-\alpha\right).
    \end{aligned}
\end{equation}
The size is maximized. After this moment, the size begins to decrease. Here, we choose the ending moment as
\begin{equation}
	\begin{aligned}
		T_2= \frac{2}{\omega_x}\left(\frac{\pi}{2}-\alpha\right)+T_1.
	\end{aligned}
\end{equation}
At this moment, 
\begin{align}
	&X_2^+=\frac{A_0}{\omega_x^2}\left\{1-\cos\left(\omega_x\tau_2\right)\right\}\notag\\
	&\qquad+\frac{\dot{X}_1^++\dot{X}_1^-}{2\omega_x}\sin\left(\omega_x\tau_2\right)+\frac{1}{2}\Delta X_1,\label{capxposition2}\\
	&X_2^-=\frac{A_0}{\omega_x^2}\left\{1-\cos\left(\omega_x\tau_2\right)\right\}\notag\\
	&\qquad+\frac{\dot{X}_1^++\dot{X}_1^-}{2\omega_x}\sin\left(\omega_x\tau_2\right)-\frac{1}{2}\Delta X_1,\\
	&\dot{X}_2^+=\frac{A_0}{\omega_x}\sin\left(\omega_x\tau_2\right)
		+\frac{\dot{X}_1^++\dot{X}_1^-}{2}\cos\left(\omega_x\tau_2\right)-\frac{1}{2}\Delta\dot{X}_1,\\
	&\dot{X}_2^-=\frac{A_0}{\omega_x}\sin\left(\omega_x\tau_2\right)+\frac{\dot{X}_1^++\dot{X}_1^-}{2}\cos\left(\omega_x\tau_2\right)+\frac{1}{2}\Delta\dot{X}_1,\\
	&\Delta X_2:=\Delta x_2(T_2)=X_2^+-X_2^-=\Delta X_1,\\
	&\Delta\dot{X}_2:=\Delta\dot{x}_2(T_2)=\dot{X}_2^+-\dot{X}_2^-=-\Delta\dot{X}_1.\label{capdeltav2}
\end{align}


\subsection{Stage-3}\label{recombination}
At the beginning of the recombination stage, the magnetic-field gradient is reversed relative to Stage 2 and restored to the Stage-1 configuration, namely $\eta\rightarrow -\eta$. Hence, the equation of motion has the same form as that in Stage 1, while the initial positions and velocities are Eqs. (\ref{capxposition2}-\ref{capdeltav2}).
\begin{align}
	x_3^\pm(t)=&-\frac{A_\pm}{\omega_x^2}+\left(X_2^\pm+\frac{A_\pm}{\omega_x^2}\right)\cos\left[\omega_x(t-T_2)\right]\notag\\
	&\quad+\frac{\dot{X}_2^\pm}{\omega_x}\sin\left[\omega_x(t-T_2)\right],\\
	x_3(t)=&-\frac{A_0}{\omega_x^2}+\left[\frac{X_2^++X_2^-}{2}+\frac{A_0}{\omega_x^2}\right]
	\cos\left[\omega_x(t-T_2)\right]\notag\\
	&+\frac{\dot{X}_2^++\dot{X}_2^-}{2\omega_x}\sin\left[\omega_x(t-T_2)\right],\\
	\dot{x}_3^\pm(t)=&-\omega_x\left(X_2^\pm+\frac{A_\pm}{\omega_x^2}\right)\sin\left[\omega_x(t-T_2)\right]\notag\\
	&\quad+\dot{X}_2^\pm\cos\left[\omega_x(t-T_2)\right],\\
	\dot{x}_3(t)=&-\omega_x\left[\frac{X_2^++X_2^-}{2}+\frac{A_0}{\omega_x^2}\right]\sin\left[\omega_x(t-T_2)\right]\notag\\
	&+\frac{\dot{X}_2^++\dot{X}_2^-}{2}\cos\left[\omega_x(t-T_2)\right],\\
	\Delta x_3(t)=&-\frac{\Delta A}{\omega_x^2}+\left(\Delta X_2+\frac{\Delta A}{\omega_x^2}\right)\cos\left[\omega_x(t-T_2)\right]\notag\\
	&\quad+\frac{\Delta\dot{X}_2}{\omega_x}\sin\left[\omega_x(t-T_2)\right],\label{deltaxposition3}\\
	\Delta\dot{x}_3(t)=&-\omega_x\left(\Delta X_2+\frac{\Delta A}{\omega_x^2}\right)\sin\left[\omega_x(t-T_2)\right]\notag\\
	&\quad+\Delta\dot{X}_2\cos\left[\omega_x(t-T_2)\right].
\end{align}
Eq. (\ref{deltaxposition3}) can be rewritten as,
\begin{equation}
	\Delta x_3(t)=R_3\sin\left[\omega_x(t-T_2)+\beta\right]-\frac{\Delta A}{\omega_x^2},
\end{equation}
where, 
\begin{align}
	R_3&=\sqrt{\left(\Delta X_2+\frac{\Delta A}{\omega_x^2}\right)^2+\left(\frac{\Delta\dot{X}_2}{\omega_x}\right)^2},\\
	\beta&=\arctan\left[\frac{\Delta X_2+\dfrac{\Delta A}{\omega_x^2}}{\dfrac{\Delta\dot{X}_2}{\omega_x}}\right].
\end{align}
one can find that at the moment
\begin{equation}
	\tau_3=\tau_2+\frac{1}{\omega_x}\left(\frac{3\pi}{2}-\beta\right),
\end{equation}
the two trajectories are closed in both position and momentum space, satisfying
\begin{equation}
	\Delta x_3(\tau_3)=0,\qquad\Delta \dot{x}_3(\tau_3)=0.
	\end{equation}
For a perfectly linear magnetic field, trajectory closure is strictly guaranteed by Eq.~(\ref{closure1}). However, as discussed in the main text, the magnetic field generated by the wires is accurately linear only in the vicinity of $x=0$. Consequently, the trajectories calculated numerically using the simulated magnetic field exhibit small deviations from the analytical trajectories.

In Fig.~\ref{nuandan1}, we illustrate this discrepancy. In this figure, the particle is initialized at $x=-x_0=-30~\mu\mathrm{m}$. At the final time $\tau_3$, the recombination position obtained from the numerically simulated magnetic field differs slightly from that predicted by the analytical model. This discrepancy mainly arises from the higher-order spatial corrections to the simulated magnetic field. It should be noted that a relatively large initial displacement $x_0$ is deliberately chosen here to make this difference more visible. The magnetic-field gradient is $\eta \approx 106.7~\mathrm{T/m}$, and the corresponding current-switching sequence is given in Tab.~\ref{tab}.

To characterise the closure of the numerical trajectories, we plot the corresponding phase-space trajectories in Figs.~\ref{example}(a), (b), and (c). In these three plots, the positions and velocities of particles with masses $m=10^{-19},~10^{-18},~10^{-17}~\mathrm{kg}$ are normalized by their corresponding ground-state position uncertainty $\sigma_x=\sqrt{\hbar/(2m\omega_x)}$ and ground-state velocity uncertainty $\sigma_v=\sqrt{\hbar\omega_x/(2m)}$ of the c.o.m., respectively, where for each stage $\omega_x\approx 7.5$ rad/s. In the insets of these three plots, we mark the unit region in the normalised phase space, defined by $\sqrt{(\Delta x/\sigma_x)^2+(\Delta \dot{x}/\sigma_v)^2}=1$. It can be seen that, for all three particle masses, the trajectories finally return to this region, indicating good phase-space closure. Moreover, the degree of closure improves as the particle mass increases.

\section{Recombination contrast for a Gaussian motional state with finite switching}
	\label{app:thermalcontrast}
	\label{finiteoccupationcontrast}
	
	For the longitudinal dynamics considered here, the motional states associated with the two spin states can be described within a Gaussian phase-space framework. As will be shown in the following subsection, the quadratic dynamics considered in this work lead to the same wave-packet deformation for the two spin states, while their c.o.m. trajectories can retain a residual displacement in phase space. It is therefore useful to begin with the general Gaussian expression for the contrast \cite{Roura:2014vwa},
	\begin{align}
		&C(\tau_3)\label{gaussiancontrast}\\
		&=\exp\left[-\frac{1}{2\hbar^2}\Delta\boldsymbol{\chi}^{T}(\tau_3)\mathbf{J}^{T}\mathbf{\Sigma}_{\mathrm{th}}(\tau_3)\mathbf{J}\Delta\boldsymbol{\chi}(\tau_3)\right] \notag\\
		&=\exp\left[\frac{-1}{2\hbar^2}
		\begin{pmatrix}
			\Delta X_3\\
			\Delta P_3
		\end{pmatrix}^{T}\mathbf{J}^{T}
		\begin{pmatrix}
			\sigma_{x,\mathrm{th}}^2 & C_{xp,\mathrm{th}}\\
			C_{xp,\mathrm{th}} & \sigma_{p,\mathrm{th}}^2
		\end{pmatrix}\mathbf{J}
		\begin{pmatrix}
			\Delta X_3\\
			\Delta P_3
		\end{pmatrix}\right] \notag\\
		&=\exp\left[-\frac{\sigma_{p,\mathrm{th}}^2\Delta X_3^2+\sigma_{x,\mathrm{th}}^2\Delta P_3^2-2C_{xp,\mathrm{th}}\Delta X_3\Delta P_3}{2\hbar^2}\right].\notag
	\end{align}
	For compactness, the explicit final-time dependence of the quantities entering Eq.~\eqref{gaussiancontrast} is suppressed in its expanded forms. The Gaussian wave-packet parameters introduced below are defined at an arbitrary time $t$, while the recombination contrast is evaluated at the end of Stage 3, $t=\tau_3$.
	
	The vector $\Delta\boldsymbol{\chi}(t)$ describes the instantaneous phase-space displacement between the two c.o.m. trajectories,
	\begin{equation}
		\begin{aligned}
			\Delta\boldsymbol{\chi}(t)&=
			\begin{pmatrix}
				\Delta x(t)\\
				\Delta p(t)
			\end{pmatrix}.
		\end{aligned}\label{phasevector}
	\end{equation}
	Here, $\Delta x(t)=x^+(t)-x^-(t)$ and $\Delta p(t)=p^+(t)-p^-(t)=m[\dot{x}^+(t)-\dot{x}^-(t)]$ denote the instantaneous position and momentum differences between the two c.o.m. trajectories. Following the notation used for the stage-end positions $X_i^\pm$ and velocities $\dot{X}_i^\pm$ in Appendix~\ref{protocol}, we additionally define the corresponding stage-end momenta as $P_i^\pm=m\dot{X}_i^\pm$. The phase-space differences at the end of Stage $i$ are then denoted by $\Delta X_i=X_i^+-X_i^-$ and $\Delta P_i=P_i^+-P_i^-$. In particular, the residual phase-space displacement entering Eq.~\eqref{gaussiancontrast} is $\Delta\boldsymbol{\chi}(\tau_3)=(\Delta X_3,\Delta P_3)^T$.
	
	In the present treatment, $x^\pm(t)$, $p^\pm(t)$, and $\Delta\boldsymbol{\chi}(t)$ describe the continuous evolution through the complete protocol, including the two finite-switching intervals. The stage index $i=1,2,3$ is retained only for the quantities evaluated at the end of each stage. In the instantaneous-switching limit, $T_{\rm sw}\rightarrow0$, these stage-end quantities reduce to those obtained from the three-stage dynamics described in Appendix~\ref{protocol}.
	
	The matrix $\mathbf{\Sigma}_{\mathrm{th}}(t)$ is the covariance matrix of the thermal Gaussian motional state and characterizes its shape and orientation in phase space,
\begin{equation}
\begin{aligned}
	\mathbf{\Sigma}_{\mathrm{th}}(t)&=
	\begin{pmatrix}
		\sigma_{x,\mathrm{th}}^2(t) & C_{xp,\mathrm{th}}(t)\\
		C_{xp,\mathrm{th}}(t) & \sigma_{p,\mathrm{th}}^2(t)
	\end{pmatrix}.
\end{aligned}\label{shapematrix}
\end{equation}
The elements of the covariance matrix in Eq.~\eqref{shapematrix} are defined by
\begin{align}
		\sigma_{x,\mathrm{th}}^2(t)&=\left\langle [\delta x(t)]^2\right\rangle,\label{sigmax}\\
		\sigma_{p,\mathrm{th}}^2(t)&=\left\langle [\delta p(t)]^2\right\rangle,\label{sigmap}\\
		C_{xp,\mathrm{th}}(t)&=\frac{1}{2}\left\langle\delta x(t)\delta p(t)+\delta p(t)\delta x(t)\right\rangle.\label{covarianceelements}
\end{align}
Here, $\delta x(t)$ and $\delta p(t)$ describe the position and momentum fluctuations around the corresponding c.o.m. trajectory. The quantities $\sigma_{x,\mathrm{th}}^2(t)$ and $\sigma_{p,\mathrm{th}}^2(t)$ characterize the position and momentum spreads of the motional state, respectively, while $C_{xp,\mathrm{th}}(t)$ describes the correlation between the two fluctuations. In particular, a nonzero $C_{xp,\mathrm{th}}(t)$ corresponds to a Gaussian distribution tilted with respect to the $x$ and $p$ axes in phase space.
	
Finally, $\mathbf{J}$ is the symplectic matrix for the one-dimensional phase space,
\begin{equation}
	\begin{aligned}
		\mathbf{J}&=
		\begin{pmatrix}
				0 & 1\\
				-1 & 0
		\end{pmatrix}.
	\end{aligned}\label{symplectic}
\end{equation}
	
\subsection{Gaussian wave-packet dynamics under finite switching}
Considering the continuously varying magnetic-field gradient during the finite-switching process, the longitudinal Hamiltonian can be written as
\begin{align}
	H_\pm(x,p,t)&=\frac{p^2}{2m}+\frac{1}{2}m\omega_x^2(t)x^2-mA_\pm(t)x+E_\pm,\label{longham}
\end{align}
where
\begin{align}
	\omega_x^2(t)&=-\frac{\chi_\rho}{\mu_0}\eta_S^2(t),\\
	A_\pm(t)&=\left(\frac{\chi_\rho B_0}{\mu_0}-\frac{\hbar\gamma_eS_x}{m}\right)\eta_S(t),\\
	E_\pm&=-\frac{\chi_\rho mB_0^2}{2\mu_0}+\hbar\gamma_eB_0S_x+\hbar D S_x^2,
	\label{hamparams}
\end{align}
The gradient $\eta_S(t)$ follows the finite-switching sequence defined in Eqs.~(\ref{etaswitcha})--(\ref{etaswitche}), consisting of three constant-gradient stages connected by two finite-switching intervals. The time-dependent frequency $\omega_x(t)$ is determined by the instantaneous magnetic-field gradient. For the nominal finite-switching sequence considered here, $|\eta_S(t)|=\eta$ so that $\omega_x(t)=\omega_x=\eta\sqrt{-\chi_\rho/\mu_0}$ during the three constant-gradient stages, whereas $\omega_x(t)$ varies continuously during the two switching intervals. The quantity $A_\pm(t)$ gives the spin-dependent acceleration associated with the magnetic-field gradient. In the instantaneous-switching limit, $T_{\rm sw}\rightarrow0$, $A_\pm(t)$ reduces in each stage to the corresponding constant acceleration $A_\pm$ defined in Eq.~\eqref{apm}. Finally, $E_\pm$ collects the terms that are independent of the c.o.m. position. Since $E_\pm$ does not contribute to the spatial force, it does not affect the c.o.m. trajectories or the wave-packet deformation considered below.

At an arbitrary time $t$, the canonical position and momentum variables governed by the Hamiltonian in Eq.~\eqref{longham} obey
\begin{align}
	\dot{x}(t)&=\frac{p(t)}{m},\label{fulleq1}\\
	\dot{p}(t)&=-m\omega_x^2(t)x(t)+mA_\pm(t).\label{fulleq2}
\end{align}
For the two spin states, the centres of the corresponding Gaussian motional states are denoted by $x^\pm(t)$ and $p^\pm(t)$. Since the force generated by Eq.~\eqref{longham} is linear in $x$, their c.o.m. dynamics take the same form as Eqs.~\eqref{fulleq1} and \eqref{fulleq2}. These c.o.m. trajectories are propagated through the three stages and the two finite-switching intervals and determine the phase-space displacement $\Delta\boldsymbol{\chi}(t)$.

Using the position and momentum fluctuations $\delta x(t)$ and $\delta p(t)$ introduced in Eqs.~(\ref{sigmax},~\ref{sigmap}), the corresponding c.o.m. equations can be subtracted from the full equations of motion Eqs. (\ref{fulleq1}, \ref{fulleq2}), giving
\begin{align}
	\delta\dot{x}(t)&=\frac{\delta p(t)}{m},\label{deformationeq1}\\
	\delta\dot{p}(t)&=-m\omega_x^2(t)\delta x(t).\label{deformationeq2}
\end{align}
The spin-dependent acceleration $A_\pm(t)$ therefore cancels from the fluctuation dynamics. Consequently, although the two motional states follow different c.o.m. trajectories, their internal fluctuations obey the same equations of motion. The two spin components are initially prepared with the same motional Gaussian state, so that they have the same initial covariance matrix before the spin-dependent separation begins. Since the quadratic curvature $\omega_x^2(t)$ is also common to the two spin states, their covariance matrices subsequently undergo the same evolution. In phase space, the two Gaussian motional states therefore undergo the same deformation about their respective c.o.m. trajectories. Finite switching can modify their common position and momentum spreads and generate a nonzero position--momentum correlation, but these changes occur identically for the two spin states. The recombination contrast in Eq.~\eqref{gaussiancontrast} is therefore still governed primarily by the residual separation of the two wave-packet centres in phase space, $(\Delta X_3,\Delta P_3)$. However, since $C_{xp,\mathrm{th}}(\tau_3)$ does not vanish after the finite-switching evolution, the cross term $C_{xp,\mathrm{th}}(\tau_3)\Delta X_3\Delta P_3$ also contributes to the final contrast.

Using Eqs.~\eqref{sigmax}--\eqref{covarianceelements} together with Eqs.~\eqref{deformationeq1} and \eqref{deformationeq2}, the three independent elements of the covariance matrix form a closed set of coupled equations. For the position variance,
\begin{align}
	\frac{d\sigma_{x,\mathrm{th}}^2(t)}{dt}&=\frac{d}{dt}\left\langle [\delta x(t)]^2\right\rangle \notag\\
	&=\left\langle\delta\dot{x}(t)\delta x(t)+\delta x(t)\delta\dot{x}(t)\right\rangle \notag\\
	&=\frac{1}{m}\left\langle\delta p(t)\delta x(t)+\delta x(t)\delta p(t)
	\right\rangle \notag\\
	&=\frac{2}{m}C_{xp,\mathrm{th}}(t).\label{momentx}
\end{align}
Similarly, for the momentum variance,
\begin{align}
	\frac{d\sigma_{p,\mathrm{th}}^2(t)}{dt}&=\frac{d}{dt}\left\langle [\delta p(t)]^2\right\rangle \notag\\
	&=\left\langle\delta\dot{p}(t)\delta p(t)+\delta p(t)\delta\dot{p}(t)\right\rangle \notag\\
	&=-m\omega_x^2(t)\left\langle\delta x(t)\delta p(t)+\delta p(t)\delta x(t)\right\rangle \notag\\
	&=-2m\omega_x^2(t)C_{xp,\mathrm{th}}(t).\label{momentp}
\end{align}
Finally, differentiating the position--momentum correlation gives
\begin{align}
	\frac{dC_{xp,\mathrm{th}}(t)}{dt}&=\frac{1}{2}\frac{d}{dt}\left\langle\delta x(t)\delta p(t)+\delta p(t)\delta x(t)\right\rangle \notag\\
	&=\frac{1}{2}\left\langle\delta\dot{x}(t)\delta p(t)+\delta x(t)\delta\dot{p}(t)\right. \notag\\
	&\qquad\left.+\delta\dot{p}(t)\delta x(t)+\delta p(t)\delta\dot{x}(t)\right\rangle \notag\\
	&=\frac{\sigma_{p,\mathrm{th}}^2(t)}{m}-m\omega_x^2(t)\sigma_{x,\mathrm{th}}^2(t).\label{momentxp}
\end{align}

Eqs.~\eqref{momentx}--\eqref{momentxp} completely determine the covariance evolution once the initial covariance elements and the time dependence of $\omega_x(t)$ are specified, where $\omega_x^2(t)=-\chi_\rho\eta_S^2(t)/\mu_0$. These equations are applicable to both the finite-switching analysis in Sec.~\ref{4211} and the independent current-deviation analysis in Sec.~\ref{422}. In the latter case, the modified gradients defined in Eq.~\eqref{etaerror} are substituted into the piecewise switching sequence given by Eqs.~\eqref{etaswitcha}--\eqref{etaswitche}, and the same covariance evolution equations are solved numerically for each realization. The following stage-by-stage discussion focuses on the nominal finite-switching sequence, for which $\eta_1=-\eta$, $\eta_2=+\eta$, and $\eta_3=-\eta$.

We take the initial longitudinal motional state to be a displaced thermal Gaussian state centred at $x=-x_0$ with vanishing mean momentum. The displacement of the c.o.m. does not affect the covariance matrix, whose initial elements are \cite{zhou2025spin}

\begin{align}
	\sigma_{x,\mathrm{th}}^2(0)&=(2\bar n_x+1)\frac{\hbar}{2m\omega_x},
	\label{thermalinitial1}\\
	\sigma_{p,\mathrm{th}}^2(0)&=(2\bar n_x+1)\frac{\hbar m\omega_x}{2},
	\label{thermalinitial2}\\
	C_{xp,\mathrm{th}}(0)&=0.\label{thermalinitial3}
\end{align}
Here, $\bar n_x$ denotes the mean thermal occupation number of the longitudinal c.o.m. motional mode along the $x$ direction and therefore characterizes only the thermal occupation associated with the longitudinal motion considered here. The frequency $\omega_x$ appearing in Eqs.~\eqref{thermalinitial1} and \eqref{thermalinitial2} denotes the longitudinal trapping frequency of the nominal constant-gradient stages. Since $|\eta_1|=|\eta_2|=|\eta_3|=\eta$ for the nominal protocol, it is given by $\omega_x=\eta\sqrt{-\chi_\rho/\mu_0}$. It should be distinguished from the time-dependent frequency $\omega_x(t)$ during the finite-switching intervals. The initial condition $C_{xp,\mathrm{th}}(0)=0$ corresponds to a Gaussian motional state whose principal axes are initially aligned with the position and momentum axes in phase space.

The initial thermal variances further satisfy
\begin{equation}
	\frac{\sigma_{p,\mathrm{th}}^2(0)}{m}=m\omega_x^2\sigma_{x,\mathrm{th}}^2(0).\label{initialbalance}
\end{equation}
Thus, the initial covariance matrix is matched to the harmonic curvature associated with $\omega_x$. Since the fluctuation dynamics depend on $\omega_x^2(t)\propto\eta_S^2(t)$, they depend on the magnitude of the magnetic-field gradient rather than its sign. Consequently, for the nominal values $\eta_1=-\eta$, $\eta_2=+\eta$, and $\eta_3=-\eta$, the three constant-gradient stages have the same harmonic curvature $\omega_x^2$, whereas the curvature varies only during the two finite-switching intervals.

Starting from the initial covariance elements in Eqs.~\eqref{thermalinitial1}--\eqref{thermalinitial3}, Eqs.~\eqref{momentx}--\eqref{momentxp} can therefore be solved sequentially through the three constant-gradient stages and the two finite-switching intervals. The covariance elements obtained at the end of each interval serve as the initial conditions for the subsequent evolution. We discuss this propagation in the order defined by Eqs.~(\ref{etaswitcha})--(\ref{etaswitche}).

\begin{itemize}
	
\item \textit{Stage 1: $0\leq t\leq\tau_1$.}
	
During Stage 1, $\eta_S(t)=\eta_1=-\eta$ for the nominal protocol, and hence $\omega_x(t)=\omega_x$. Since the initial covariance matrix satisfies Eq.~\eqref{initialbalance} and $C_{xp,\mathrm{th}}(0)=0$, Eqs.~\eqref{momentx}--\eqref{momentxp} give
\begin{equation}
		\frac{d\sigma_{x,\mathrm{th}}^2}{dt}=\frac{d\sigma_{p,\mathrm{th}}^2}{dt}=\frac{dC_{xp,\mathrm{th}}}{dt}=0.
\end{equation}
	The covariance matrix therefore remains unchanged throughout Stage 1, and at $t=\tau_1$,
\begin{align}
	\sigma_{x,\mathrm{th}}^2(\tau_1)&=(2\bar n_x+1)\frac{\hbar}{2m\omega_x},\label{sigmaxth1}\\
		\sigma_{p,\mathrm{th}}^2(\tau_1)&=(2\bar n_x+1)\frac{\hbar m\omega_x}{2},\label{sigmapth1}\\
		C_{xp,\mathrm{th}}(\tau_1)&=0.\label{cxpth1}
\end{align}
Thus, although the two spin states follow different c.o.m. trajectories, their Gaussian wave-packet shapes remain stationary during Stage 1.
	
	\item \textit{First finite-switching interval: $\tau_1<t\leq\tau_1+T_{\rm sw}$.}
	
Introducing the local time $u_1=t-\tau_1$, with $0<u_1\leq T_{\rm sw}$, Eq.~\eqref{etaswitchb} gives $\eta_S(u_1)=\eta_1+(\eta_2-\eta_1)u_1/T_{\rm sw}$. For the nominal values $\eta_1=-\eta$ and $\eta_2=+\eta$, the corresponding trapping frequency becomes
\begin{equation}
	\omega_x^2(u_1)=\omega_x^2\left(1-\frac{2u_1}{T_{\rm sw}}\right)^2.\label{omegasw1}
\end{equation}
	The harmonic curvature therefore decreases continuously from $\omega_x^2$ to zero and subsequently increases back to $\omega_x^2$ during the switching interval.
	
	The covariance elements are obtained by solving
\begin{align}
	\frac{d\sigma_{x,\mathrm{th}}^2(u_1)}{du_1}&=\frac{2}{m}C_{xp,\mathrm{th}}(u_1),\\
	\frac{d\sigma_{p,\mathrm{th}}^2(u_1)}{du_1}&=-2m\omega_x^2\left(1-\frac{2u_1}{T_{\rm sw}}
	\right)^2C_{xp,\mathrm{th}}(u_1),\\
	\frac{dC_{xp,\mathrm{th}}(u_1)}{du_1}&=\frac{\sigma_{p,\mathrm{th}}^2(u_1)}{m}-m\omega_x^2
	\left(1-\frac{2u_1}{T_{\rm sw}}\right)^2\sigma_{x,\mathrm{th}}^2(u_1).\label{firstswitch}
\end{align}
	In Eqs.~\eqref{firstswitch}, $\sigma_{x,\mathrm{th}}^2(u_1)$, $\sigma_{p,\mathrm{th}}^2(u_1)$, and $C_{xp,\mathrm{th}}(u_1)$ are all functions of the local switching time $u_1$. Their initial values at $u_1=0$ are given by the corresponding covariance elements at the end of Stage 1, namely $\sigma_{x,\mathrm{th}}^2(\tau_1)$, $\sigma_{p,\mathrm{th}}^2(\tau_1)$, and $C_{xp,\mathrm{th}}(\tau_1)$, as given in Eqs.~\eqref{sigmaxth1}--\eqref{cxpth1}. The time dependence of $\omega_x(u_1)$ breaks the stationary balance in Eq.~\eqref{initialbalance}, so that a nonzero $C_{xp,\mathrm{th}}(u_1)$ is generated. This correlation subsequently drives the evolution of both $\sigma_{x,\mathrm{th}}^2(u_1)$ and $\sigma_{p,\mathrm{th}}^2(u_1)$, corresponding to a deformation of the Gaussian distribution in phase space.
	
The coupled covariance equations are solved numerically over $0\leq u_1\leq T_{\rm sw}$. The resulting values of $\sigma_{x,\mathrm{th}}^2$, $\sigma_{p,\mathrm{th}}^2$, and $C_{xp,\mathrm{th}}$ at $t=\tau_1+T_{\rm sw}$ characterize the Gaussian state at the end of the first switching interval and are subsequently used as the initial conditions for Stage 2.

\item \textit{Stage 2: $\tau_1+T_{\rm sw}<t\leq\tau_2$.}

During Stage 2, $\eta_S(t)=\eta_2=+\eta$ for the nominal protocol, and therefore $\omega_x(t)=\omega_x$ is again constant. Since $|\eta_2|=|\eta_1|=\eta$, the harmonic curvature is the same as that in Stage 1. The covariance elements obtained at the end of the first switching interval are used as the initial conditions for this stage. Although the covariance matrix has generally been deformed during the first switching interval, with $C_{xp,\mathrm{th}}\neq0$ and modified position and momentum variances, the covariance elements continue to evolve according to Eqs.~\eqref{momentx}--\eqref{momentxp}.

Numerical integration over Stage 2 gives the covariance elements $\sigma_{x,\mathrm{th}}^2(\tau_2)$, $\sigma_{p,\mathrm{th}}^2(\tau_2)$, and $C_{xp,\mathrm{th}}(\tau_2)$ at the end of the stage. These values provide the initial conditions for the second finite-switching interval.
	
\item \textit{Second finite-switching interval: $\tau_2<t\leq\tau_2+T_{\rm sw}$.}

Introducing the local time $u_2=t-\tau_2$, with $0<u_2\leq T_{\rm sw}$, Eq.~\eqref{etaswitchd} gives $\eta_S(u_2)=\eta_2+(\eta_3-\eta_2)u_2/T_{\rm sw}$. For the nominal values $\eta_2=+\eta$ and $\eta_3=-\eta$, the corresponding trapping frequency becomes
\begin{equation}
	\omega_x^2(u_2)=\omega_x^2\left(1-\frac{2u_2}{T_{\rm sw}}\right)^2.\label{omegasw2}
\end{equation}
The covariance elements during the second switching interval therefore satisfy
\begin{align}
	\frac{d\sigma_{x,\mathrm{th}}^2(u_2)}{du_2}&=\frac{2}{m}C_{xp,\mathrm{th}}(u_2),\\
	\frac{d\sigma_{p,\mathrm{th}}^2(u_2)}{du_2}&=-2m\omega_x^2
	\left(1-\frac{2u_2}{T_{\rm sw}}\right)^2C_{xp,\mathrm{th}}(u_2),\\
	\frac{dC_{xp,\mathrm{th}}(u_2)}{du_2}&=\frac{\sigma_{p,\mathrm{th}}^2(u_2)}{m}
	-m\omega_x^2\left(1-\frac{2u_2}{T_{\rm sw}}\right)^2\sigma_{x,\mathrm{th}}^2(u_2).\label{secondswitch}
\end{align}
The initial conditions at $u_2=0$ are given by the covariance elements obtained at the end of Stage 2,
\begin{align}
	\sigma_{x,\mathrm{th}}^2(u_2=0)
	&=\sigma_{x,\mathrm{th}}^2(\tau_2),\\
	\sigma_{p,\mathrm{th}}^2(u_2=0)
	&=\sigma_{p,\mathrm{th}}^2(\tau_2),\\
	C_{xp,\mathrm{th}}(u_2=0)
	&=C_{xp,\mathrm{th}}(\tau_2).
\end{align}
For the nominal symmetric sequence, the two switching intervals have the same time dependence of $\omega_x^2$; however, their effects on the covariance matrix are generally different because the covariance entering the second switching interval has already evolved through the first switching interval and Stage 2. The coupled covariance equations are therefore solved numerically over $0\leq u_2\leq T_{\rm sw}$. The resulting covariance elements at $t=\tau_2+T_{\rm sw}$ are subsequently used as the initial conditions for Stage 3.
	
	\item \textit{Stage 3: $\tau_2+T_{\rm sw}<t\leq\tau_3$.}
	
	During Stage 3, $\eta_S(t)=\eta_3=-\eta$ for the nominal protocol, and therefore $\omega_x(t)=\omega_x$ is again constant. The covariance elements obtained at the end of the second finite-switching interval,
	$\sigma_{x,\mathrm{th}}^2(\tau_2+T_{\rm sw})$,
	$\sigma_{p,\mathrm{th}}^2(\tau_2+T_{\rm sw})$, and
	$C_{xp,\mathrm{th}}(\tau_2+T_{\rm sw})$,
	are used as the initial conditions for this stage. As in Stage 2, the covariance matrix is generally no longer stationary because it has already been modified by the preceding switching and constant-gradient evolution. The three covariance elements therefore continue to evolve according to Eqs.~\eqref{momentx}--\eqref{momentxp}.
	
	These equations are numerically integrated over Stage 3 up to the final time $t=\tau_3$, yielding the final covariance elements $\sigma_{x,\mathrm{th}}^2(\tau_3)$, $\sigma_{p,\mathrm{th}}^2(\tau_3)$, and $C_{xp,\mathrm{th}}(\tau_3)$. They form the final covariance matrix
\begin{equation}
	\mathbf{\Sigma}_{\mathrm{th}}(\tau_3)=
	\begin{pmatrix}
		\sigma_{x,\mathrm{th}}^2(\tau_3) & C_{xp,\mathrm{th}}(\tau_3)\\
		C_{xp,\mathrm{th}}(\tau_3) & \sigma_{p,\mathrm{th}}^2(\tau_3)
	\end{pmatrix},\label{finalcovariance}
\end{equation}
	which enters directly into the recombination contrast in Eq.~\eqref{gaussiancontrast}.
\end{itemize}

It is useful first to consider the instantaneous-switching limit, $T_{\rm sw}\rightarrow0$. For the nominal sequence, $|\eta_1|=|\eta_2|=|\eta_3|=\eta$. In this limit, the gradient therefore changes discontinuously between the three nominal values while its magnitude remains constant, so that $\omega_x(t)=\omega_x$ throughout the sequence. Since the initial covariance matrix is matched to this constant trapping frequency, the covariance elements remain unchanged throughout the evolution,
\begin{align}
	\sigma_{x,\mathrm{th}}^2(\tau_3)&=\sigma_{x,\mathrm{th}}^2(0)=(2\bar n_x+1)\frac{\hbar}{2m\omega_x},\\
	\sigma_{p,\mathrm{th}}^2(\tau_3)&=\sigma_{p,\mathrm{th}}^2(0)=(2\bar n_x+1)\frac{\hbar m\omega_x}{2},
\end{align}
\begin{align}
	C_{xp,\mathrm{th}}(\tau_3)=0.
\end{align}
Substituting these expressions into Eq.~\eqref{gaussiancontrast}, the recombination contrast reduces to
\begin{align}
	C(\bar n_x,0)&=\exp\left[-\frac{2\bar n_x+1}{8}\left(\frac{\Delta X_3^2}{\sigma_{x,0}^2}
	+\frac{\Delta P_3^2}{\sigma_{p,0}^2}\right)\right]
	\notag\\
	&=\exp\left[-\frac{2\bar n_x+1}{8}\left(\frac{\Delta X_3^2}{\sigma_{x,0}^2}+
	\frac{\Delta\dot{X}_3^2}{\sigma_{v,0}^2}\right)\right],
	\label{thermalinstant}
\end{align}
where $\Delta P_3=m\Delta\dot{X}_3$ and $\Delta\dot{X}_3=\dot{X}_3^+-\dot{X}_3^-$. Equation~\eqref{thermalinstant} shows that a finite thermal occupation enhances the contrast-loss exponent by the factor $2\bar n_x+1$. In the further limit $\bar n_x=0$, the expression reduces to the standard Humpty--Dumpty recombination contrast given in Ref.~\cite{PhysRevB.49.15122}.

For a finite switching model considered here, the covariance evolution described above generally produces changes in $\sigma_{x,\mathrm{th}}^2(t)$ and $\sigma_{p,\mathrm{th}}^2(t)$ together with a nonzero position--momentum correlation $C_{xp,\mathrm{th}}(t)$. Since Eqs.~\eqref{momentx}--\eqref{momentxp} are linear and homogeneous in the covariance elements, the common thermal factor contained in the initial covariance matrix is preserved throughout the evolution,
\begin{equation}
	\mathbf{\Sigma}_{\mathrm{th}}(t)=(2\bar n_x+1)\mathbf{\Sigma}_0(t),\label{thermalscaling}
\end{equation}

where $\mathbf{\Sigma}_0(t)$ denotes the covariance matrix obtained for $\bar n_x=0$ under the same finite-switching sequence. Substituting Eq.~\eqref{thermalscaling} into the general Gaussian overlap expression in Eq.~\eqref{gaussiancontrast} directly yields Eqs.~\eqref{thermalcontrast} and \eqref{closureexponent} in the main text. The thermal occupation enters through the factor $2\bar n_x+1$, while the finite-switching dynamics affect both the residual phase-space displacement and the evolved covariance matrix. In particular, finite switching can generate a nonzero position--momentum covariance $C_{xp,0}(\tau_3)$, which contributes to the contrast-loss exponent through the cross term in Eq.~\eqref{closureexponent}.

\color{black}{
\section{ESTIMATE OF TRANSLATIONAL HEATING}
In this section, we estimate the motional heating of the diamond particle along the $z$ direction and derive the corresponding current-stability requirement for the levitation wires. Analogously to Eq.~(\ref{deltaeta}), we assume that the levitation current
$I_{\mathrm L}$ is subject to white-noise fluctuations,
\begin{equation}
	\xi_{\mathrm L}(t) \equiv \frac{\delta\eta_{\mathrm L}(t)}{\eta_{\mathrm L}}
	= \frac{\delta I_{\mathrm L}(t)}{I_{\mathrm L}}. \label{deltaetal}
\end{equation}
We model $\xi_{\mathrm L}(t)$ as a zero-mean stationary white-noise
process satisfying 

\begin{align}
	\left\langle \xi_{\mathrm L}(t) \right\rangle &= 0, \qquad \left\langle \xi_{\mathrm L}(t)\xi_{\mathrm L}(t') \right\rangle = S_{\xi_{\mathrm L}}\delta(t-t'), \label{xiL}
\end{align}
where $S_{\xi_{\mathrm L}}$ denotes the power spectral density of the fractional levitation-current noise. Substituting Eq.~(\ref{deltaetal}) into Eq.~(\ref{zeom}) and imposing the equilibrium condition between the diamagnetic force and gravity, one obtains the perturbed levitation position,
\begin{equation}
	\begin{aligned}
		z_{\mathrm L}'(t)&=
		\frac{z_{\mathrm L}}{\left[1+\xi_{\mathrm L}(t)\right]^2}\\
		&\approx z_{\mathrm L}\left[1-2\xi_{\mathrm L}(t)\right]\\
		&=	z_{\mathrm L}+\delta z_{\mathrm L}(t),
	\end{aligned}\label{zlprime}
\end{equation}
where $z_{\mathrm L}'(t)$ is the perturbed levitation height, $z_{\mathrm L}$ is the unperturbed levitation height defined in Eq.~(\ref{levitationheight}), and
\begin{equation}
	\delta z_{\mathrm L}(t)=-2z_{\mathrm L}\xi_{\mathrm L}(t).\label{deltaZL}
\end{equation}
Near the unperturbed levitation position, the motion along the $z$ direction can be approximated as that of a harmonic oscillator with trapping frequency $\omega_z$, as given in Eq.~(\ref{omegaz}). A fluctuation of the levitation position therefore acts as a stochastic driving force on the particle. To first order in the current fluctuation, the force fluctuation associated with the displacement of the trap centre is
\begin{equation}
	\delta F_{\mathrm L}(t)=m\omega_z^2\delta z_{\mathrm L}(t).\label{deltaFL}
\end{equation}
Using Eq.~(\ref{deltaFL}), the corresponding two-sided force-noise power spectral density is defined as\footnote{A two-sided power spectral density is defined on the full Fourier-frequency axis, $\omega\in(-\infty,\infty)$. The negative-frequency part is a mathematical consequence of the Fourier representation and does not correspond to a separate physical oscillation. For a real classical stationary noise process, the spectrum is symmetric, $S_{F_{\mathrm L}}(-\omega)=S_{F_{\mathrm L}}(\omega)$.} \cite{brownnutt2015ion} 

\begin{equation}
	\begin{aligned}
		S_{F_L}(\omega)&=
		\int_{-\infty}^{\infty}\mathrm{d}t~\text{e}^{i\omega(t-t')}\left\langle\delta F_L(t)\delta F_L(t')\right\rangle\\
		&=
		m^2\omega_z^4\int_{-\infty}^{\infty}\mathrm{d}t~\text{e}^{i\omega(t-t')}\left\langle\delta z_L(t)\delta z_L(t')\right\rangle .\label{forcePSDposition}
	\end{aligned}
\end{equation}
where $t'$ is an arbitrary fixed reference time. For a stationary noise process, the correlation function depends only on the time difference $t-t'$, and the resulting spectrum is independent of the choice of $t'$.

Using Eqs.~(\ref{xiL}) and~(\ref{deltaZL}), the correlation function of the levitation-position fluctuations is
\begin{equation}
	\left\langle\delta z_{\mathrm L}(t)\delta z_{\mathrm L}(t')\right\rangle
	=4z_{\mathrm L}^2S_{\xi_{\mathrm L}}\delta(t-t').\label{positioncorrL}
\end{equation}
Substituting Eq.~(\ref{positioncorrL}) into Eq.~(\ref{forcePSDposition}) gives
\begin{align}
	S_{F_{\mathrm L}}(\omega_z)&=4m^2\omega_z^4z_{\mathrm L}^2S_{\xi_{\mathrm L}}\int_{-\infty}^{\infty}\text{d}t~\text{e}^{i\omega_z(t-t')}\delta(t-t')
	\nonumber\\
	&=4m^2\omega_z^4z_{\mathrm L}^2S_{\xi_{\mathrm L}}.\label{forcePSDwhite}
\end{align}
For a harmonic oscillator driven by a stationary fluctuating force, the growth rate of the mean motional occupation number is determined by the force-noise power spectral density at the trapping frequency. With the two-sided power spectral density convention adopted in Eq.~(\ref{forcePSDposition}), the heating rate is ~\cite{turchette2000heating}
\begin{equation}
	\begin{aligned}
	\dot{\bar n}_z&=\frac{S_{F_{\mathrm L}}(\omega_z)}{2m\hbar\omega_z}\\
	&=\frac{2m\omega_z^3z_{\mathrm L}^2}{\hbar}S_{\xi_{\mathrm L}}.
	\end{aligned}\label{generalHeatingRate}
\end{equation}
The increase in the mean motional occupation number over the total
interferometric duration $\tau_3=T_1+T_2+T_3$ is therefore
\begin{equation}
	\Delta\bar n_z=\dot{\bar n}_z\tau_3=
	\frac{2m\omega_z^3z_{\mathrm L}^2\tau_3}{\hbar}S_{\xi_{\mathrm L}}.\label{addedphononL}
\end{equation}
Requiring the current-noise-induced heating to add less than one
motional quantum, $\Delta\bar n_z<1$, yields the following
requirement on the fractional current-noise amplitude,
\begin{equation}
	\sqrt{S_{\xi_{\mathrm L}}}<\sqrt{\frac{\hbar}{2m\omega_z^3z_{\mathrm L}^2\tau_3}}.
	\label{levitationNoiseRequirement}
\end{equation}
Similar to Eq. (\ref{xieff}), for a 0.1 second duration, we estimate the root-mean-square levitation current fluctuation $\xi^\text{eff}_\text{L}$ as,
\begin{equation}
	\xi^\text{eff}_\text{L}=\sqrt{\frac{S_{\xi_\text{L}}}{0.1}}.
	\label{xileff}
\end{equation}
The estimated $\sqrt{S_{\xi_\text{L}}}$ and $\xi^\text{eff}_\text{L}$ is given in Tab. \ref{tablev} for the three masses considered in this work.
\begin{table}[htbp]
	\centering
	\caption{Current-noise requirements for different particle masses.}
	\label{tablev}
	\begin{tabular}{|c|c|c|c|}
		\hline\hline\noalign{\vskip 2pt}
		$m~(kg)=$
		& $10^{-19}$
		& $10^{-18}$
		& $10^{-17}$ \\
		\hline\noalign{\vskip 2pt}
		$\sqrt{S_{\xi_{\mathrm L}}}~(\text{Hz}^{-1/2})\leq$
		&$5.77\times 10^{-7}$
		&$1.82\times 10^{-7}$
		&$5.77\times 10^{-8}$\\
		\hline\noalign{\vskip 2pt}
		$\displaystyle
		\xi^\text{eff}_\text{L}\leq$
		& $1.82\times 10^{-6}$
		& $5.77\times 10^{-7}$
		& $1.82\times 10^{-7}$\\
		\hline\hline
	\end{tabular}
\end{table}
These results show that the stability requirement on the levitation current becomes more stringent as the mass of the diamond particle increases. The values reported in the table therefore impose challenging stability requirements on the current supplied to the levitation wires.
}

\color{black}
\end{appendices}

\pagebreak

\bibliographystyle{apsrev4-1}
\bibliography{1b.bib}

\end{document}